\documentclass[traditabstract]{aa}

\usepackage{txfonts}

\usepackage{color}

\usepackage[pdftex]{graphicx}
\usepackage{rotating}

\usepackage{amsfonts}

\usepackage{subfig}

\usepackage{aalongtable}

\usepackage{breqn}

\usepackage{natbib}
\bibpunct{(}{)}{;}{a}{}{,} 

\newcommand{\rhk}{$\log R'_{HK}$~}
\newcommand{\iha}{$\log I_{\mathrm{H}\alpha}$~}

\newcommand{\ha}{H$\alpha$~}

\begin{document}

\title{On the long-term correlation between the flux in the \ion{Ca}{ii} H \& K and H$\alpha$ lines for FGK stars}
\subtitle{}

\author{J. Gomes da Silva\inst{1,2}
	\and N.C. Santos\inst{1,2}
	\and I. Boisse\inst{1}
	\and X. Dumusque\inst{1,3}
	\and C. Lovis\inst{3}
	}

\institute{Centro de Astrof\'isica, Universidade do Porto, Rua das Estrelas, 4150-762 Porto, Portugal \\ \email{Joao.Silva@astro.up.pt}
	\and Departamento de F\'isica e Astronomia, Faculdade de Ci\^encias, Universidade do Porto, Portugal
	\and Observatoire de Gen\`eve, Universit\'e de Gen\`eve, 51 ch. des Maillettes, CH-1290 Versoix, Switzerland
	}

\abstract{
The re-emission in the cores of the \ion{Ca}{ii} H \& K and H$\alpha$ lines, 
are well known proxies of stellar activity.
However, these activity indices probe different activity phenomena, the first being more sensitive to plage variation, while the other one being more sensitive to filaments.
In this paper we study the long-term correlation between $\log R'_{HK}$ and $\log I_{H\alpha}$, two indices based on the \ion{Ca}{ii} H \& K and H$\alpha$ lines respectively, for a sample of 271 FGK stars using measurements obtained over a $\sim$9 year time span.
Because stellar activity is one of the main obstacles to the detection of low-mass and long-period planets, understanding further this activity index correlation can give us some hints about the optimal target to focus on, and ways to correct for these activity effects.

We found a great variety of long-term correlations between $\log R'_{HK}$ and $\log I_{H\alpha}$.
Around 20\% of our sample has strong positive correlation between the indices while about 3\% show strong negative correlation.
These fractions are compatible with those found for the case of early-M dwarfs.
Stars exhibiting a positive correlation have a tendency to be more active when compared to the median of the sample, while stars showing a negative correlation are more present among higher metallicity stars.

There is also a tendency for the positively correlated stars to be more present among the coolest stars, a result which is probably due to the activity level effect on the correlation.
Activity level and metallicity seem therefore to be playing a role on the correlation between $\log R'_{HK}$ and $\log I_{H\alpha}$.
Possible explanations based on the influence of filaments for the diversity in the correlations between these indices are discussed in this paper.
As a parallel result, we show a way to estimate the effective temperature of FGK dwarfs exhibiting a low activity level by using the H$\alpha$ index.}

\date{Received date / Accepted date}

\maketitle

\section{Introduction}\label{sec:intro}
Stellar activity is one of the main limitations to the detection of low-mass and/or long-period planets using the radial-velocity method \citep[e.g.][]{saar1997a,santos2000,queloz2001,boisse2009,boisse2011,dumusque2011b,lovis2011,gomesdasilva2012}.
Fortunately, the radial-velocity noise induced by these effects can, in some cases, be corrected for example if the activity is simultaneously measured using activity indices \citep[e.g.][]{dumusque2011a,dumusque2012}.
Therefore, understanding the behaviour of activity indices and their relation with radial-velocity is vital to reduce the impact of activity in radial-velocity measurements and thus improve its sensitivity to planetary signals.

The re-emission in the \ion{Ca}{ii} H \& K lines are widely used proxies of activity induced signals in radial-velocity measurements.
However, for solar-type stars, the relation between this index and H$\alpha$ is not well understood.
Since these two activity indices are affected by different activity phenomena in different ways (the emission in the centre of the \ion{Ca}{ii} and H$\alpha$ lines are not formed at the same temperature in the chromosphere), understanding their relationship and differences might bring new insights not only to stellar physics but also to the detection and characterisation of extrasolar planets.

It is known that there is a long-term correlation between the emission in the \ion{Ca}{ii} H \& K and H$\alpha$ lines that follow the Sun's 11-year activity cycle \citep{livingston2007}.
Other authors have suggested that the correlation is also present in other stars \citep{giampapa1989,robinson1990,strassmeier1990,pasquini1991,montes1995}.
However, when \citet{cincunegui2007b} measured simultaneously the flux in the two lines for a sample of 109 southern FGK and M stars, they found a large scatter in correlations, from very strong positive correlations to negative ones.
They also suggested that the mean values of the flux in the \ion{Ca}{ii} and H$\alpha$ lines are correlated due to the effect of stellar colour on both fluxes.

\citet{meunier2009} studied the contribution of plages and filaments to the emission in \ion{Ca}{ii} and H$\alpha$ lines during a solar cycle.
In their work, plages contributes to an increase in emission in both fluxes while filaments increases absorption in H$\alpha$ only.
They found that the contribution of filaments to H$\alpha$ can be responsible for the decrease in the correlation coefficient between the two fluxes depending on their spatial distribution and contrast compared to those of plages.
They also noted that at higher activity levels (e.g. cycle maxima), the filament filling factor saturates and the correlation between the two fluxes increases.
Other factors contributing to a decrease in the measured correlation can be the time-span of observations, cycle phase at which they are measured, and stellar inclination angle.
For example, if the time-span is less than the cycle period (or the activity range is not well spanned) the correlation will probably be underestimated.

\citet{santos2010} studied the long-term activity of 8 FGK stars using the \ion{Ca}{ii} H \& K based $S_{MW}$ and H$\alpha$ indices and found a general long-term correlation between the two.
However their sample was not large enough to have any statistical significance.
\citet{gomesdasilva2011} expanded the comparison between these two activity sensitive lines to early-M dwarfs.
Similarly to \citet{cincunegui2007b} they detected a large variety of correlation coefficients, including anti-correlations for the least active stars in their sample.
The most active stars were all, however, positively correlated.
They also found hints that in some cases the H$\alpha$ index was following an "anti-cycle" relative to their $S$-index, i.e., the maxima and minima measured in the two indices were anti-correlated.
However, their time-span was not long enough to detect full cycles and confirm this effect.

In this paper, we analyse the behaviour of the flux in \ion{Ca}{ii} H \& K and H$\alpha$ lines in FGK stars via two activity indices corrected for the effects of photospheric flux.
We describe our sample and data in Sect. \ref{sec:sample}.
The activity indices derivation, statistics, correlations between mean values, and activity cycle detectability are presented in Sect. \ref{sec:indices} and Appendix \ref{a1}.
The correlations between the two indices are discussed in Sect. \ref{sec:results1}.
The distribution of the correlations in mean values of activity are discussed in Sect. \ref{sec:results2}.
The effects of metallicity on the correlation are studied in Sect. \ref{sec:results3}, and the distribution of the correlations in effective temperature is presented in Sect. \ref{sec:teff}.
We discuss possible causes for the existence of positive correlations and anti-correlations, and compare our results with those found for early-M dwarfs in Sect. \ref{discussion}, and finally conclude in Sect. \ref{conclusions}.
A possible use of the H$\alpha$ index to estimate the effective temperature of low activity level FGK dwarfs is proposed in Appendix \ref{a3}.

\section{Sample and data} \label{sec:sample}
The sample comes from the $\sim$400 FGK stars HARPS (spectral resolution = 115 000) high-precision sample already used by \citet{lovis2011} to study the long-term activity of FGK stars and its effect on the measurement of precise radial velocities.
A description of the sample is presented in their paper.
The spectra used in this work were obtained between February 2003 and February 2012.
We used effective temperature, metallicity, and surface gravity that were already calculated for this sample by \citet{sousa2008}.
Absolute magnitude and luminosity were both obtained from the Hipparcos catalogue.

We selected only spectra with $S/N \geq 100$ at spectral order 56 ($\sim$5870 \AA), and nightly averaged our measurements.
Only stars with 10 or more nights of observations were selected. 
Then, we selected just the Main Sequence (MS) stars as in \citet{lovis2011}: we fitted a straight line through the H-R diagram and then excluded all stars with luminosity greater than +0.25 dex above that line.

We ended up with 271 MS stars with a median time span of $\sim$7 years that we used for the rest of this work.
This sample is comprised of 11,432 data points, with a median of 23 nights of observations per star (and a maximum of 279).
The sample ranges in spectral type from F8 to K6, in effective temperature from 4595 to 6276 K, and in metallicity from $-0.84$ to $+0.39$ dex.

\section{The activity indices}\label{sec:indices}
The $\log R'_{HK}$ index, which is already corrected for the photospheric flux \citep{noyes1984}, and respective errors were directly obtained from the HARPS DRS.
This index is based on the $S$-index which is calculated as the sum of the flux in two 0.6 \AA~bands centered at the calcium H (3968.47 \AA) and K (3933.66 \AA) lines divided by two 20 \AA~reference bands centered at 3900 and 4000 \AA~\citep[see e.g.][]{boisse2009}.

The H$\alpha$ index and errors were calculated as in \citet{gomesdasilva2011}.
We used a 1.6 \AA~band centered at 6562.808 \AA~and divided the flux in the central line by the flux in two reference bands of 10.75 and 8.75 \AA~centered at 6550.87 and 6580.31 \AA, respectively.
The flux errors were calculated as the photon noise in the line core, $\sqrt{N}$, where $N$ is the number of photons in the band.
The activity indices errors were obtained via error propagation.
The calibration of H${\alpha}$ for the effects of photospheric flux is presented in Appendix \ref{a1} and results in the $I_{H\alpha}$ index.

\subsection{Statistics of the $\log R'_{HK}$ index}
Our sample, which is biased towards inactive stars in order to increase the chances of finding low-mass planets, has a median $\log R'_{HK}$ of $-4.948$ and a mean of $-4.923$.
In this 271-star sample, only 22 (around 8\%) are considered active stars, with $\log R'_{HK} \geq -4.75$, lying on the higher activity region above the "Vaughan-Preston gap" \citep{vaughan1980}.

The star with the highest activity level is HD224789, with $\log R'_{HK} = -4.433$ and the most inactive star is HD181433 with $\log R'_{HK} = -5.144$.
The median of the errors obtained for the $\log R'_{HK}$ index is 0.003, or in relative terms, 0.06\% around the mean.
In terms of variability, the median standard deviation of the sample is 0.0154 (0.3\% around the mean), with HD177758 being the least variable star with $\sigma(\log R'_{HK}) = 0.0035$ (0.07\% around the mean) and HD7199 the star that varies the most with $\sigma(\log R'_{HK}) = 0.08$ (1.6\% around the mean).

\subsection{Statistics of the $\log I_{H\alpha}$ index}
In terms of $\log I_{H\alpha}$, our sample has a median value of $-1.7129$ and a mean of $-1.7118$.
The star with the highest $\log I_{H\alpha}$ mean value is HD85119, with an activity level of $-1.6562$ and the most inactive star is HD82516 with $\log I_{H\alpha} = -1.7299$.
The median of the errors we obtained for the $\log I_{H\alpha}$ index is 0.0002, or in relative terms, 0.01\% around the mean.
As stated before, we are only considering photon noise as a source of errors, and since the H$\alpha$ line is in a brighter area of the spectrum compared to the \ion{Ca}{ii} H \& K lines, we expect the photon noise to be lower for $I_{H\alpha}$ than for $R'_{HK}$.
In terms of variability, the median standard deviation of the sample is 0.0019 (0.11\% around the mean), with HD74014 being the least variable star with $\sigma(\log I_{H\alpha}) = 0.0008$ (0.05\% around the mean) and HD224789 the star that varies the most with $\sigma(\log I_{H\alpha}) = 0.0063$ (0.4\% around the mean).

From these simple statistics we can see that the $\log R'_{HK}$ is more sensitive to activity variations than $\log I_{H\alpha}$.
While $\log R'_{HK}$ has a median standard deviation of 0.3\% of the mean, $\log I_{H\alpha}$ only has a median standard deviation of 0.1\% of the mean, which means that $\log R'_{HK}$ will have a more noticeable variation.

\subsection{Mean activity level correlations}

\begin{figure}[tbp]
\begin{center}
\resizebox{\hsize}{!}{\includegraphics{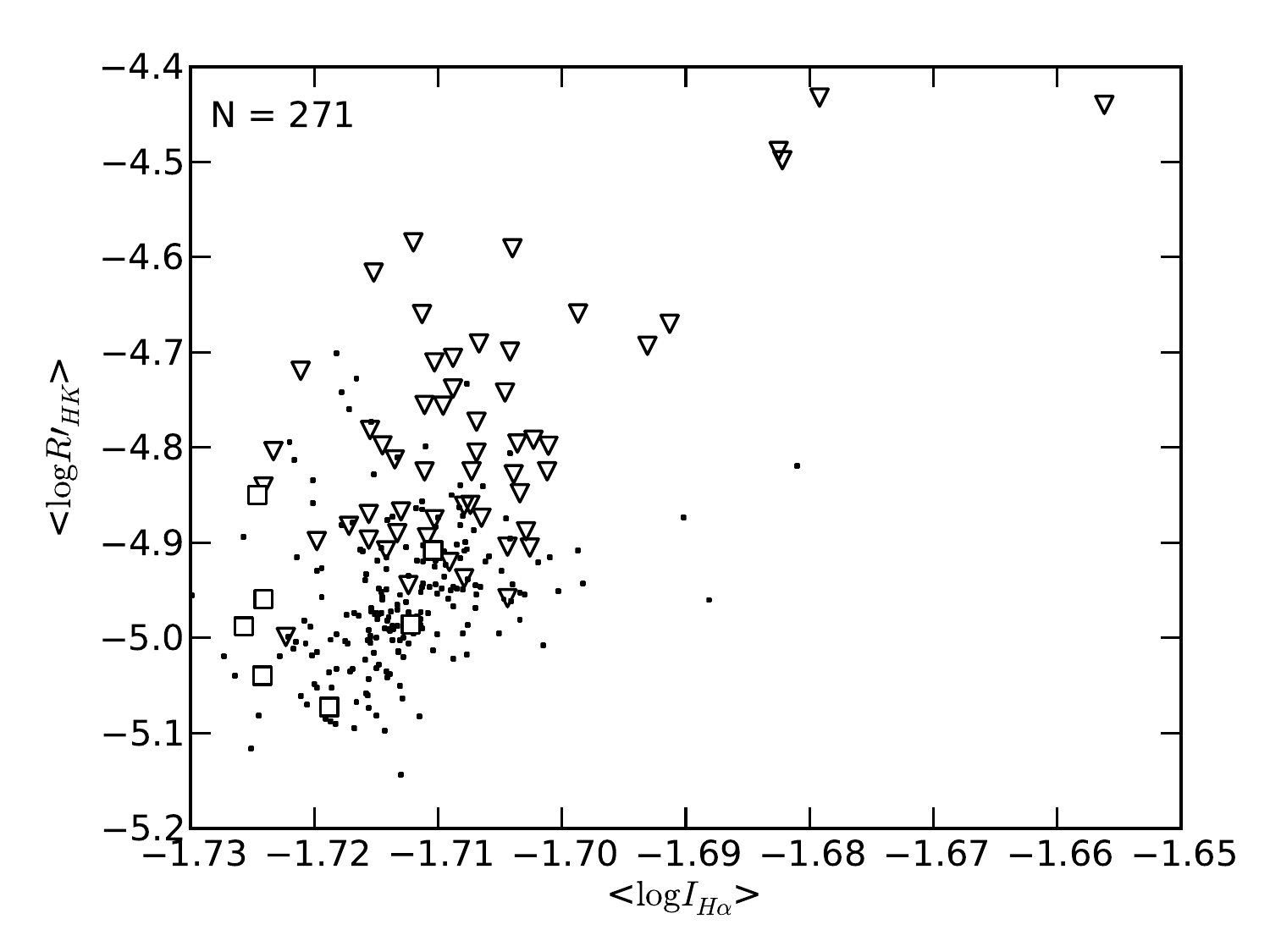}}
\resizebox{\hsize}{!}{\includegraphics{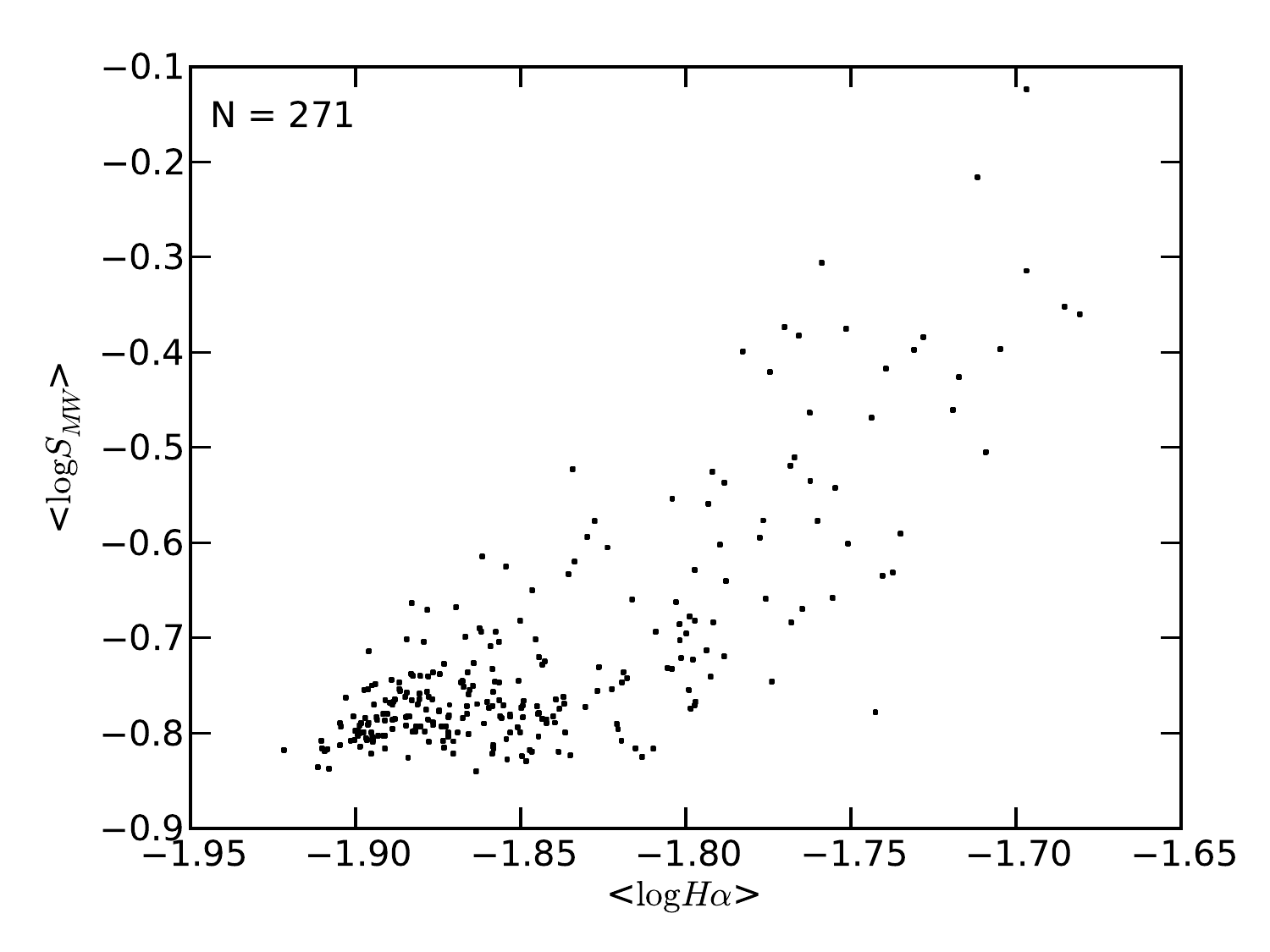}}
\caption{\textit{Upper panel:} Relationship between $\log R'_{HK}$ and $\log I_{H\alpha}$ mean activity levels. Open triangles are stars with positive correlation between the two indices with $\rho \geq$ 0.5, open squares stars with negative correlation with $\rho \leq -0.5$, and dots stars with no correlations. \textit{Lower panel:} Relationship between the logarithms of $S_{MW}$ and H$\alpha$ indices.}
\label{mean_rhk_iha}
\end{center}
\end{figure}

Our activity indices are corrected for the effects of photospheric flux, and can, if they are not dependent on other factors other than chromospheric flux, be used to compare the activity levels between different stars.
Figure \ref{mean_rhk_iha} (upper panel) shows the correlation between the mean values of $\log R'_{HK}$ and $\log I_{H\alpha}$.
These mean values were calculated by averaging the two indices over all our nightly measurements, and represent the average activity level of each star.
Open triangles are stars with correlation coefficient, $\rho \geq 0.5$, squares are stars with $\rho \leq -0.5$, and dots stars with no strong correlations.
There is a correlation between the indices, with a correlation coefficient of $0.53$, but the scatter is large and the relation appears not to be linear \citep[c.f.][Fig. 12]{cincunegui2007b}.
However, if we choose only the positively correlated stars (open triangles), they show a slightly more well defined relationship for the mean values with a correlation coefficient of $0.65$.
When \citet{cincunegui2007b} studied the correlation between the mean values of the flux in \ion{Ca}{ii} and H$\alpha$ they concluded that the correlation between them is due to the dependence of the mean fluxes on stellar colour.
Indeed, when we plot the logarithm of the mean indices $S_{MW}$ vs. H$\alpha$ (without colour correction), we have a stronger correlation with $\rho = 0.79$ (Fig. \ref{mean_rhk_iha}, lower panel).
We can therefore confirm that stellar colour is playing a role in the correlation between the mean flux levels of the \ion{Ca}{ii} and H$\alpha$ lines.

\subsection{Activity cycles: detectability} \label{cycles}
To detect activity cycles we fitted sinusoids to the time-series of the two activity indices.
The significance of the fitting process was addressed by using an F-test where $F=\sigma_{const}^2/\sigma_{sin}^2$ to compare the fitting of a sinusoid with that of a constant model with $\sigma$ being the standard deviation of the residuals of the fitted model.
The probability $p(F)$ will give the probability that the data is better fitted by a constant model than a sinusoidal function.
We selected stars with cycles as the ones where probabilities, $p(F)_{HK}$ and $p(F)_{H\alpha}$, are lower than 0.05 and, similarly to \citet{lovis2011}, we searched for periods in the region between 2 and 11 years.

Based on this selection criteria and using $\log R'_{HK}$, we detected 69 stars (26\%) with significant activity cycles with periods varying between 2.0 and 10.8 years.
The $\log I_{H\alpha}$ index, however, is not so sensitive at detecting magnetic cycles.
Only 9 stars (3.3\%) showed significant cycles with periods varying between 3.9 and 9.5 years.
As a comparison, \citet{robertson2013} detected activity cycles with periods longer than one year in 5\% of their sample of 93 K5-M5 stars using an H$\alpha$ index similar to ours.
In their study of activity cycles based on this sample but with a different selection criteria, \citet{lovis2011} found that, out of their 284-star sample, 99 stars (35\%\footnote{We should note that for 165 stars they do not find cycles but they cannot exclude cycles either. In their conclusions they arrive at a final value of 61\% of stars with cycles when they exclude these stars from the fraction.}) showed long-term activity cycles in their $\log R'_{HK}$ index.
Their slightly higher fraction of stars with cycles is probably due to the fact that they use a different selection criteria with a different restriction on the number of data points (some of their stars with detected cycles have less than 10 observations), we use only data with $S/N \geq 100$, and we have more data points.

\section{Correlations between $\log R'_{HK}$ and $\log I_{H\alpha}$} \label{sec:results1}

\begin{figure}[tbp]
\begin{center}
\resizebox{\hsize}{!}{\includegraphics{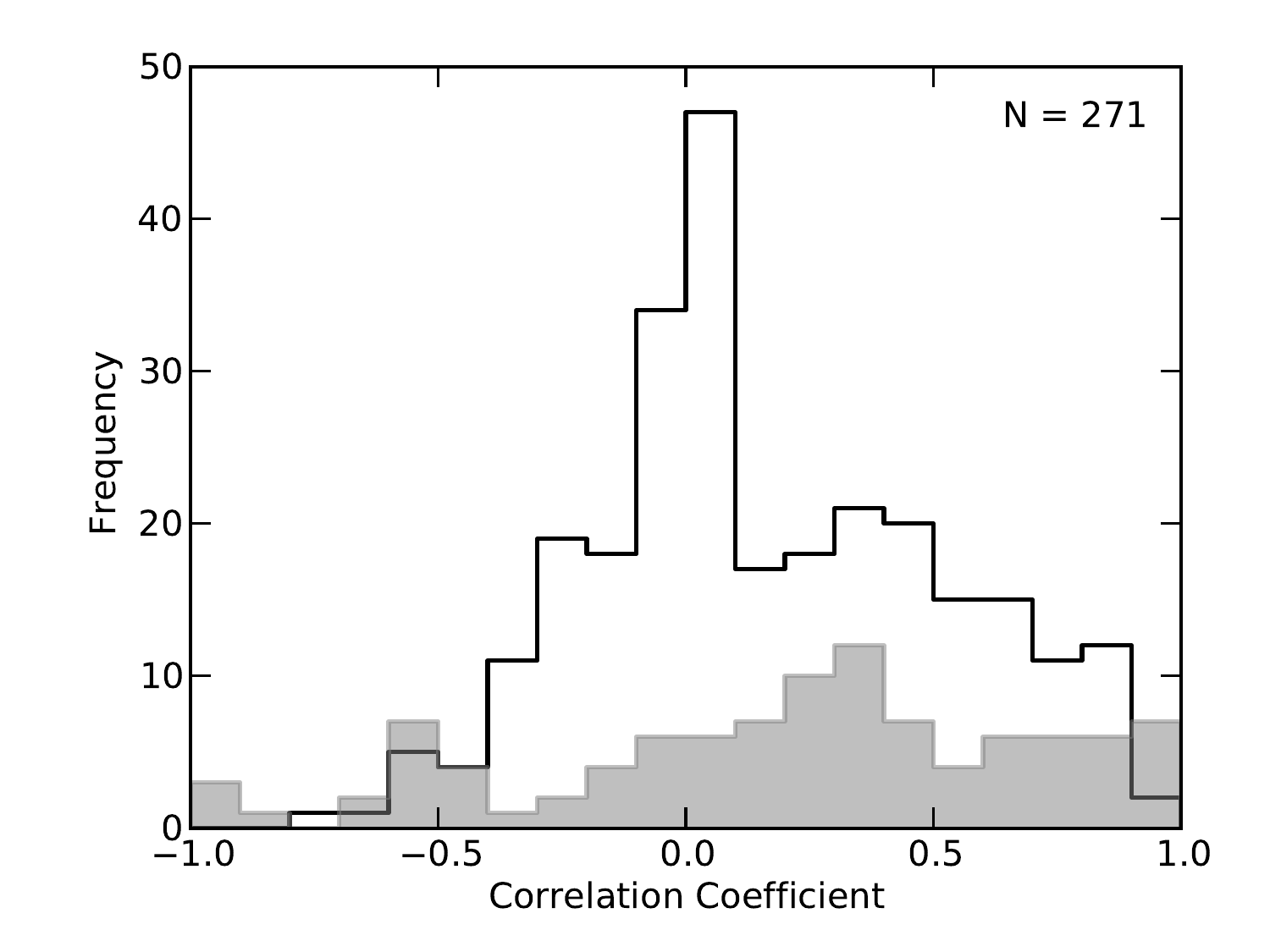}}
\caption{Distribution of correlation coefficients between $\log R'_{HK}$ and $\log I_{H\alpha}$ for the whole sample (black line). The grey filled histogram shows the distribution of correlations for the 101 stars in \protect\citet{cincunegui2007b} sample that have their correlations calculated.}
\label{hist_r}
\end{center}
\end{figure}

For all stars we calculated the Pearson correlation coefficient between $\log R'_{HK}$ and $\log I_{H\alpha}$.
As was detected by \citet{cincunegui2007b} for the flux in the \ion{Ca}{ii} H \& K and H$\alpha$ lines, we also find a great variety of correlation coefficients between $\log R'_{HK}$ and $\log I_{H\alpha}$, in the range $-0.78 \leq \rho \leq 0.95$ (Fig. \ref{hist_r}).
Although there is a tendency for the stronger correlations to be positive, we found a few cases of anti-correlations with $\rho \leq -0.5$.

Since we are interested in studying the cases of strong long-term correlations between the flux in the \ion{Ca}{ii} H \& K and H$\alpha$ lines, we made a new selection of stars with good quality data that we are going to describe in the following section.

\subsection{Stars with "strong" long-term correlations} \label{strong_sel}

\begin{table*}[tbp]
\caption{Variability and correlations using binned data for the stars with strong long-term correlations.}
\label{table:sel}
\centering
\begin{tabular}{l c c | c c | c c c | c c c} \\
\hline
\hline
\multicolumn{1}{l}{Star} &
\multicolumn{1}{c}{$N_{bins}$} &
\multicolumn{1}{c}{$T_{span}$} &
\multicolumn{1}{c}{$\rho$} &
\multicolumn{1}{c}{FAP} &
\multicolumn{3}{c}{$\log R'_{HK}$} &
\multicolumn{3}{c}{$\log I_{H\alpha}$} \\
\multicolumn{1}{c}{} &
\multicolumn{1}{c}{} &
\multicolumn{1}{c}{[days]} &
\multicolumn{1}{c}{} &
\multicolumn{1}{c}{} &
\multicolumn{1}{c}{$\sigma_e$} &
\multicolumn{1}{c}{$\langle\sigma_i\rangle$} &
\multicolumn{1}{c}{$P(F)$} &
\multicolumn{1}{c}{$\sigma_e$} &
\multicolumn{1}{c}{$\langle\sigma_i\rangle$} &
\multicolumn{1}{c}{$P(F)$} \\
\hline
HD100508&	4&	766&	$-0.93$&	0.047&		0.0199&	0.0034&		0.0082&		0.00102&	0.00031&		0.040 \\
HD13808&	13&	2245&	0.97&	0.0001&		0.0794&	0.0043&		$< 10^{-5}$&	0.00302&	0.00065&		$< 10^{-5}$ \\
HD154577&	7&	2117&	0.92&	0.0014&		0.0302&	0.0030&		0.00001&		0.00242&	0.00047&		0.00046 \\
HD209100&	7&	829&	0.92&	0.0021&		0.0278&	0.0048&		0.00022&		0.00282&	0.00088&		0.0062 \\
HD215152&	9&	1160&	0.83&	0.0034&		0.0322&	0.0028&		$< 10^{-5}$&	0.00144&	0.00045&		0.0016 \\
HD4915&		5&	646&	0.98&	0.0046&		0.0343&	0.0041&		0.00062&		0.00185&	0.00062&		0.029 \\
HD63765&	8&	1936&	0.88&	0.0024&		0.0374&	0.0077&		0.00023&		0.00290&	0.00068&		0.00052 \\
HD71835&	10&	2618&	0.85&	0.0010&		0.0403&	0.0048&		$< 10^{-5}$&	0.00165&	0.00046&		0.00040 \\
HD7199&		11&	2237&	$-0.82$&	0.0009&		0.0758&	0.0071&		$< 10^{-5}$&	0.00207&	0.00048&		0.00003 \\
HD78612&	4&	2902&	0.88&	0.050&		0.0135&	0.0036&		0.029&		0.00187&	0.00057&		0.042 \\
HD85512&	12&	2906&	0.93&	$< 10^{-4}$&	0.0495&	0.0026&		$< 10^{-5}$&	0.00395&	0.00052&		$< 10^{-5}$ \\
HD88742&	5&	1729& 	0.93&	0.015&		0.0323&	0.0043&		0.00089&		0.00308&	0.00086&		0.015 \\
\hline
\end{tabular}
\end{table*}

\begin{table*}[tbp]
\caption{Stellar parameters of the stars with strong long-term correlations.}
\label{table:param}
\centering
\begin{tabular}{l c c c c c c c c} \\
\hline
\hline
\multicolumn{1}{l}{Star} &
\multicolumn{1}{c}{$\langle\log R'_{HK}\rangle$} &
\multicolumn{1}{c}{$\langle\log I_{H\alpha}\rangle$} &
\multicolumn{1}{c}{[Fe/H]} &
\multicolumn{1}{c}{$T_{\mathrm{eff}}$} &
\multicolumn{1}{c}{$\log g$} &
\multicolumn{1}{c}{$M_V$} &
\multicolumn{1}{c}{$B-V$} &
\multicolumn{1}{c}{$P_{rot}$} \\
\multicolumn{1}{c}{} &
\multicolumn{1}{c}{} &
\multicolumn{1}{c}{} &
\multicolumn{1}{c}{} &
\multicolumn{1}{c}{[K]} &
\multicolumn{1}{c}{[cm\,s$^{-2}$]} &
\multicolumn{1}{c}{} &
\multicolumn{1}{c}{} &
\multicolumn{1}{c}{[days]} \\
\hline
\textbf{HD100508}&	$-5.055$&		$-1.7198$	&	$0.39 \pm 0.05$&	$5449 \pm 61$	&	$4.42 \pm 0.09$&	5.16&	0.83&	48.4 \\
HD13808&		$-4.892$&		$-1.7138$	&	$-0.20 \pm 0.03$&	$5087 \pm 41$	&	$4.40 \pm 0.08$&	6.08&	0.87&	42.8 \\
HD154577&		$-4.878$&		$-1.7019$	&	$-0.70 \pm 0.02$&	$4900 \pm 37$	&	$4.52 \pm 0.08$&	6.70&	0.89&	41.3 \\
HD209100&		$-4.781$&		$-1.7153$	&	$-0.20 \pm 0.04$&	$4754 \pm 89$	&	$4.45 \pm 0.19$&	6.89&	1.06&	37.2 \\
HD215152&		$-4.871$&		$-1.7157$	&	$-0.10 \pm 0.04$&	$4935 \pm 76$	&	$4.40 \pm 0.14$&	6.45&	0.97&	42.0 \\
HD4915&			$-4.798$&		$-1.7038$	&	$-0.21 \pm 0.01$&	$5658 \pm 13$	&	$4.52 \pm 0.03$&	5.26&	0.66&	20.4 \\
HD63765&		$-4.741$&		$-1.7044$	&	$-0.16 \pm 0.01$&	$5432 \pm 19$	&	$4.42 \pm 0.03$&	5.53&	0.74&	25.0 \\
HD71835&		$-4.889$&		$-1.7194$	&	$-0.04 \pm 0.02$&	$5438 \pm 22$	&	$4.39 \pm 0.04$&	5.38&	0.77&	35.2 \\
\textbf{HD7199}&	$-4.946$&		$-1.7270$	&	$0.28 \pm 0.03$&	$5386 \pm 45$	&	$4.34 \pm 0.08$&	5.29&	0.85&	45.9 \\
HD78612&		$-5.004$&		$-1.7154$	&	$-0.24 \pm 0.01$&	$5834 \pm 14$	&	$4.27 \pm 0.02$&	4.06&	0.61&	21.7 \\
HD85512&		$-4.898$&		$-1.7023$	&	$-0.32 \pm 0.03$&	$4715 \pm 102$&	$4.39 \pm 0.28$&	7.43&	1.16&	47.3 \\
HD88742&		$-4.688$&		$-1.7031$	&	$-0.02 \pm 0.01$&	$5981 \pm 13$	&	$4.52 \pm 0.02$&	4.60&	0.59&	11.4 \\
\hline
\end{tabular}
\tablefoot{The average values of $\log R'_{HK}$ and $\log I_{H\alpha}$ were calculated using the binned data.}
\end{table*}

We are interested in measuring the long-term Pearson correlation coefficient ($\rho$) between the $\log R'_{HK}$ and $\log I_{H\alpha}$ indices.
We need therefore to ensure that we have (a) a long time-span to certify that we are measuring long-term variations\footnote{Since this sample derives from a planet hunt selection of stars, active stars with $\log R'_{HK} \geq -4.7$ were monitored early and only rarely measured. Therefore, stars with higher activity will have fewer measurements and possibly a lower time-span of observations. This selection will thus reduce even more the number of active stars in the sample.}, (b) variability in the long-term so that we are not measuring correlations due to noise, (c) no short-term variations that can interfere with or hide the long-term ones, (d) enough quantity of points to calculate a significant $\rho$, and (e) strong correlations.
To achieve this, we perform the following selection criteria on our 271-star sample:

\begin{enumerate}

\item
All data was binned into 100-day averages, each bin with at least three nights of observations, where the errors were calculated as the standard error on the mean, $\sigma/\sqrt(N)$, where $\sigma$ is the standard deviation of the observations and $N$ the number of observations.
This will reduce the variation induced by short-term activity modulated by stellar rotation.

\item
We selected stars with at least four bins. This selection ensures that we have enough points to calculate $\rho$ and that the time span is at least 400 days.

\item
Only stars that showed long-term variability in $\log R'_{HK}$ were selected. This will ensure that we are not detecting random variations due to noise. We performed an $F$-test on the binned data where $F = \sigma_e^2/\langle\sigma_i\rangle^2$, with $\sigma_e$ the standard deviation of the binned data and $\langle\sigma_i\rangle$ the mean of the errors on the bins \citep[e.g.][]{zechmeister2009}. We calculated the probability of the $F$-test, $P(F)$, that the variations are due to the internal errors of the binned data, and selected stars with $P(F) \leq 0.05$ (95\% probability that the variability in not due to the internal errors).

\item
We also applied the variability $F$-test for the $\log I_{H\alpha}$ index in a similar way as described above.

\item
To select significant correlation coefficients between $\log R'_{HK}$ and $\log I_{H\alpha}$ we calculated the False Alarm Probability (FAP) of having absolute values of $\rho$ higher than the ones obtained for each star by bootstrapping the binned data and calculating the fraction of cases with higher $\lvert \rho \rvert$ values. We used 10000 permutations per star to calculate the FAP values. Only stars with FAP $\leq 0.05$ (95\% significance level) were selected.

\item
Stars with strong correlations were selected as the ones having $\lvert \rho \rvert \geq 0.70$. 
\end{enumerate}

From the 129 stars that passed selections (1) and (2), 95 stars (73.6\%) show long-term variability in $\log R'_{HK}$, 51 stars (39.5\%) show long-term variability in $\log I_{H\alpha}$, and 45 stars (34.9\%) show long-term variability on both indices.
Out of the 45 stars that show variability on both indices, 12 stars (26.7\%) show strong positive correlations between the indices, 10 of them (22.2\%) having positive correlations while two (4.4\%) having anti-correlations.

Table \ref{table:sel} shows the variability and correlations data for the 12 stars with strong long-term correlations, where $N_{bins}$ the number of bins for each star, $\rho$ the correlation coefficient value, FAP the false alarm probability of $\rho$, and the parameters of the $F$-tests for both activity indices.
The time series of $\log R'_{HK}$, $\log I_{H\alpha}$, and their respective correlations for these 12 stars are shown in Fig. \ref{all_strong_plots}.
We also tried to fit sinusoids to these stars (see Sect. \ref{cycles}) using the binned data of both indices to check if these stars have significant activity cycles.
These fits appear in Fig. \ref{all_strong_plots} if the $p(F)_{HK}$ of the fit is lower than 0.05 (95\% significance level).
Two stars, HD100508 and HD78612, only have four bins and therefore do not have enough free parameters to calculate the probability of the fit.
From the stars with more than four bins, three have $p(F)$ values lower than 0.05 for the \rhk index, namely HD4915, HD63765, and HD88742.
These are all stars with strong positive correlations.
The 7 stars with significant cycles in \rhk have periods in the range 1528 to 10665 days, and 5 of them could be fitted in \iha with the same period found for \rhk and a $p(F)_{H\alpha}$ value lower than 0.05 (HD13808, HD154577, HD215152, HD7199, and HD85512).
For this sample, no star showed a period in \iha that was not found also in \rhk, and at a higher significance.

To try to understand why some stars have positive correlations while others have negative, we compared the correlations with the basic stellar parameters shown in Table \ref{table:param}.
The two stars with negative correlations are shown in bold.
First, we observe that the two stars with the negative correlations are two of the most inactive in terms of both $\log R'_{HK}$ and $\log I_{H\alpha}$.
Second, while all the stars with positive correlation coefficient have negative metallicity (median value of $-0.20$ dex), the two stars with negative correlations have positive metallicity (median value of 0.34 dex).

Although we can se hints that activity level and metallicity could be influencing the correlation between the two indices, the small number of stars we are using is insufficient to clearly show a solid trend between these parameters. We therefore chose to relax our selection criteria to increase the number of stars in our sample and check if the trends with activity level and metallicity are maintained.

\subsection{"Relaxed" selection of stars with correlations} \label{relaxed_sel}
To increase the number of stars in our study we discarded the variability tests, FAPs on the correlation coefficients and used the full data sets based on the nightly averaged data.
The correlation coefficient limit was also decreased to $\lvert \rho \rvert \geq  0.5$.
This produced a larger sample which will include weaker correlations that can be due to a lower number of data points, shorter time-spans, and/or due to short-term variations.
We shall therefore take this part of the study as an indication and not as a proof.
However we will now be able to do statistical tests to this sample.

Using this selection, we found that out of the 271 stars in our original sample, 58 (21.4\% of the sample) have positive correlations between $\log R'_{HK}$ and $\log I_{H\alpha}$, and 8 (3.0\% of the sample) have anti-correlations.
Table \ref{table3} shows the 66 stars with $\lvert \rho \rvert \geq 0.5$ with their activity mean levels and standard deviations, stellar parameters, and correlation coefficient between the two indices.
Stars with correlations coefficients in the range $-0.5 < \rho < 0.5$ (no correlations) are presented in Table \ref{table4}.

All the eight stars with negative correlations ($\rho \leq -0.5$) have low $\log R'_{HK}$ activity levels with a median value of $-4.97$ and a median super-solar metallicity with a value of 0.20.
The 58 stars with positive correlations ($\rho \geq 0.5$) have \rhk with a median value of $-4.81$ and a median sub-solar metallicity with a value of $-0.16$.
This "relaxed" selection appears to maintain the trends found in Section \ref{strong_sel}.
In the next sections we will study these trends for this sample of stars.

\section{Mean activity level and correlations}\label{sec:results2}
Here we investigate the distribution of the positively and negatively correlated stars in terms of $\log R'_{HK}$ and \iha activity levels.

\begin{figure}[tbp]
\begin{center}
\resizebox{\hsize}{!}{\includegraphics{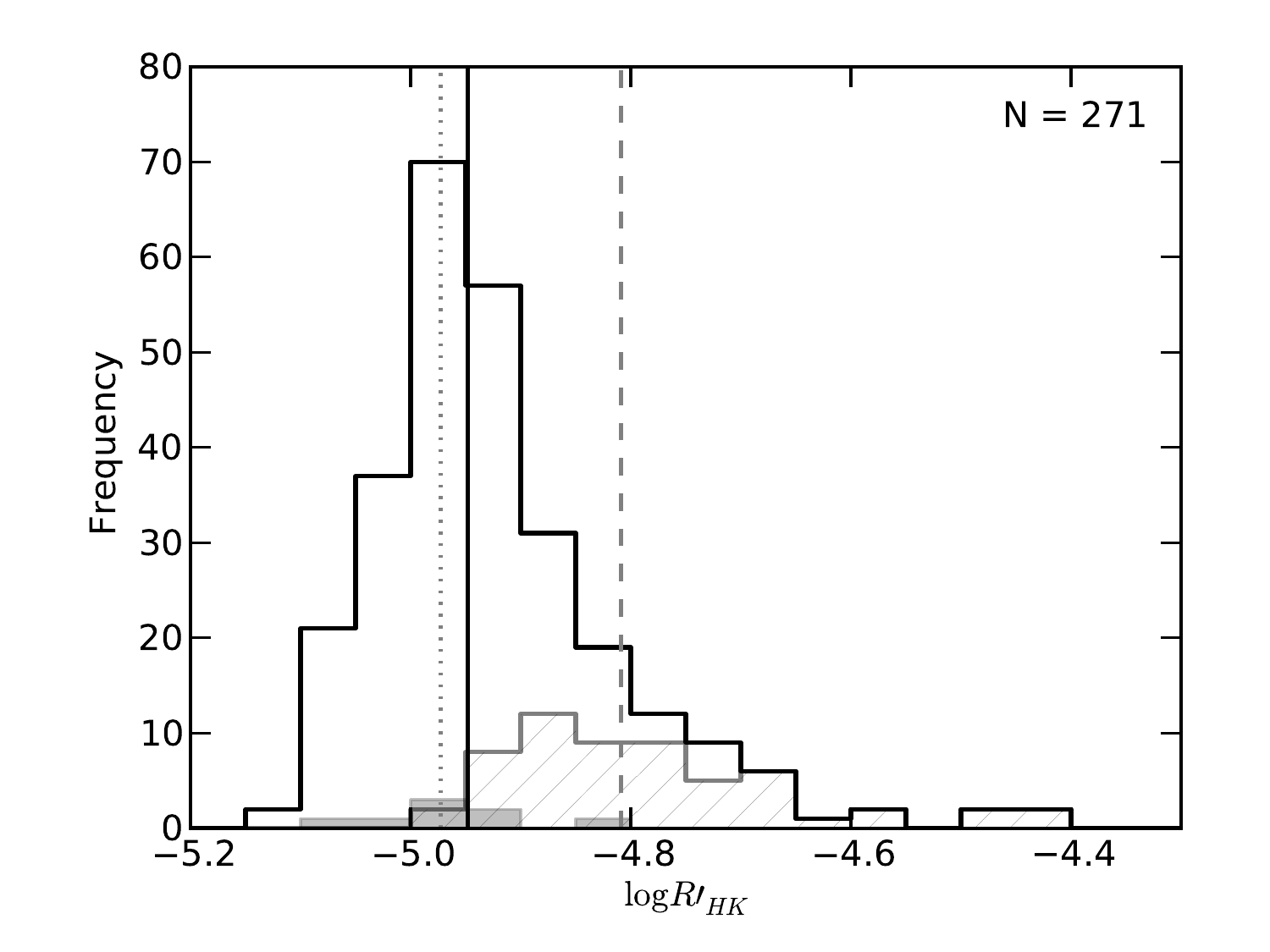}}
\resizebox{\hsize}{!}{\includegraphics{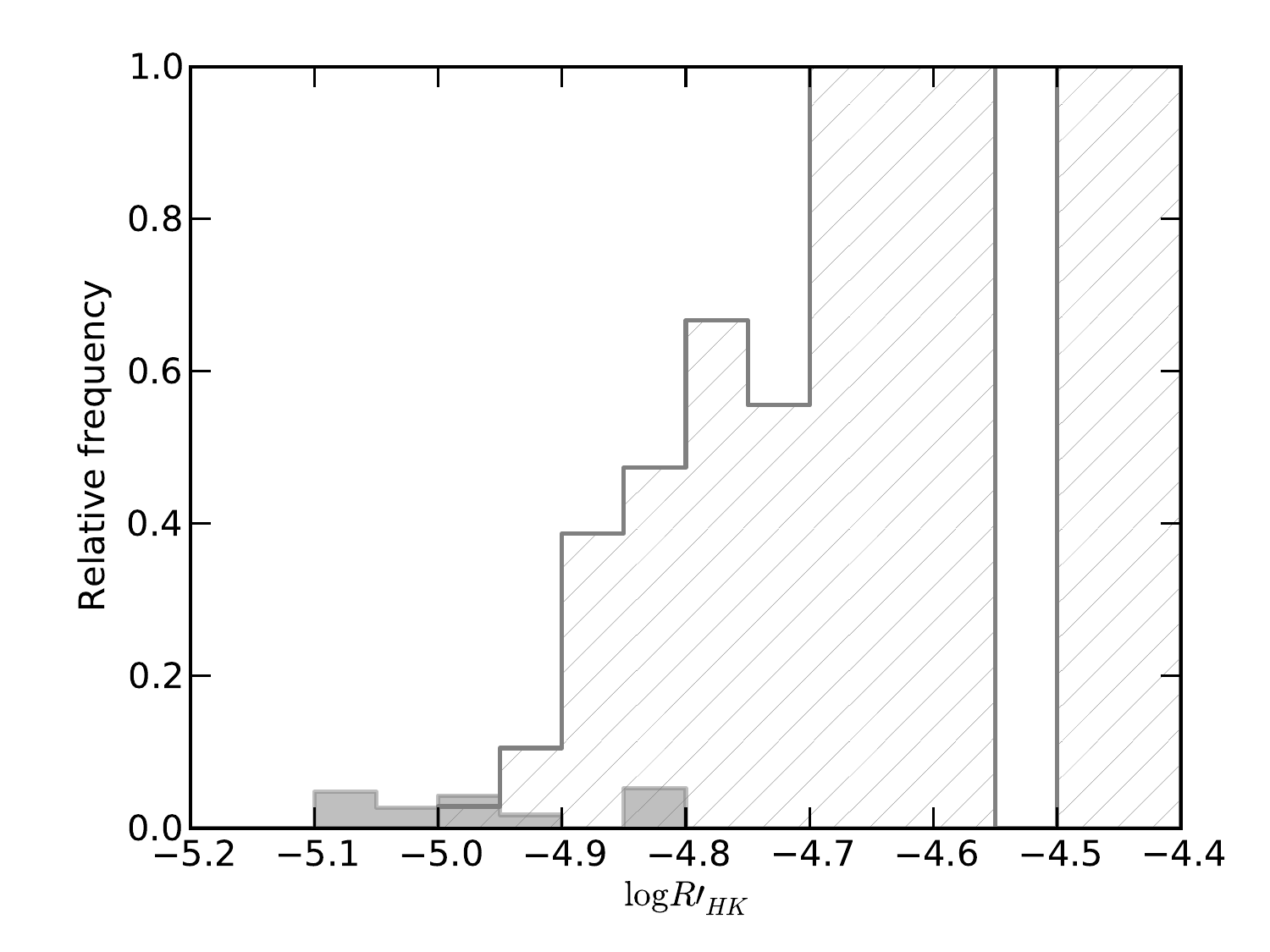}}
\caption{\textit{Upper panel:} Distribution on $\log R'_{HK}$ activity for the full sample (black), stars with positive correlation coefficient higher than 0.5 (hatched grey), and stars with negative correlation coefficient lower than $-0.5$ (filled grey). Vertical lines are the medians of the distributions with black line for the full sample, dashed line for positively correlated stars, and dotted line for negatively correlated stars. \textit{Lower panel:} Same as the upper panel but using relative distribution on $\log R'_{HK}$, i.e., the values in each bin are divided by the total number of stars in the respective bin.}
\label{hist_rhk}
\end{center}
\end{figure}

Figure \ref{hist_rhk} (upper panel) shows the distribution of activity as measured by the $\log R'_{HK}$ index.
The black line is the histogram of the selected sample of 271 main sequence stars.
We can observe the selection bias against active stars as the great majority of the sample lies between $-5.1$ and $-4.8$ dex, with a median of $-4.95$ dex.
The hatched and filled grey histograms show the distribution in average activity level of the stars with positive and negative $\log R'_{HK}$--$\log I_{H\alpha}$ correlations, respectively.
The median of the negatively correlated stars is close to the median of the full sample (but with a tendency to be less active) with a value of $-4.97$ dex, while the median of the positively correlated stars lies in a higher activity zone, with a value of $-4.81$ dex.
In general, the majority of the least active stars show no strong correlations between the two indices.
However, it is obvious from the plot that there is a tendency for the positively correlated stars to be more active in general, and all stars more active than $\log R'_{HK} = -4.7$ have positive correlations between $\log R'_{HK}$ and $\log I_{H\alpha}$.
The relative histogram in Fig. \ref{hist_rhk} (lower panel) illustrates very well this tendency.

\begin{figure}[tbp]
\begin{center}
\resizebox{\hsize}{!}{\includegraphics{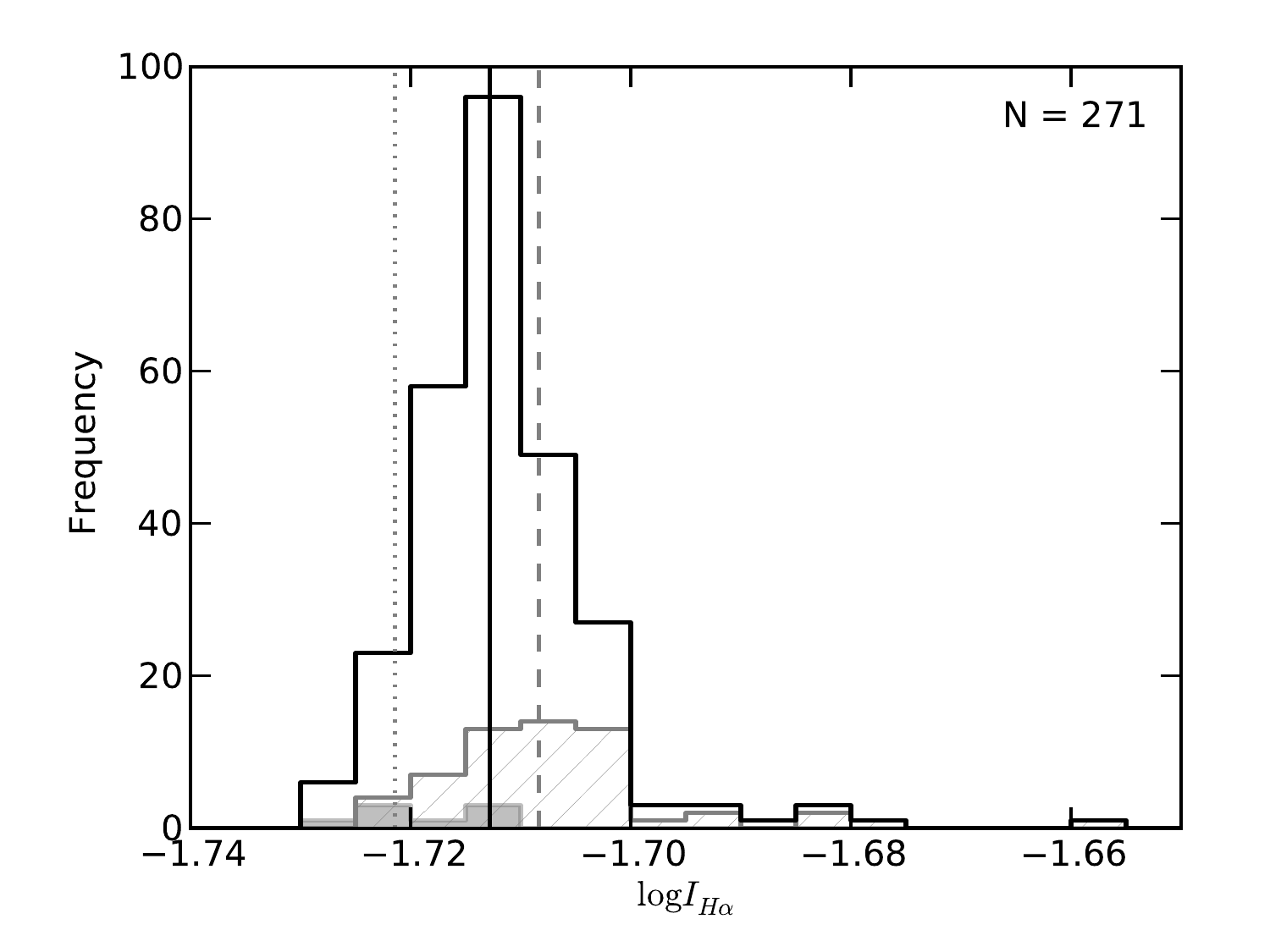}}
\resizebox{\hsize}{!}{\includegraphics{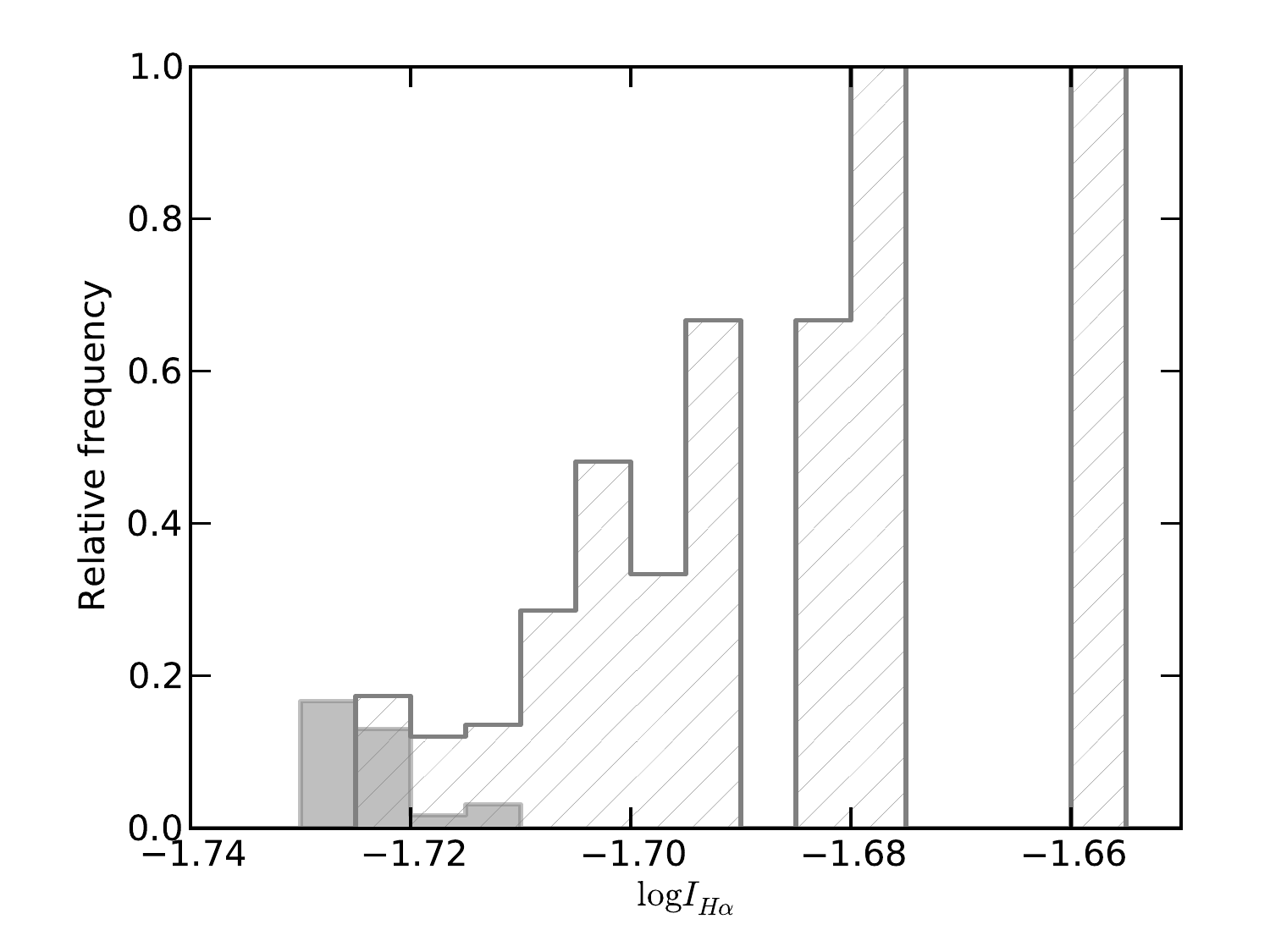}}
\caption{\textit{Upper panel:} Distribution on $\log I_{H\alpha}$ activity for the full sample (black), stars with positive correlation coefficient higher than 0.5 (hatched grey), and stars with negative correlation coefficient lower than $-0.5$ (filled grey). Vertical lines are the medians of the distributions with black line for the full sample, dashed line for positively correlated stars, and dotted line for negatively correlated stars. \textit{Lower panel:} Same as the upper panel but using relative distribution on $\log I_{H\alpha}$.}
\label{hist_iha}
\end{center}
\end{figure}

The separation between positively and negatively correlated stars is further confirmed by the Kolmogorov-Smirnov (K-S) test that shows that the two populations are distinct with a $p$-value of 0.002 and a $D$ value \footnote{The K-S $D$ value is the highest value of the difference between the cumulative distributions of the two populations. The $p$-value gives the probability that the two populations come from the same parent distribution.} of 0.664.
A similar distribution was found for $\log I_{H\alpha}$ (Fig. \ref{hist_iha}).
The correlation between the two indices have different distributions according to activity level, with negatively correlated stars being the least active ones and the positively correlated stars increasing in number with $I_{H\alpha}$ activity level.
In this case, the K-S test have a $D = 0.513$ and $p$-value = 0.03. 
The histograms also show that the values in $\log I_{H\alpha}$ are very well constrained between $-1.73$ and $-1.70$, and only a few cases of higher activity stars exists beyond these values.
Note that, in the relative histogram (lower panel) the "hole" in the region between $-1.675$ and $-1.660$ is due to lack of data.

\section{Metallicity and correlations}\label{sec:results3}
\begin{figure}[tbp]
\begin{center}
\resizebox{\hsize}{!}{\includegraphics{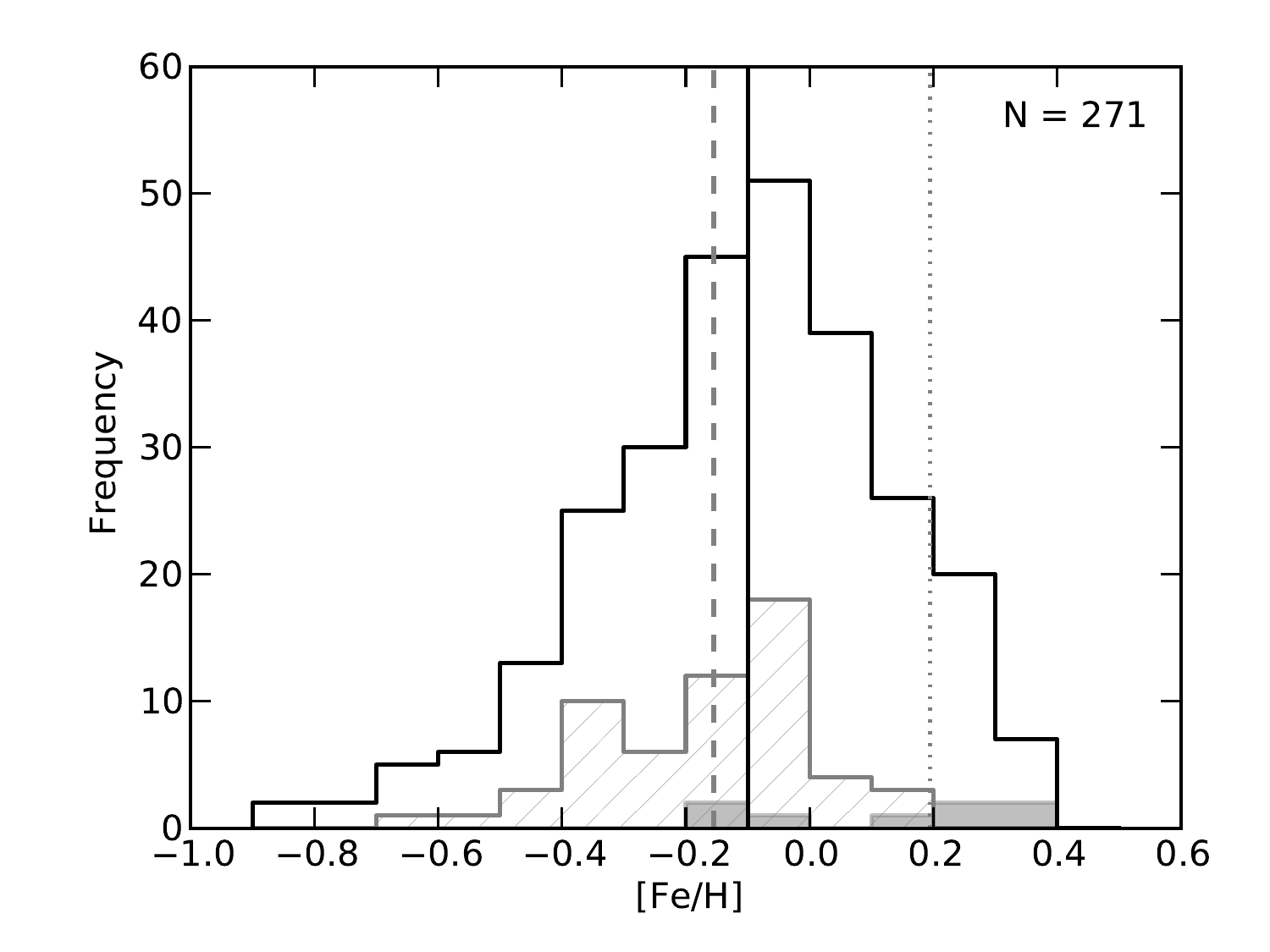}}
\resizebox{\hsize}{!}{\includegraphics{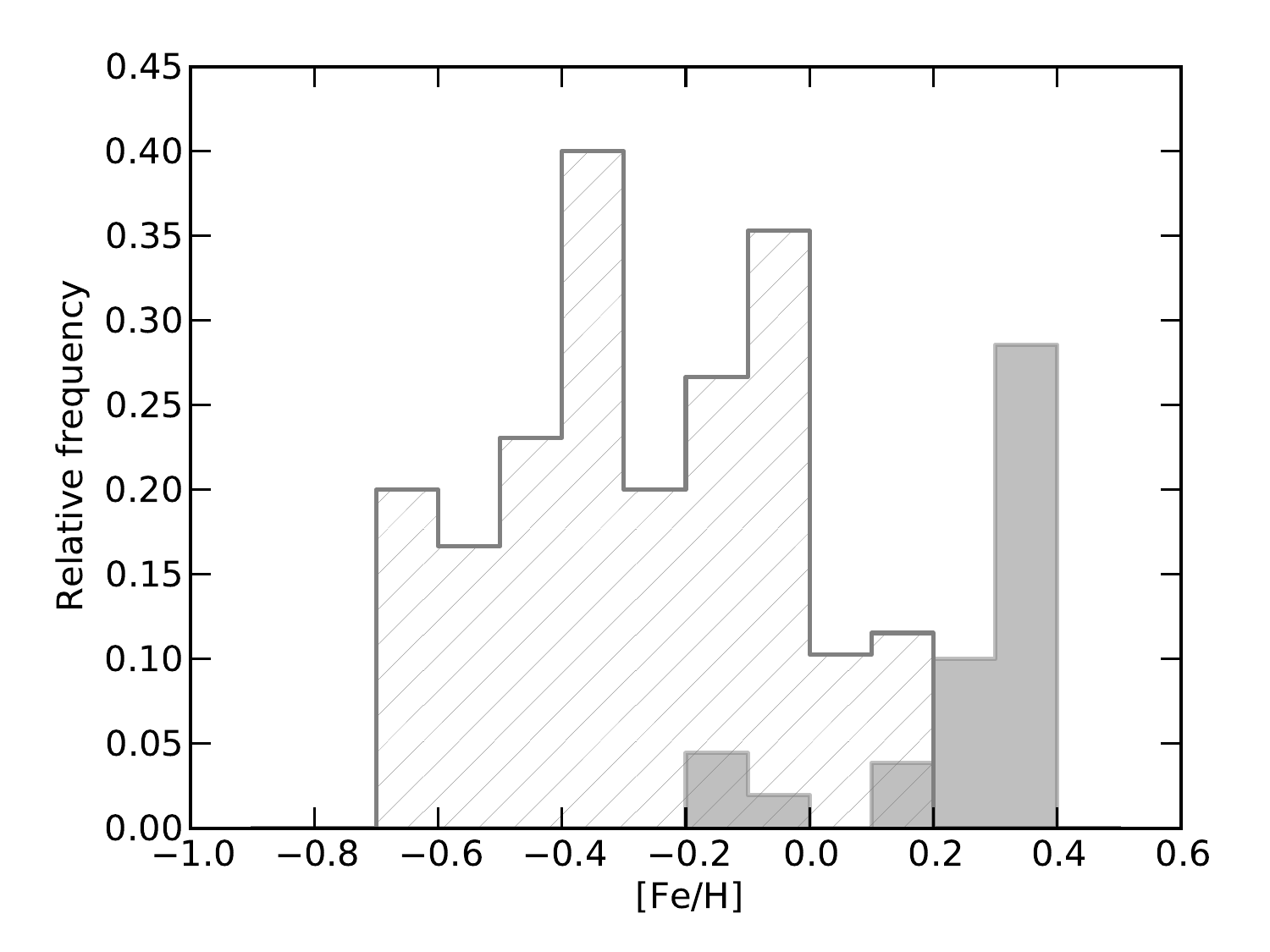}}
\caption{\textit{Upper panel:} Distribution of metallicity for the full sample (black), stars with positive correlation coefficient higher than 0.5 (hatched grey), and stars with negative correlation coefficient lower than $-0.5$ (filled grey). Black vertical line is the median of the full sample, dashed vertical line the median of the positively correlated stars, and dotted line the median of the negatively correlated stars. \textit{Lower panel:} Same as the top panel but for relative distributions. The K-S test gives a p-value of 0.01\% for the probability that the two populations are drawn from the same distribution.}
\label{hist_feh}
\end{center}
\end{figure}

\begin{figure}[tbp]
\begin{center}
\resizebox{\hsize}{!}{\includegraphics{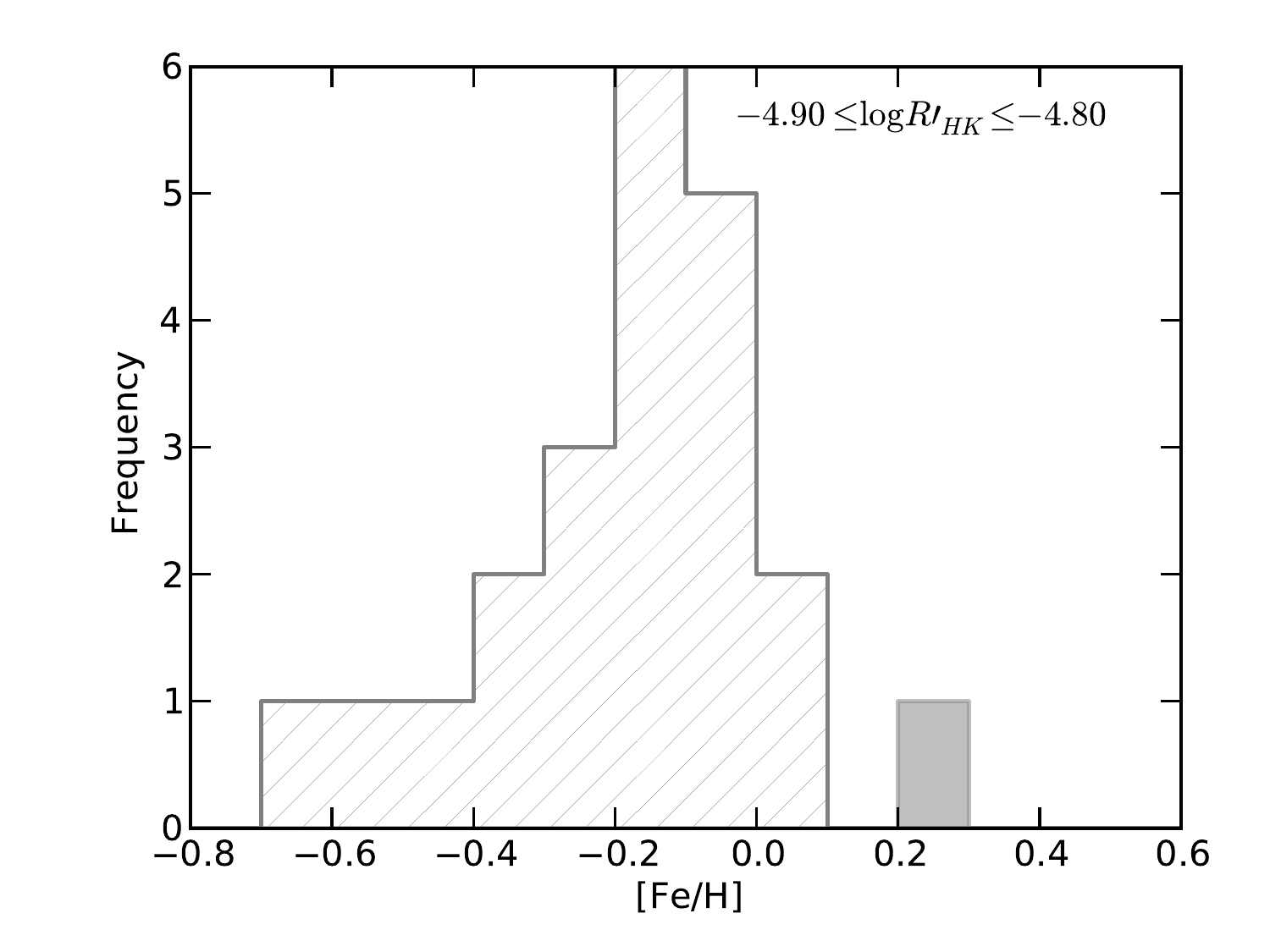}}
\resizebox{\hsize}{!}{\includegraphics{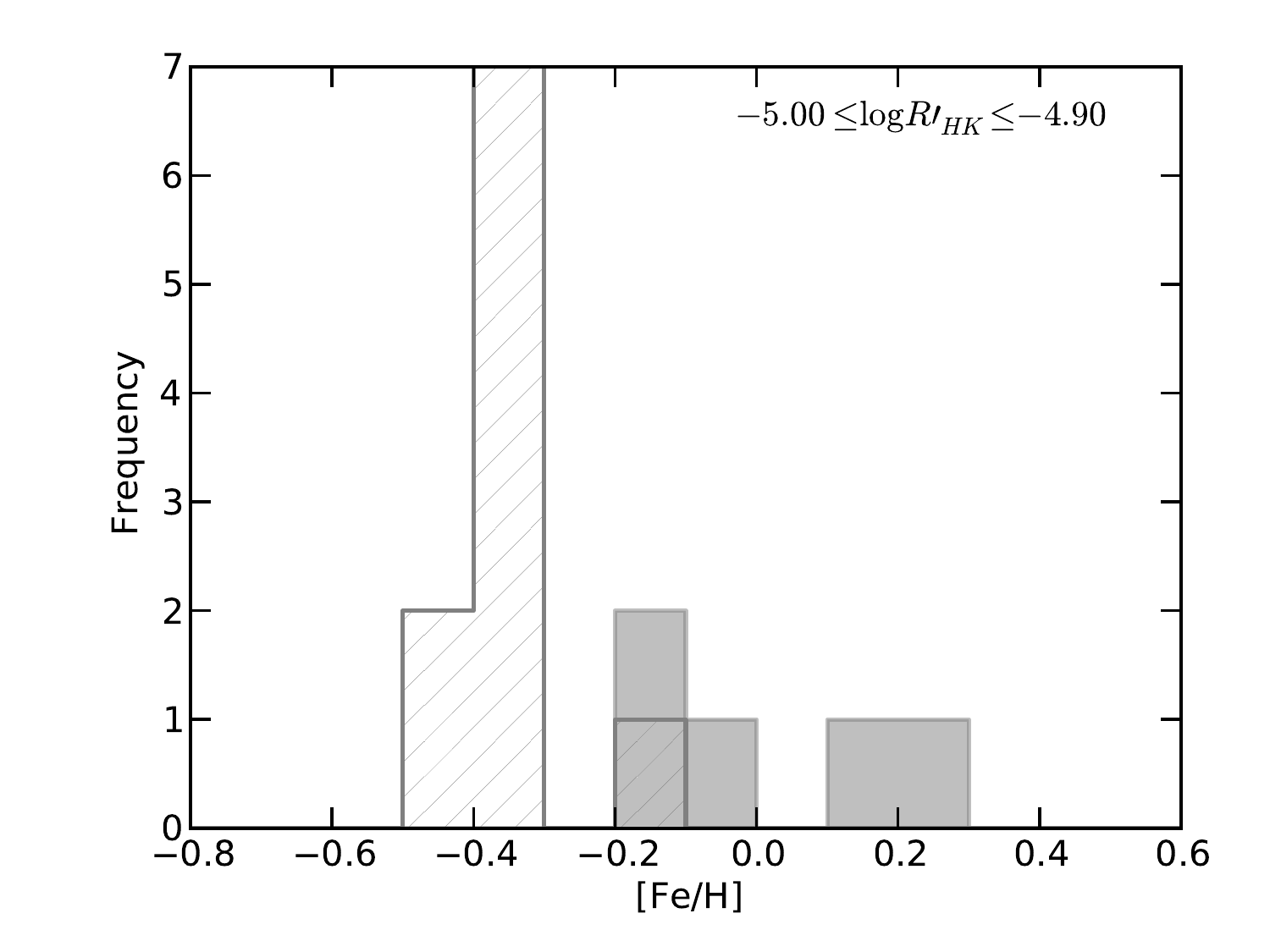}}
\caption{\textit{Upper panel:} Distribution of metallicity for stars with positive correlation coefficient higher than 0.5 (hatched grey), and stars with negative correlation coefficient lower than $-0.5$ (filled grey) with activity in the range $-4.9 \leq \log R'_{HK} \leq -4.8$. \textit{Lower panel:} Same as the top panel but for stars with activity in the range $-5.0 \leq \log R'_{HK} \leq -4.9$.}
\label{hist_feh_bins}
\end{center}
\end{figure}

Is stellar activity the only variable playing a role in the definition of the correlation or anti-correlation observed?
In Table \ref{table3}, it is noticeable that there is a tendency for the eight stars with negative correlation between the $\log R'_{HK}$ and $\log I_{H\alpha}$ indices to have super-solar metallicity.
We plotted the histogram of the two populations, the ones with a positive and a negative correlation, against metallicity (Fig. \ref{hist_feh}).
Symbols and colours are the same as the ones presented Fig. \ref{hist_rhk}.
In Fig. \ref{hist_feh} (upper panel) the median of the negatively correlated stars is not coincident with de medians of both the sample and the positively correlated stars.
The histogram shows that, again, there seems to be two distinct populations of stars: the majority of the stars with positive correlations have negative metallicity while the negatively correlated stars appear to be of super-solar metallicity (mainly if compared to the overall sample).
The sample median is $-0.10$ dex, the positively correlated stars median lie at $-0.15$ dex, but the negatively correlated star's median is at a metallicity of 0.20 dex.
This is further corroborated by the K-S test, which gives a probability of 0.04\% that the two populations are indistinct (with a K-S $D$ value of 0.733).
The relative histogram of Fig. \ref{hist_feh} (lower panel) confirms this with the negatively correlated stars peaking at the super-solar metallicity while the positively correlated stars peaking at the sub-solar metallicity.
Nevertheless, there are some stars with negative correlation that have sub-solar metallicity and stars with positive correlation with super-solar metallicity.
We plotted metallicity histograms for two bins where there is superposition of positively and negatively correlated stars in activity in the region $-4.8 \leq \log R'_{HK} \leq -5.0$ (Fig. \ref{hist_feh_bins}).
The tendency for stars with higher metal content to have negative correlations is maintained in each activity bin.
In the lower panel of the figure, for the three stars with metallicity between $-0.1$ and $-0.2$ dex, the positively correlated star has [Fe/H] = $-0.20$ dex, while the two negatively correlated stars have [Fe/H] = $-0.16$ and [Fe/H] = $-0.15$ dex.
These plots show that for a given activity range, metallicity is still having an impact on the correlation between $\log R'_{HK}$ and $\log I_{H\alpha}$.

Our analysis was based on a small number of anti-correlated stars, and our conclusions can be a consequence of small-number statistics.
Also, as was stated before, this sample is not rigorous in terms of long-term variability of the stars or the significance of the correlations used.
Further studies with a larger number of metal-rich stars would be crucial to confirm or refute these results.

\section{Effective temperature and correlations}\label{sec:teff}
\begin{figure}[tbp]
\begin{center}
\resizebox{\hsize}{!}{\includegraphics{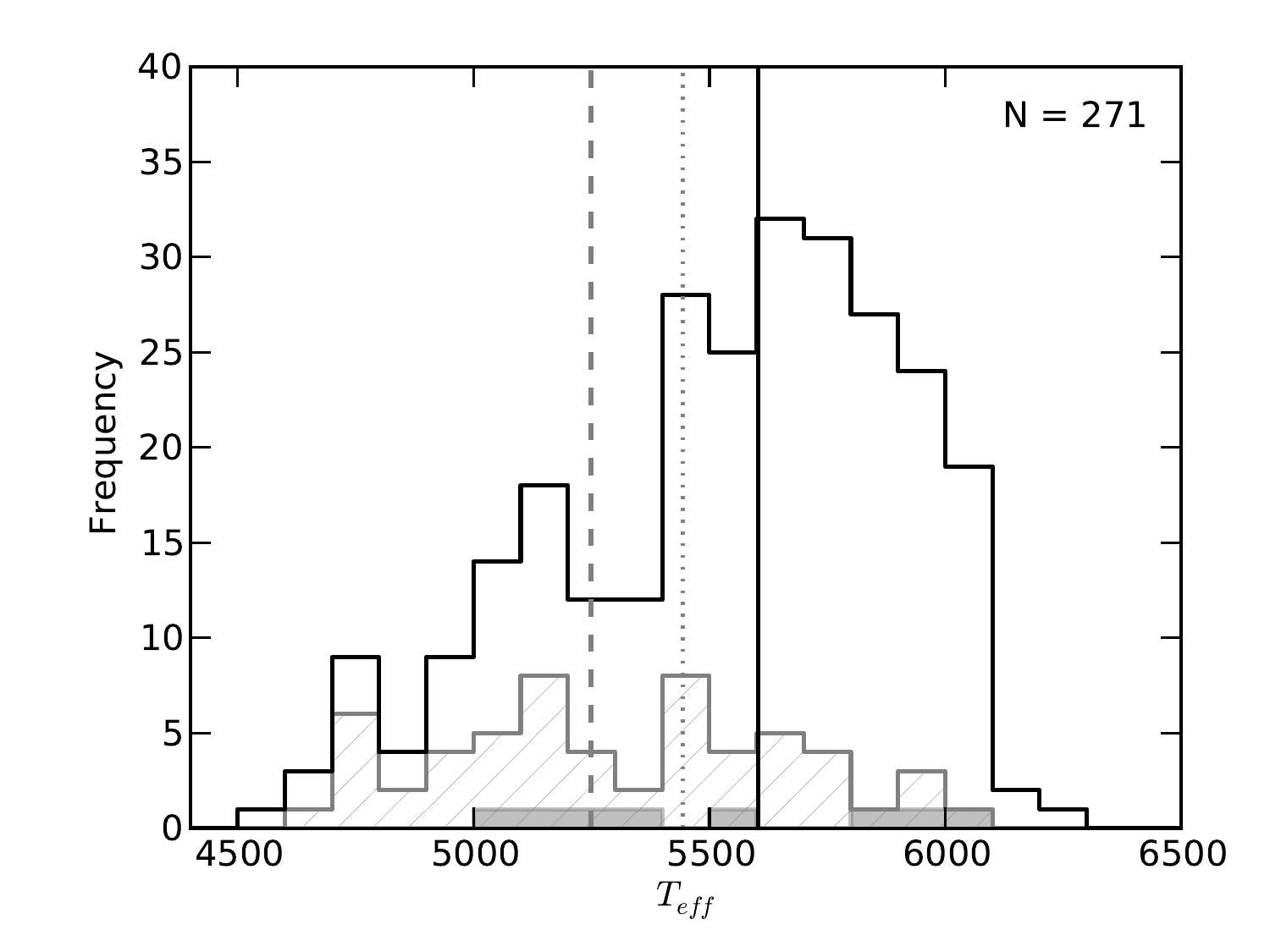}}
\resizebox{\hsize}{!}{\includegraphics{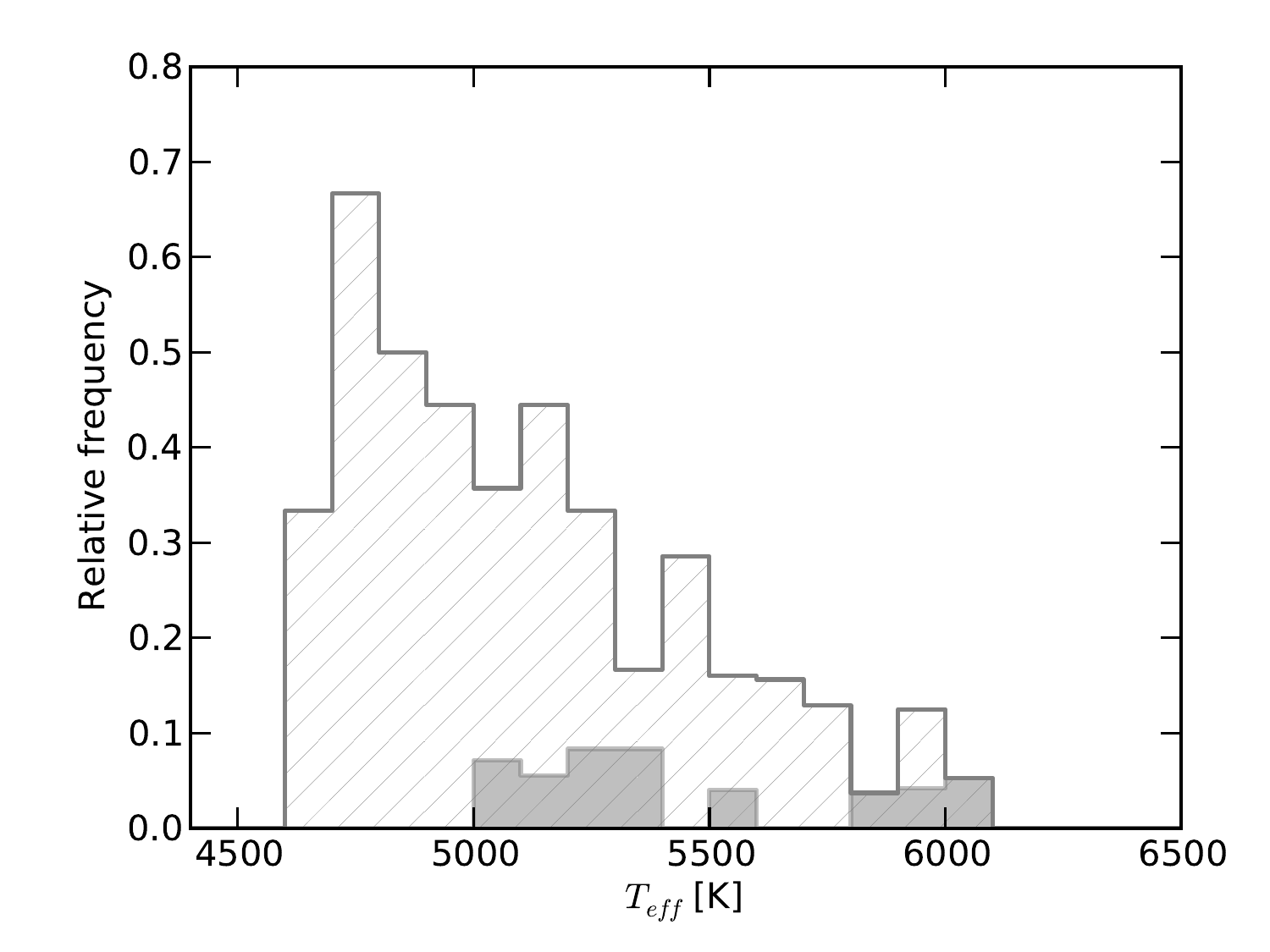}}
\caption{\textit{Upper panel:} Distribution of effective temperature for the full sample (black), stars with positive correlation coefficient higher than 0.5 (hatched grey), and stars with negative correlation coefficient lower than $-0.5$ (filled grey). Black vertical line is the median of the full sample, dashed vertical line the median of the positively correlated stars, and dotted line the median of the negatively correlated stars. \textit{Lower panel:} Same as the top panel but for relative distributions.}
\label{hist_teff}
\end{center}
\end{figure}

\begin{figure}[tbp]
\begin{center}
\resizebox{\hsize}{!}{\includegraphics{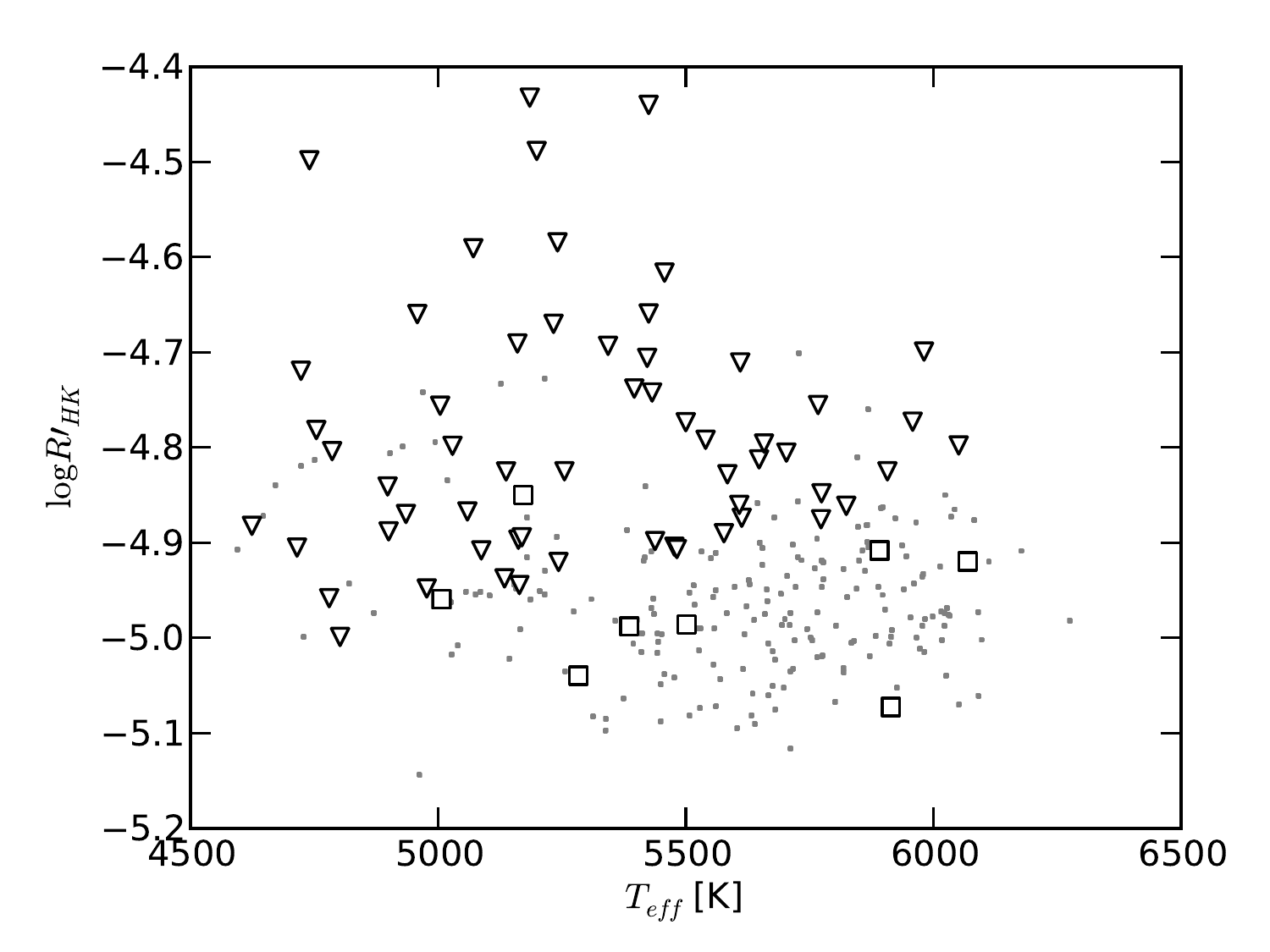}}
\caption{Activity level measured by $\log R'_{HK}$ against effective temperature. Triangles are stars with $\rho \geq 0.5$ and squares stars with $\rho \leq -0.5$}
\label{rhk_teff}
\end{center}
\end{figure}

We also analysed what would be the effect of temperature on the correlations between $\log R'_{HK}$ and $\log I_{H\alpha}$.
Figure \ref{hist_teff} (upper panel) shows the distributions of the correlations for the full sample (black), the positively correlated stars (hatched grey), and negatively correlated stars (filled grey).
Black vertical line is the median of the full sample with a value of 5604 K, dashed vertical line the median of the positively correlated stars with a value of 5243 K, and dotted line the median of the negatively correlated stars with a value of 5386 K.
There is an observational bias toward brighter stars and therefore hotter ones.
However, the positively correlated stars seem very well distributed across the temperature range, which implies that, relative to the full sample distribution, there are more cooler stars having positive correlations than hotter stars.
This can be easily observed in the lower panel of Fig. \ref{hist_teff}.
Stars with negative correlations appear also well distributed in effective temperature, but are only restricted to the range between $\sim$5000 and $\sim$6100 K.
It would be easier then to find positively correlated stars among the cooler dwarfs.
This effect is probably due to the fact that, in our sample, cooler stars have a tendency to be more active than the hotter ones (Fig. \ref{rhk_teff}).
All the stars in our sample with $\log R'_{HK} > -4.7$ have effective temperatures lower than 5500 K.
And as we saw before in Sect. \ref{sec:results2}, all stars with activity higher than $-4.7$ have positive correlations.

Since the Mount Wilson survey, it is known that stellar age and mean activity level are related: younger stars exhibit higher activity levels than their older counterparts \citep{baliunas1995}.
Stars with $0.55 < B - V < 0.9$ which are evolved have lower activity levels than non-evolved stars \citep{donascimento2003}.
Furthermore, \citet{wright2004} found that most of the stars classified as "flat" or "Maunder minimum", showing very low activity and no variability, were in fact evolved or subgiant stars.
Recently, \citet{schroder2013} showed that the mean activity level decreases with relative MS-age.
This confirms theorectical work by \citet{reiners2012}.
In other words, cooler K and M dwarfs did not had enough time to evolve (and decrease their activity level) so much as F-stars which evolve faster.
We therefore observe cooler stars at a relative younger stage, and consequently higher activity levels, than their hotter counterparts.

The tendency for more earlier types in our sample will then be a consequence of the bias towards less active stars due to the planetary search nature of this survey.

\section{Discussion} \label{discussion}

\subsection{Interpretation of the correlations via the effect of filaments and plages}
So, why we sometimes see stars with anti-correlations (and "anti-cycles") when we measure the flux in the H$\alpha$ line?
\citet{meunier2009} studied the contribution of plages and filaments to the $S_{MW}$ and H$\alpha$ indices for the case of the Sun.
They noted that the emission in the \ion{Ca}{ii} lines increases in the presence of plages but is almost unaffected by filaments (their contribution is negligible).
On the other hand, filaments contribute to the absorption in the flux of H$\alpha$ while plages contributes to emission.
However, the filling factor of filaments saturates at a given activity level while plages filling factor continues to increase as the activity level increases further.
This saturation will contribute to an increase of the correlation between the flux in the two line cores for higher activity levels.
For the Sun, the filaments are not only found in active regions.
They explain that the positive correlation between the two indices is due to the fact that as activity gets stronger (higher emission in the \ion{Ca}{ii} lines), for the H$\alpha$ index, which is more sensitive to filaments than the \ion{Ca}{ii} lines, the contribution of plages becomes more important than the contribution coming from filaments, because their contribution saturates at a certain activity level.
This will produce the observed strong positive correlation between the two indices for higher activity stars as observed in Fig. \ref{hist_rhk} and \ref{r_rhk_iha_mean_rhk}.
On the other hand, the low-activity stars with anti-correlation between the emission in \ion{Ca}{ii} and H$\alpha$, which appear in Fig. \ref{hist_rhk} and \ref{r_rhk_iha_mean_rhk}, can be explained if these stars have the filaments with a strong contrast (compared to plages) and which not reach the saturation limit.

The occurrence of positively correlated stars at higher activity levels and negatively correlated stars at lower activity levels that we observe in Section \ref{sec:results2} can then be explained by the effect of filaments on the flux of the H$\alpha$ line.

If the positively and negatively correlated stars are two different populations in terms of metallicity as discussed in Section \ref{sec:results3}, and if the ratio of the contrast/filing factor of filaments to plages is responsible for the anti-correlation between the flux in the \ion{Ca}{ii} H \& K and H$\alpha$ lines, then metallicity might have an effect on the presence of filaments (or their contrast and/or filling factor) in the stellar corona.
This could be used to predict the correlation between these two indices and to forecast the presence, contrast, and/or filling factor between plages and filaments for a given star.

\subsection{Comparison with M dwarfs}
\begin{figure}[tbp]
\begin{center}
\resizebox{\hsize}{!}{\includegraphics{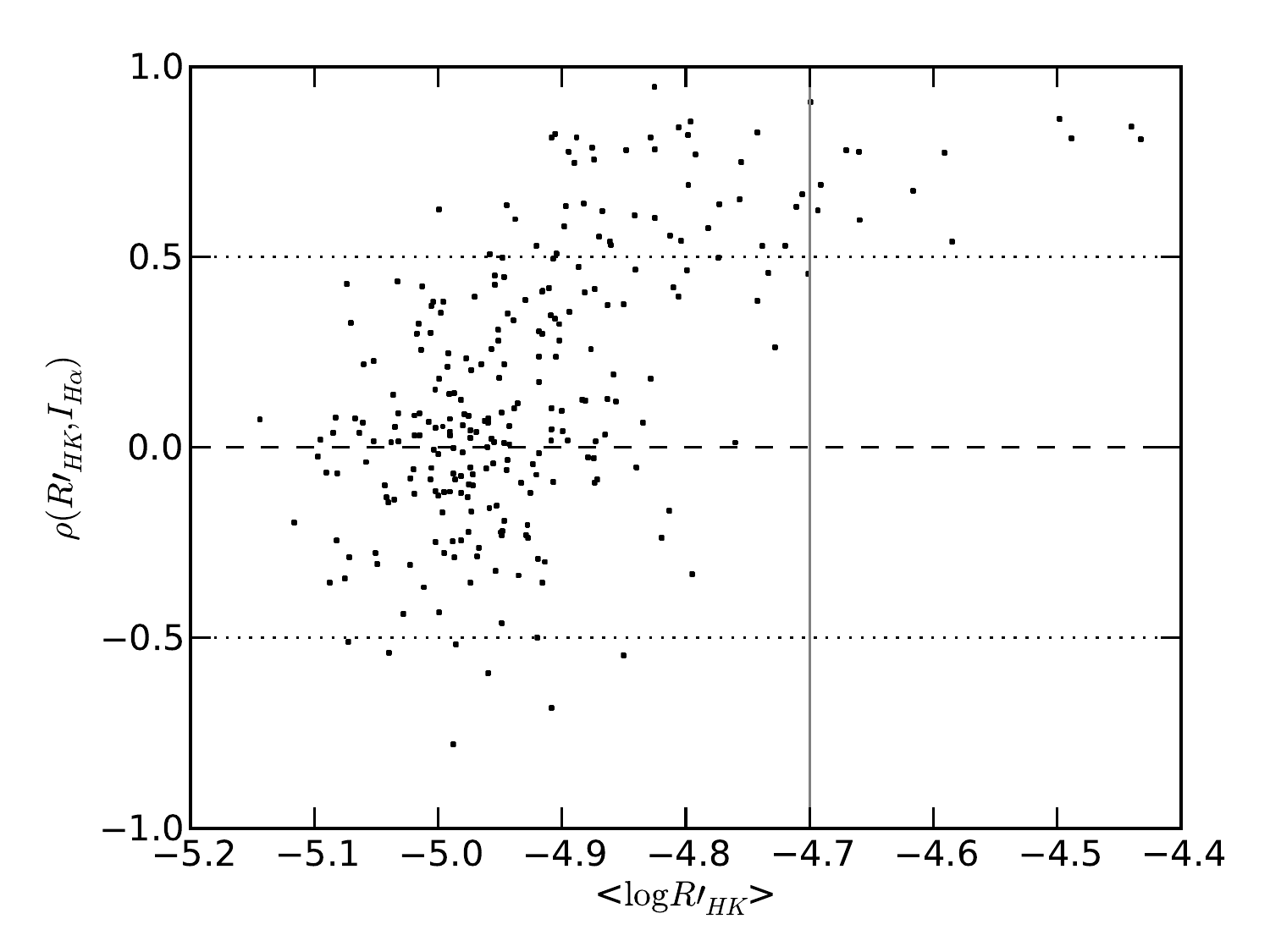}}
\caption{Correlation coefficient of the relation between $\log R'_{HK}$ and $\log I_{H\alpha}$ against mean $\log R'_{HK}$ level. The vertical line at $\log R'_{HK} = -4.7$ marks the limit after which all stars have positive correlations.}
\label{r_rhk_iha_mean_rhk}
\end{center}
\end{figure}

In a previous work, \citet{gomesdasilva2011} studied the long-term activity of 30 M0-M5 dwarfs and found hints of H$\alpha$ "anti-cycles" (inverted in comparison to the \rhk cycles) on some stars of their sample.
The potential maxima and minima of some stars were anti-correlated.
This can be an indication that the physical mechanisms responsible for the anti-correlation, and thus "anti-cycles", between the two indices are present both in solar-type stars and at least in the earlier M dwarfs.
The authors also found that after a certain value of $S$-index activity all M-dwarfs in their sample have positive correlations, and found a case of an anti-correlation with correlation coefficient value lower than $-0.5$ in the least active stars zone (see their Fig. 3).
We should note, however, that their $S$-index was not corrected for the effects of photospheric flux, and therefore there is a temperature contribution to the mean index values that will vary from star to star.
Nevertheless, their distribution of correlations is compatible with ours in the sense that after a certain level of activity all active stars have positive correlations, and there are some cases of low activity stars with anti-correlations (Fig. \ref{r_rhk_iha_mean_rhk}).
Since both FGK and early M stars have radiative cores with convective envelopes, their activity phenomena might not be too different (contrary to later M dwarfs which are fully convective).
Therefore, if the contribution of filaments to the H$\alpha$ absorption is the sole responsible to the anti-correlation between the flux in the \ion{Ca}{ii} and H$\alpha$ lines, then it is possible that this phenomenon is occurring in a similar way for the two types of stars.

Further studies of the correlations between the two indices for later M dwarfs would be interesting to understand how the behaviour of the two indices evolve in spectral type and infer about the presence of filaments in fully convective stars.

\section{Conclusions} \label{conclusions}

We studied the correlation between the flux in the \ion{Ca}{ii} H \& K and H$\alpha$ lines via two activity indices, $R'_{HK}$ and $I_{H\alpha}$, corrected for photospheric flux.
A sample of 271 low activity FGK stars, observed during $\sim$9 years, was used to this effect.
This study was the larger scale study (in both sample number and time-span) of the correlation between these two chromospheric indices for solar-type stars.

We detected significant activity cycles in 69 stars (26\% of our sample) using the $\log R'_{HK}$ index, but only in 9 stars (3.3\%) using $\log I_{H\alpha}$.
The H$\alpha$ line is not so sensitive at measuring long-term variations as the \ion{Ca}{ii} lines.
We also found a great variety of correlation coefficients, in the range $-0.78 \leq \rho \leq 0.95$, similar to what was found by \citet{cincunegui2007b}.
Possible explanations for this variety are given by \citet{meunier2009} and include the spatial distribution and difference in contrast of filaments relative to plages.

To study the correlation between the \rhk and \iha indices we first selected only the stars showing "strong" long-term correlations between the two indices by applying a rigorous selection criteria based on variability F-tests, using FAPs on the correlation coefficients and binning the data to 100-day bins.
This selection criteria returned a sample of 12 stars where two of them have anti-correlations and the rest positive correlations.
We observed that the two stars with anti-correlations have tendency to have lower activity levels and super-solar metallicity when compared to the positively correlated stars.

Since this rigorous selection returned a small number of stars, we relaxed the selection criteria to increase our sample and study the trends found with the rigorous selection.
Using this selection criteria we found that:
\begin{itemize}

\item
58 stars (21\% out of 271) have positive correlations (with $\rho \geq 0.5$) and 8 stars (3\% out of 271) show anti-correlations (with $\rho \leq -0.5$).
These numbers are compatible with those found by \citet{gomesdasilva2011} for early-M dwarfs.
Some of the stars with strong anti-correlations show "anti-cycles" measured in $\log I_{H\alpha}$: negative activity cycles when compared to those measured by $\log R'_{HK}$.

\item
The stars with positive correlation between the two indices have a tendency to be more active than those with negative correlations.
In fact, all the stars with $\log R'_{HK} \geq -4.7$ have positive correlation between the indices.
We interpret this behaviour using \citet{meunier2009} results that after a certain level of activity, the contribution to absorption in the H$\alpha$ line by filaments saturates, and only plages contribute to emission in both \ion{Ca}{ii} and H$\alpha$.

\item
We also found a tendency for the stars with negative correlations to be more metal rich than the rest of the sample and that this holds for stars of similar activity level.

\item
The distribution of the correlations in effective temperature was also studied, and we detected that, in relative terms, there are more cooler stars showing positive correlations than hotter stars.
This is because, in our sample, cooler stars are in general more active than hotter ones, and there is a tendency for the more active stars to have positive correlations.

\item
As a parallel result, we found that our \ha index can be used to estimate the effective temperature of a low-activity FGK star.
\end{itemize}

These results might affect planet detections since activity is one of the main source of errors in radial velocity (and photometric) measurements.
It would be interesting to compare the correlation between the flux in the \ion{Ca}{ii} H \& K and H$\alpha$ lines with the measured radial velocity and see if this correlation has any effect on the observed radial velocity signal.

\acknowledgements{
This work has been supported by the European Research Council/European Community under the FP7 through a Starting Grant, as well as in the form of a grant reference PTDT/CTE-AST/098528/2008, funded by Funda\c{c}\~ao para a Ci\^encia e a Tecnologia (FCT), Portugal. J.G.S. would like to thank the financial support given by FCT in the form of a scholarship, namely SFRH/BD/64722/2009. N.C.S. would further like to thank the support from FCT through a Ci\^encia 2007 contract funded by FCT/MCTES (Portugal) and POPH/FSE (EC). I.B. also acknowledges the financial support given by FCT in the form of grant reference SFRH/BPD/81084/2011.}

\bibliographystyle{aa} 
\bibliography{paper_corr} 

\Online

\appendix

\section{The $I_{H\alpha}$ hydrogen line based activity index}\label{a1}
The H$\alpha$ index is calculated from the fraction of the flux in the H$\alpha$ line centre to the flux in two continuum reference bands, one bluer other redder than the hydrogen line.
This is sufficient if we are interested in determining the activity evolution over time for a star.
However, stars with different colours have different amounts of flux in the continuum, and this will make the average H$\alpha$ level not comparable between different stars due to a systematic error introduced by the photospheric flux interference in the measurements \citep[e.g.][]{cincunegui2007b}.

\begin{figure}[tbp]
\begin{center}
\resizebox{\hsize}{!}{\includegraphics{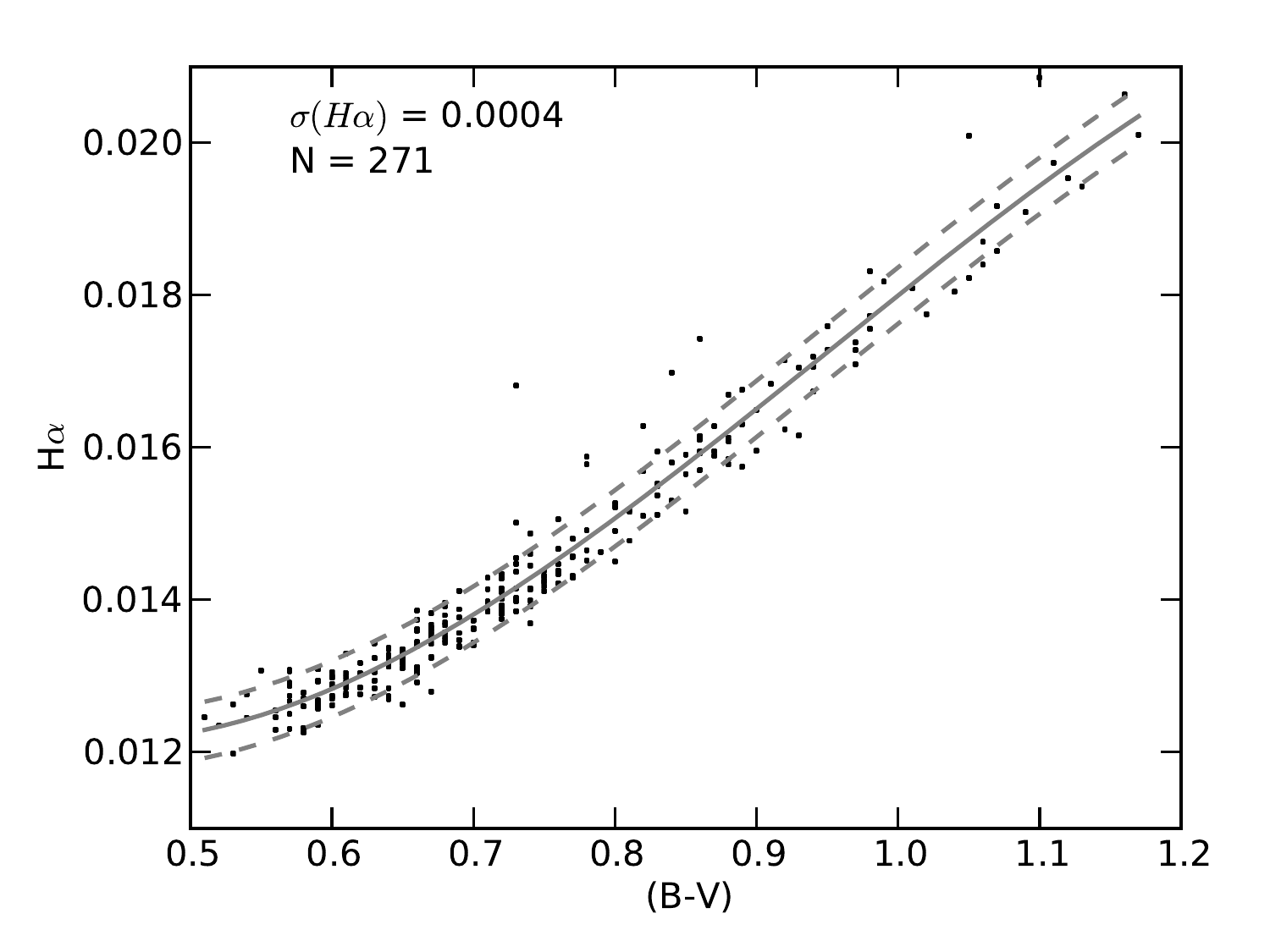}}
\caption{Calibration of H$\alpha$ index as a function of $(B-V)$ colour. The solid curve line is the best fit to the data and the dashed lines correspond to the 1--$\sigma$ limits.}
\label{ha_bv}
\end{center}
\end{figure}

\begin{figure}[tbp]
\begin{center}
\resizebox{\hsize}{!}{\includegraphics{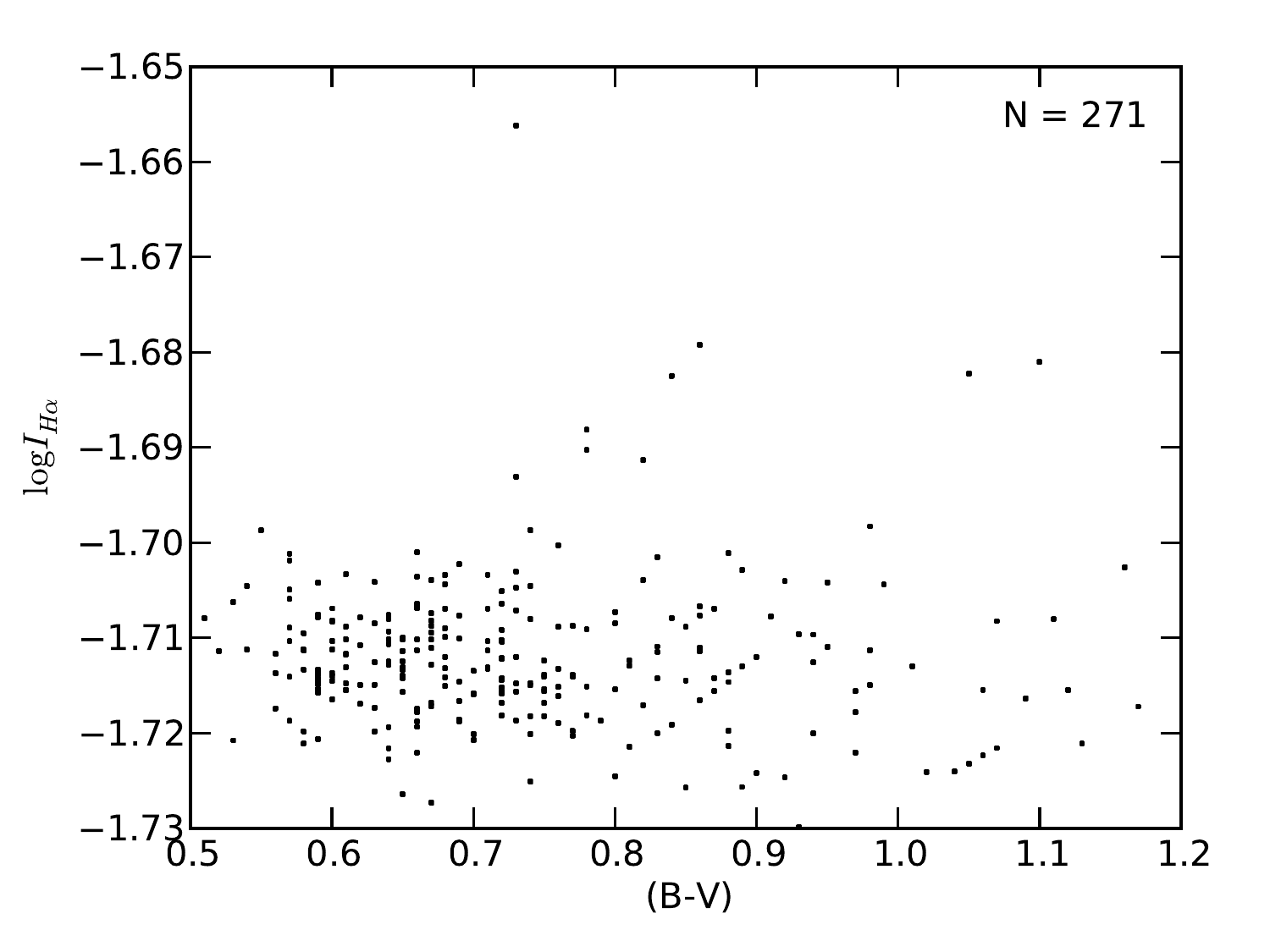}}
\caption{Dependence of the $\log I_{H\alpha}$ index on stellar colour.}
\label{iha_bv_feh}
\end{center}
\end{figure}

To be able to compare the average H$\alpha$ index between different stars the photospheric contribution to the index need to be taken into account.
Figure \ref{ha_bv} shows the calibration of H$\alpha$ to the effects of stellar colour.
We fitted H$\alpha$ to $(B-V)$ using a cubic polynomial which resulted in a standard deviation of the fit of $0.0004$.
Our corrected $I_{H\alpha}$ activity index is then
\begin{dmath}
I_{H\alpha} = \hbox{H}\alpha + 0.019(B-V)^3 - 0.054(B-V)^2 + 0.037(B-V).
\end{dmath}
Figure \ref{iha_bv_feh} shows that the resulting index is not dependent on $(B-V)$ and can therefore be used to compare the activity level of stars of different colour.
This calibration is valid for main sequence stars with $(B-V)$ colour between 0.5 and 1.2, and mean H$\alpha$ activity levels between 0.012 and 0.021.

\section{Estimating effective temperature using the flux in H$\alpha$ line}\label{a3}
The H$\alpha$ line wings are known to be a proxy of effective temperature \citep[e.g.][]{fuhrmann1993,barklem2002} and are sometimes used to confirm more accurate results by other methods.
For example, \citet{bouchy2008} used the wings of the H$\alpha$ line to derive a temperature of $5450 \pm 120$ K for the star CoRoT-Exo-2.
\citet{sozzetti2007} compared the H$\alpha$ wings to those of synthetic spectra to obtain a temperature region of 5750-6000 K for TrES-2 \citep[other authors that used the same technique as a rogue estimate of temperature include][]{santos2006,sozzetti2009}.

\begin{figure}[tbp]
\begin{center}
\resizebox{\hsize}{!}{\includegraphics{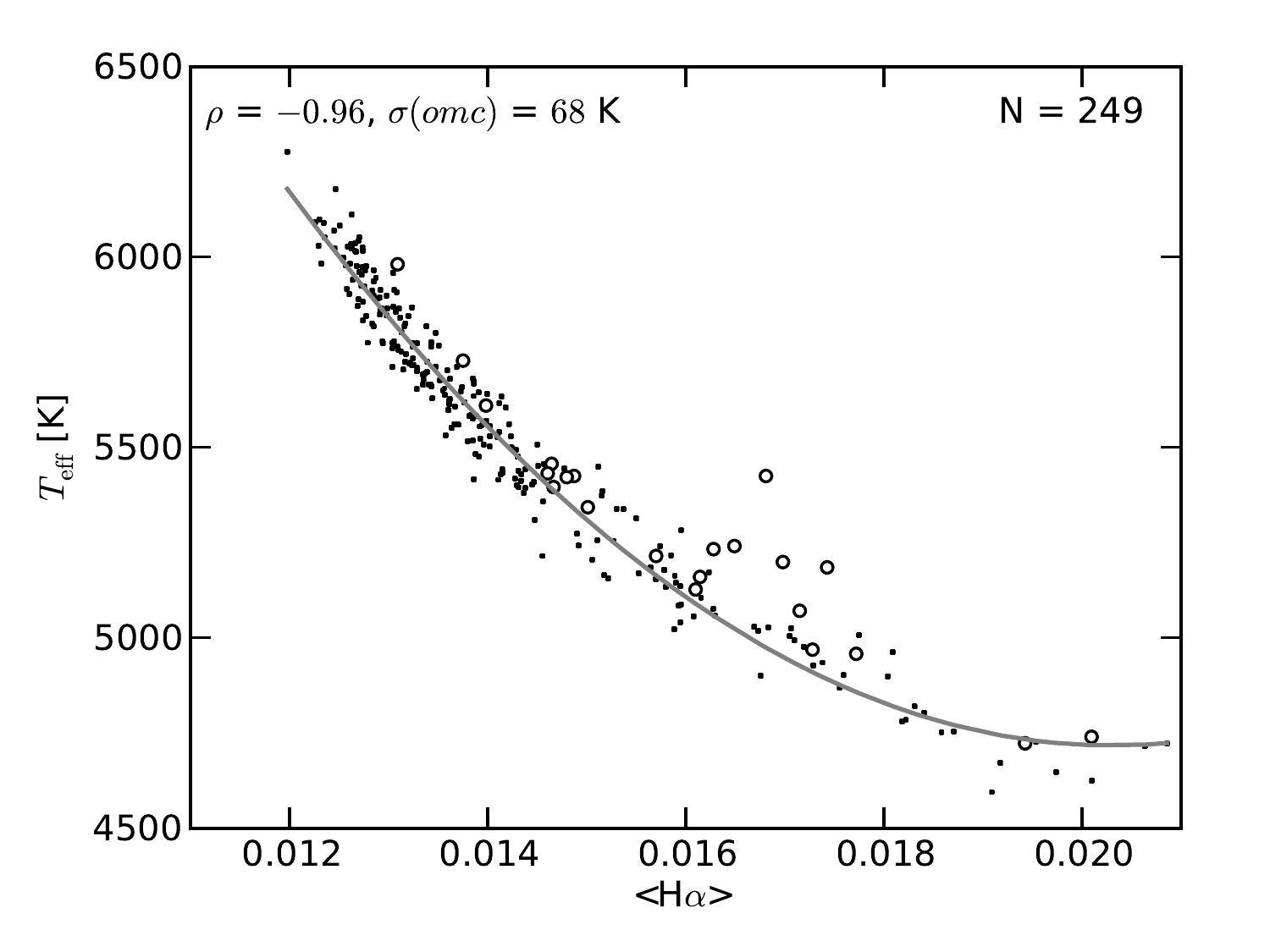}}
\caption{Calibration of $T_{\mathrm{eff}}$ by using H$\alpha$ activity index for all main sequence stars except the most active ($\log I_{HK} \geq -4.75$, open circles). The grey line is the best quadratic fit to the data.}
\label{teff_ha}
\end{center}
\end{figure}

We found that our H$\alpha$ activity index is also a good proxy of $T_{\mathrm{eff}}$.
Figure \ref{teff_ha} shows a quadratic fit to the correlation between these parameters.
Active stars (open circles) were not used due to their contribution to a larger scatter.
We obtained an rms of the $T_{\mathrm{eff}}$ residuals of $\sigma = 68$ K, and a correlation coefficient of $\rho = -0.96$.
The calibrated $T_{\mathrm{eff}}$ is of the form
\begin{dmath}
T_{\mathrm{eff}} = 10^{-4}~(2109~\hbox{H}\alpha^2 - 85.65~\hbox{H}\alpha + 1.341).
\end{dmath}
This equation can be used for dwarfs with $\log I_{HK} \leq -4.70$, mean H$\alpha$ activity in the range  $0.012 \leq \mathrm{H}\alpha \leq 0.021$, and effective temperatures in the range $4600 \leq T_{\mathrm{eff}} \leq 6280$ K.

\begin{figure}[tbp]
\begin{center}
\resizebox{\hsize}{!}{\includegraphics{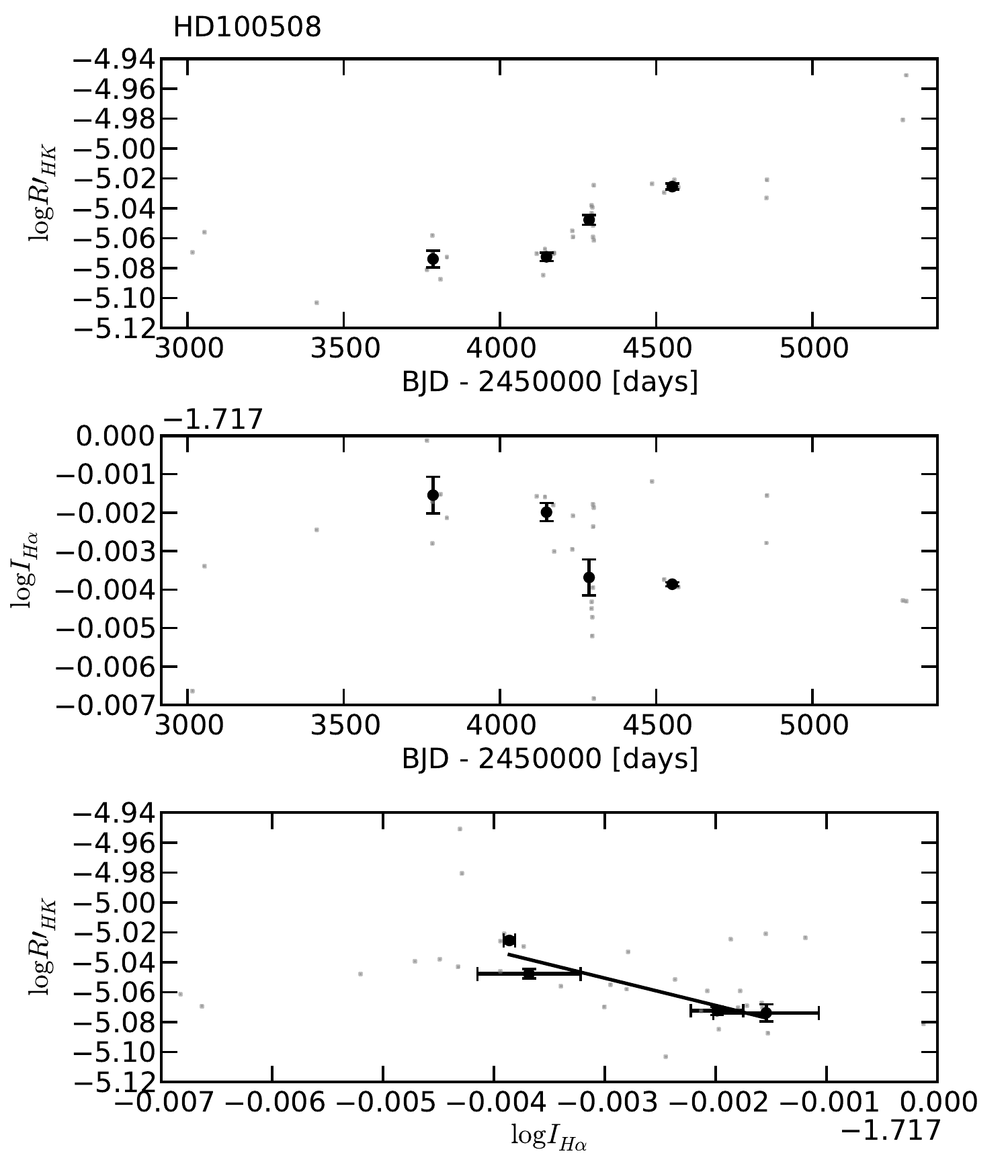}}
\resizebox{\hsize}{!}{\includegraphics{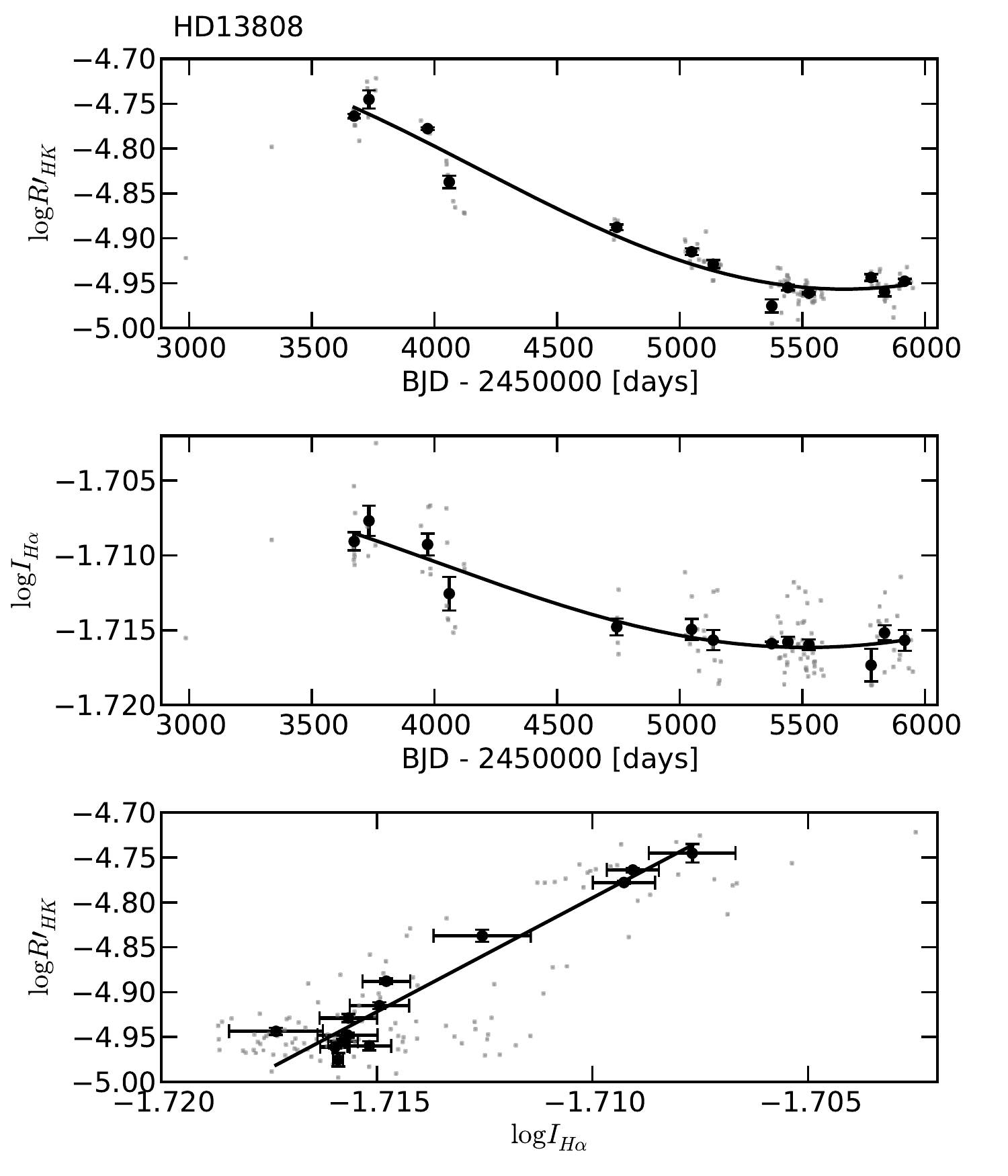}}
\caption{Time-series of $\log R'_{HK}$, $\log I_{H\alpha}$, and correlation between the two for the 12 stars with "strong" correlations. Grey dots are nightly averaged data, black points are binned data. Error bars are the standard errors on the mean. Black lines are best fit to the binned data. A sinusoid will appear in the time-series if well fitted, i.e., having $p(F) \leq 0.05$.}
\label{all_strong_plots}
\end{center}
\end{figure}

\begin{figure}[tbp]
\ContinuedFloat
\begin{center}
\resizebox{\hsize}{!}{\includegraphics{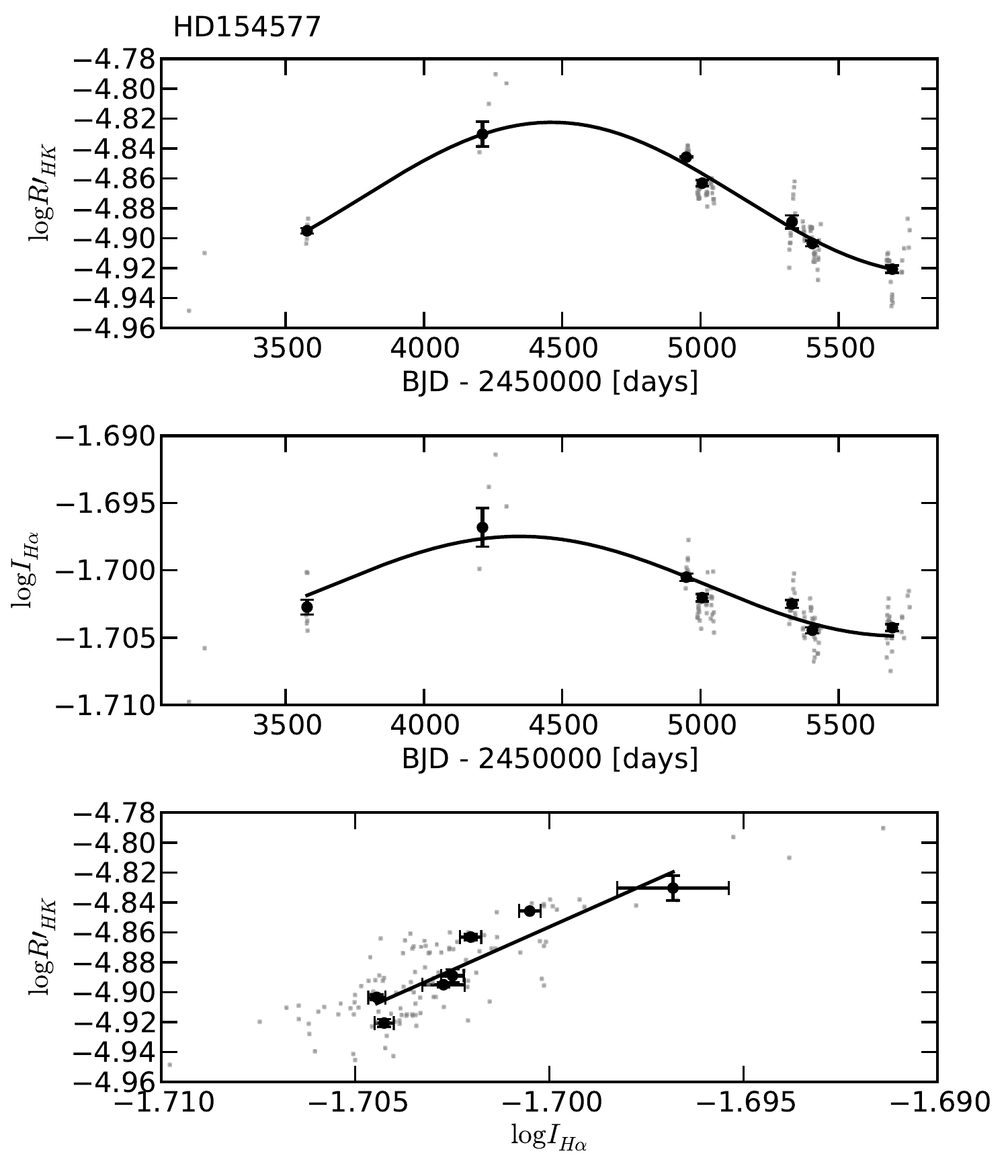}}
\resizebox{\hsize}{!}{\includegraphics{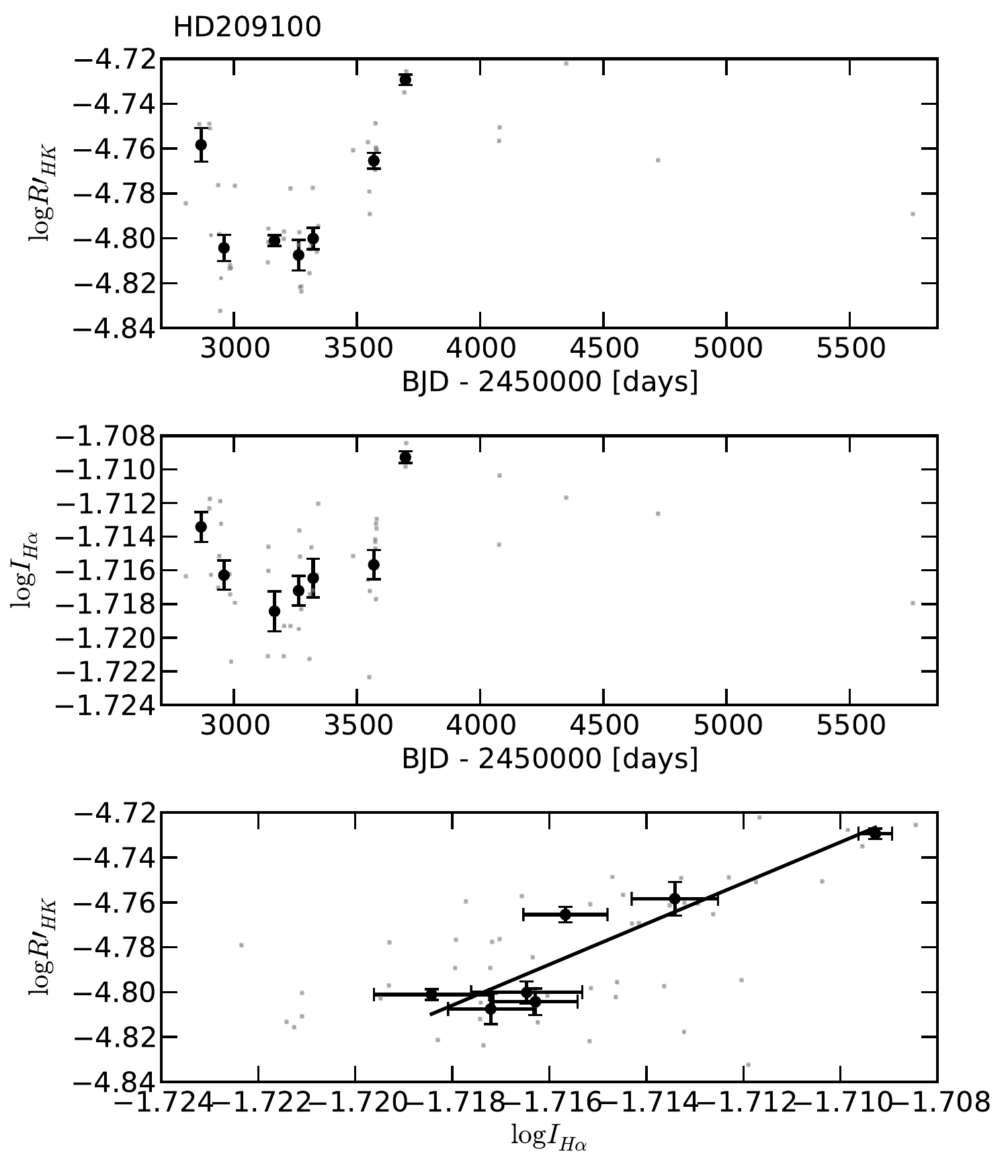}}
\caption{Continued.}
\end{center}
\end{figure}

\begin{figure}[tbp]
\ContinuedFloat
\begin{center}
\resizebox{\hsize}{!}{\includegraphics{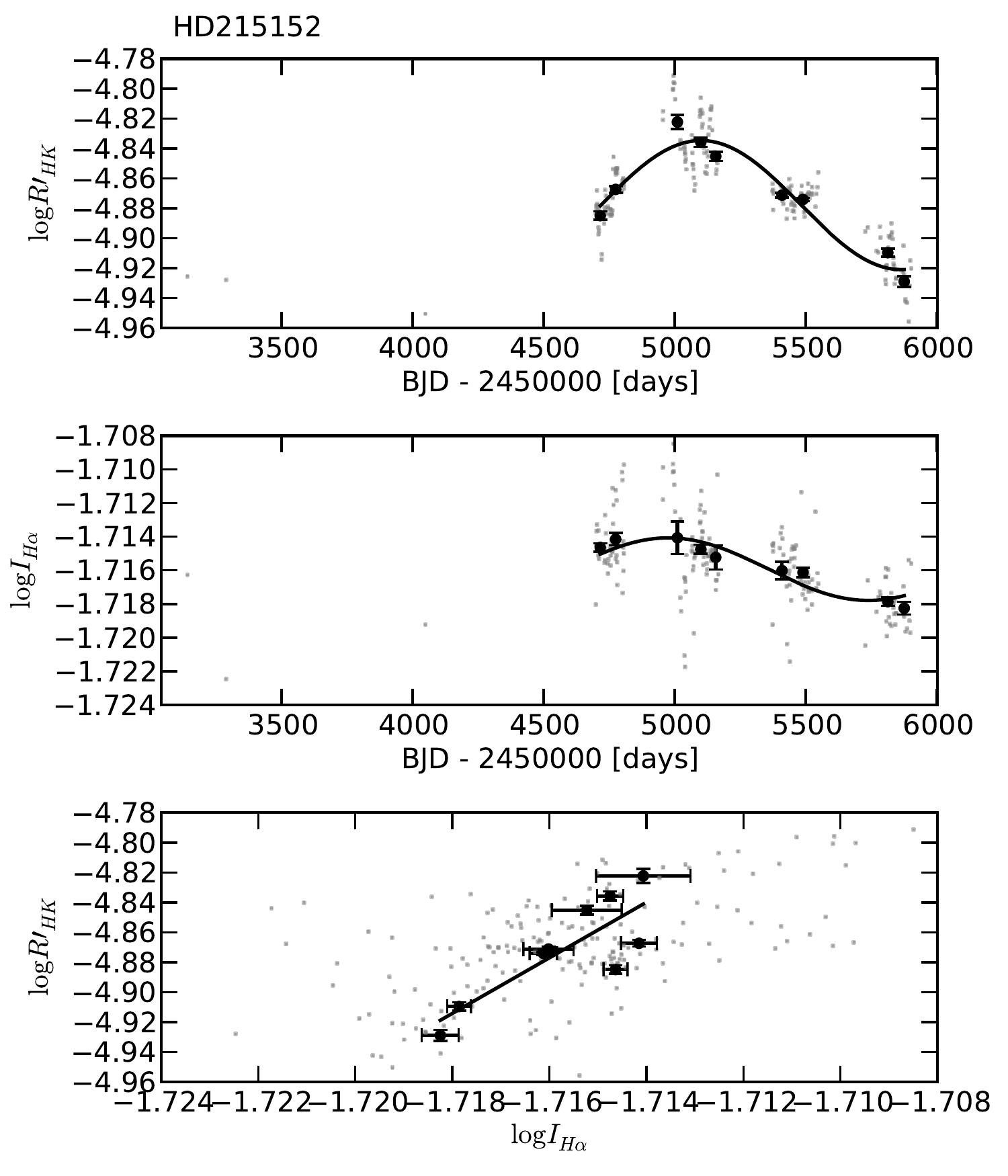}}
\resizebox{\hsize}{!}{\includegraphics{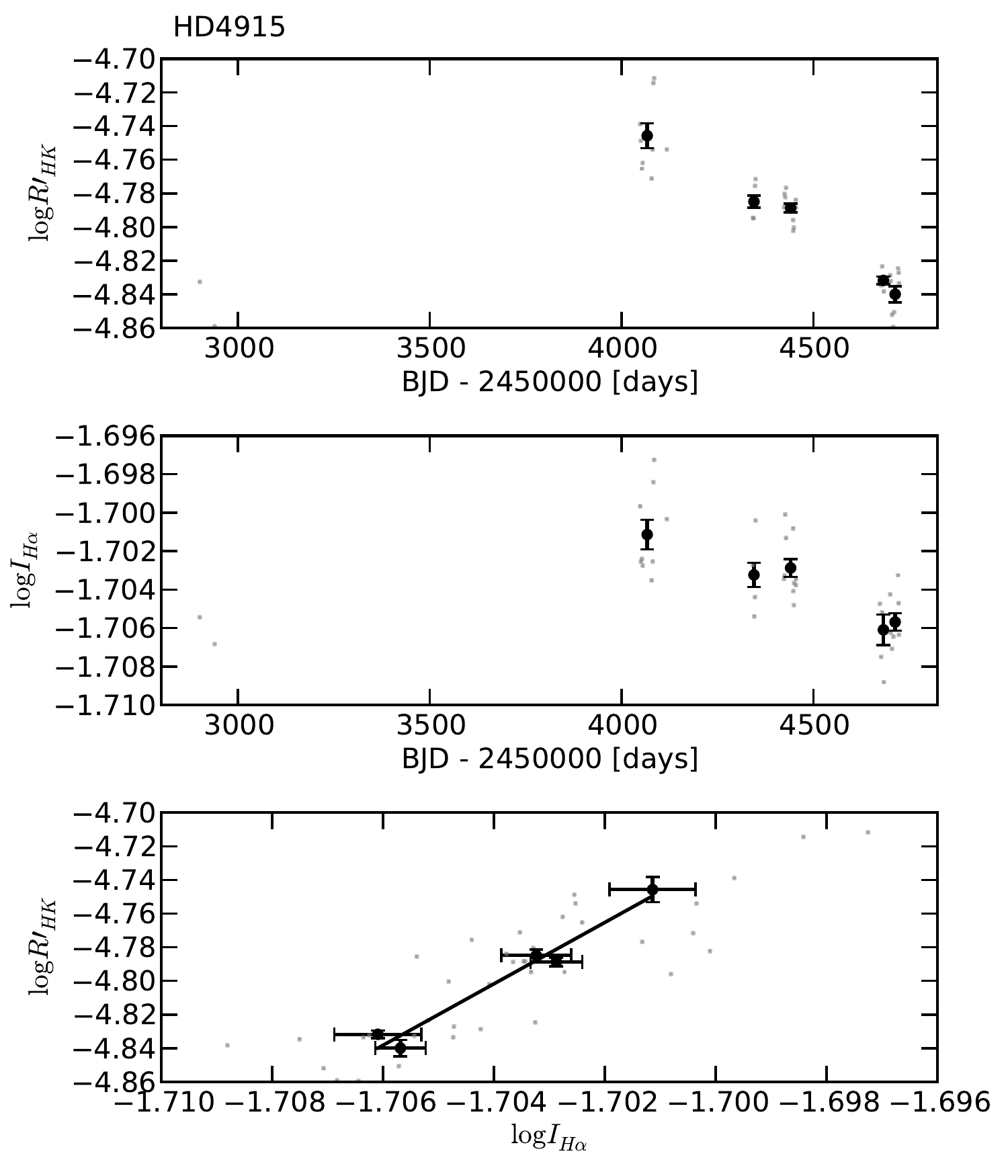}}
\caption{Continued.}
\end{center}
\end{figure}

\begin{figure}[tbp]
\ContinuedFloat
\begin{center}
\resizebox{\hsize}{!}{\includegraphics{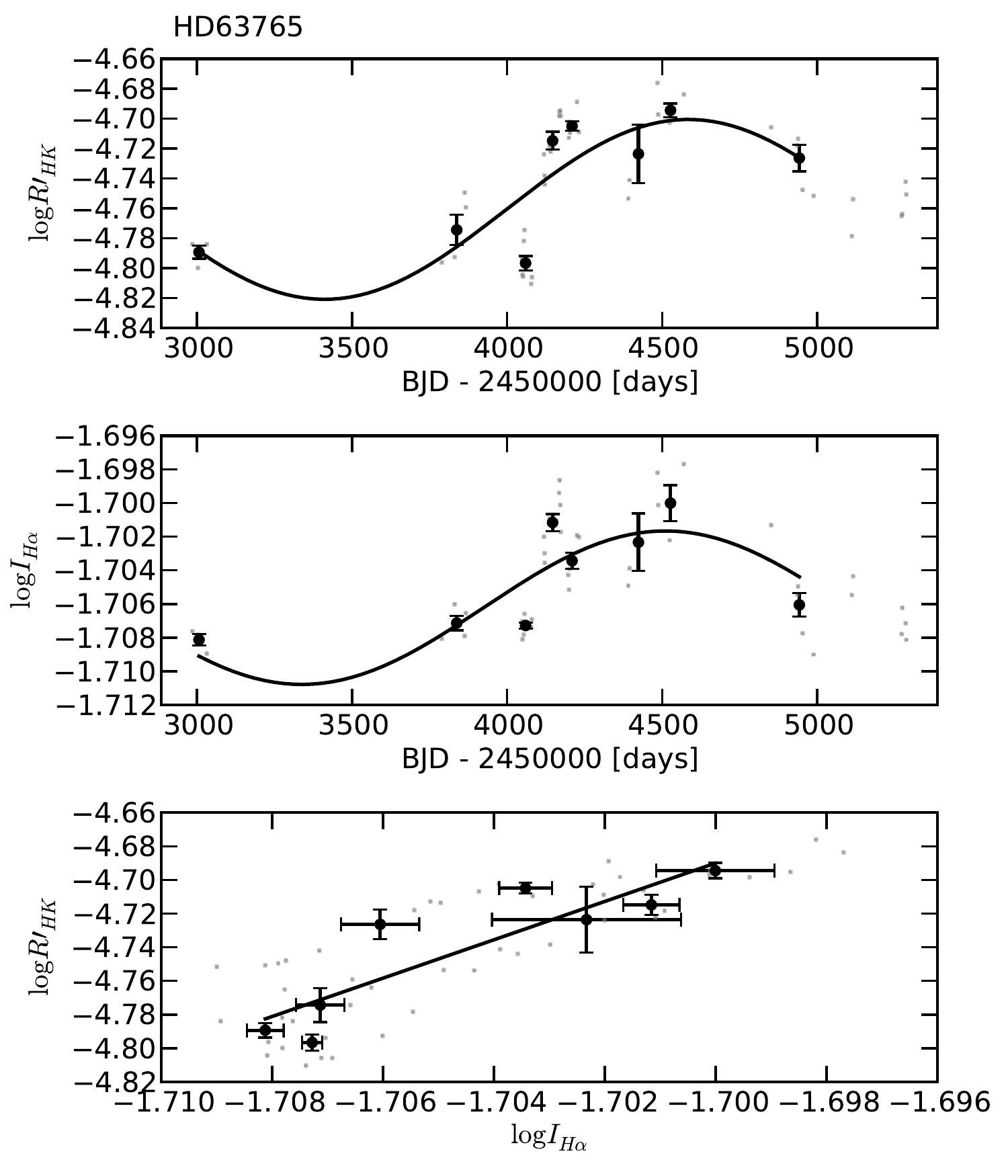}}
\resizebox{\hsize}{!}{\includegraphics{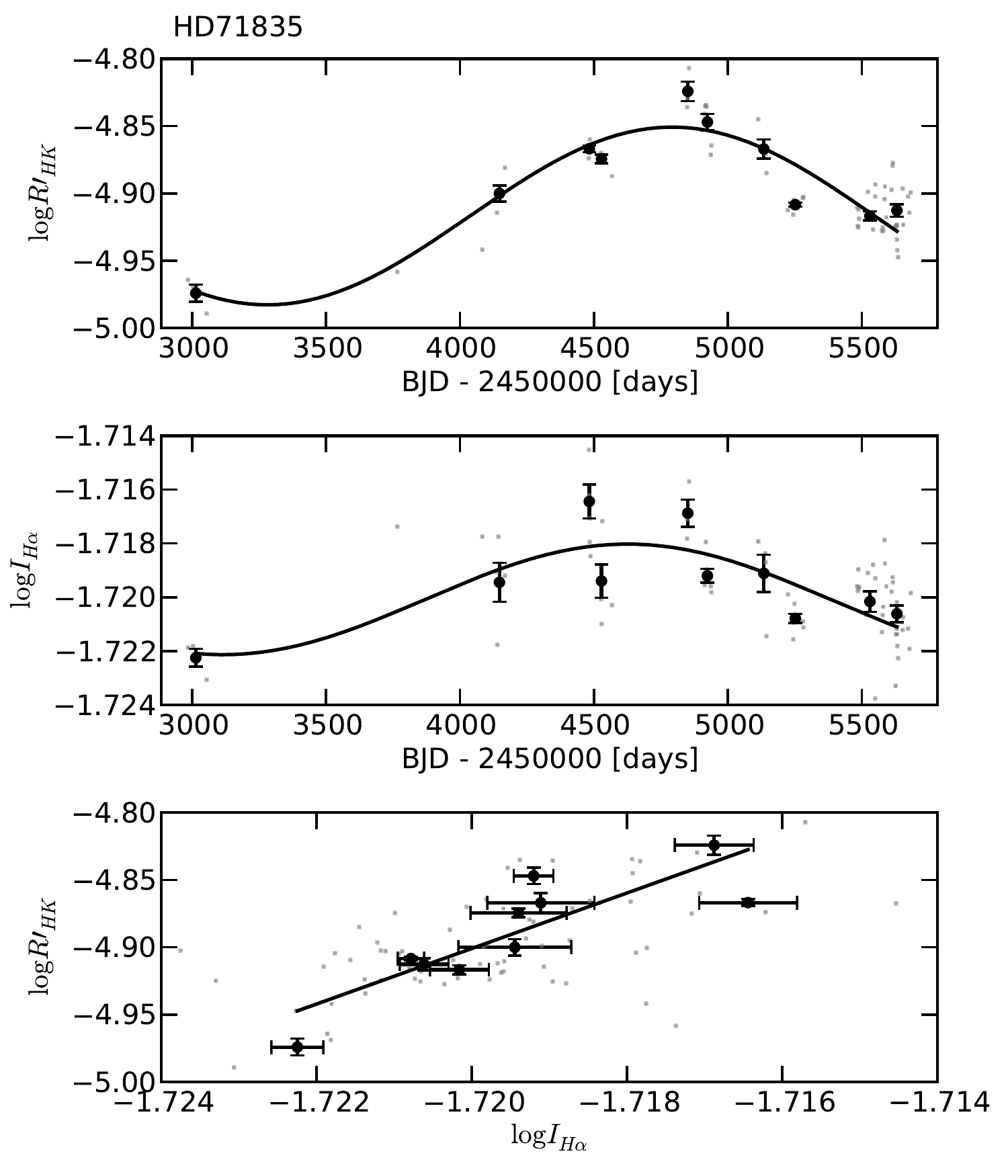}}
\caption{Continued.}
\end{center}
\end{figure}

\begin{figure}[tbp]
\ContinuedFloat
\begin{center}
\resizebox{\hsize}{!}{\includegraphics{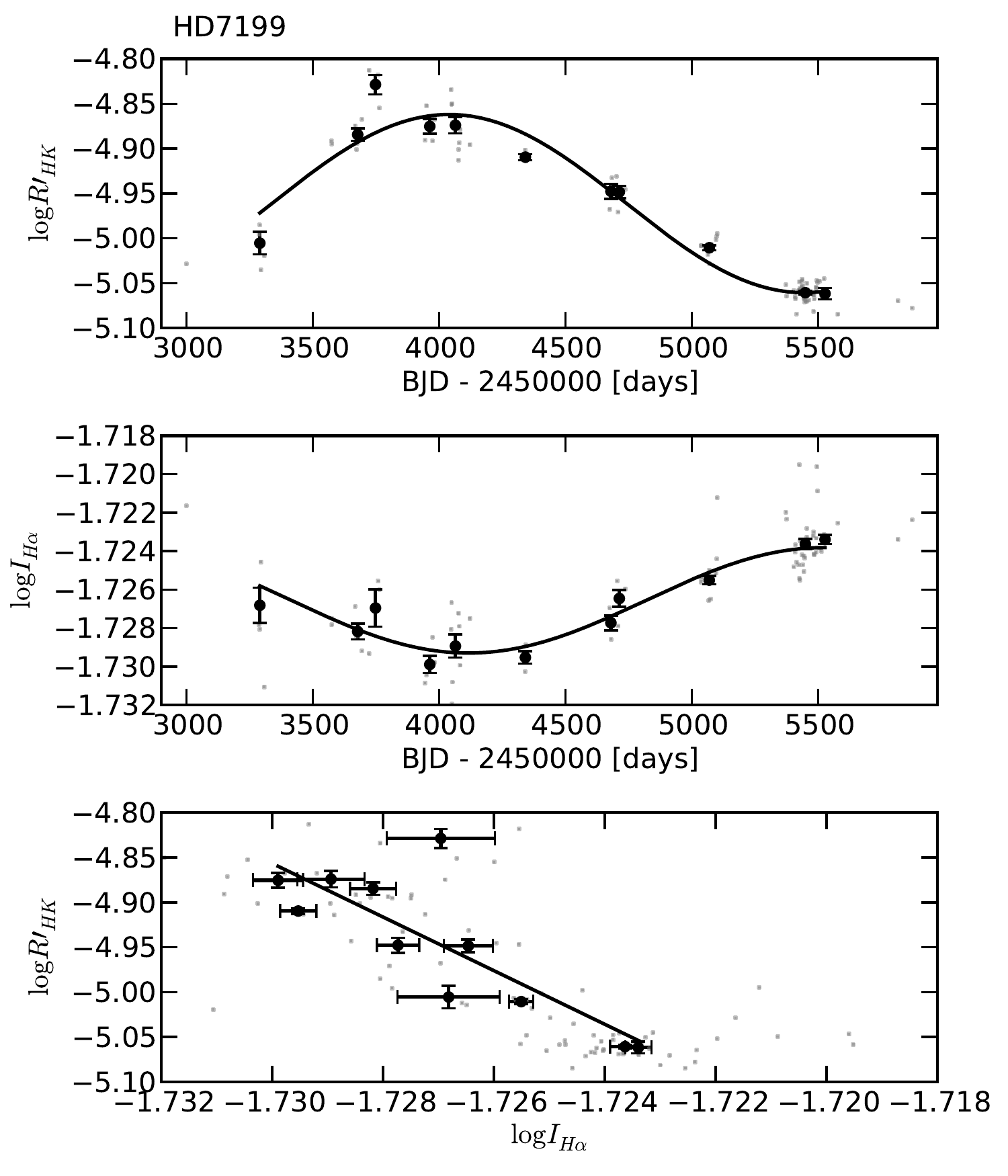}}
\resizebox{\hsize}{!}{\includegraphics{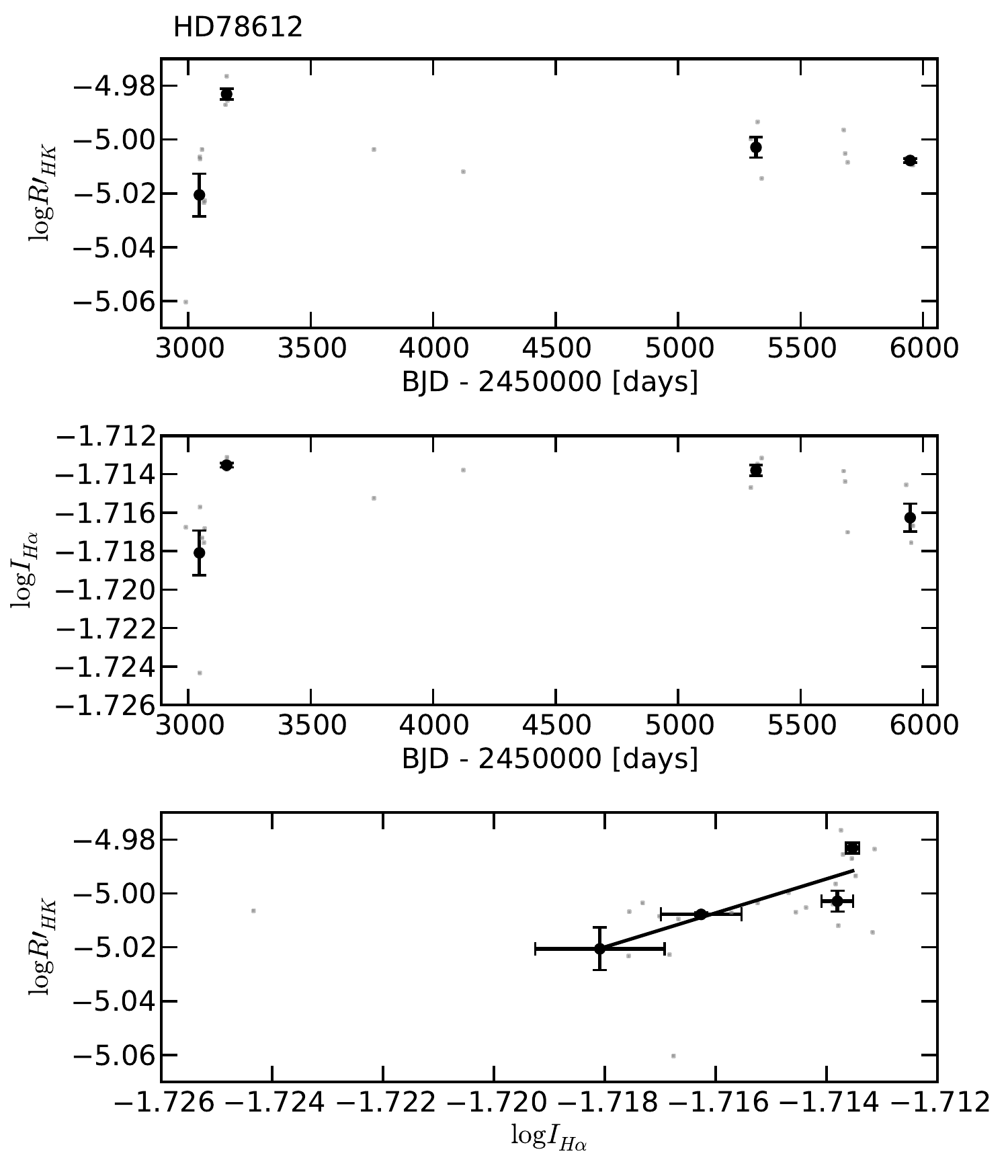}}
\caption{Continued.}
\end{center}
\end{figure}

\begin{figure}[tbp]
\ContinuedFloat
\begin{center}
\resizebox{\hsize}{!}{\includegraphics{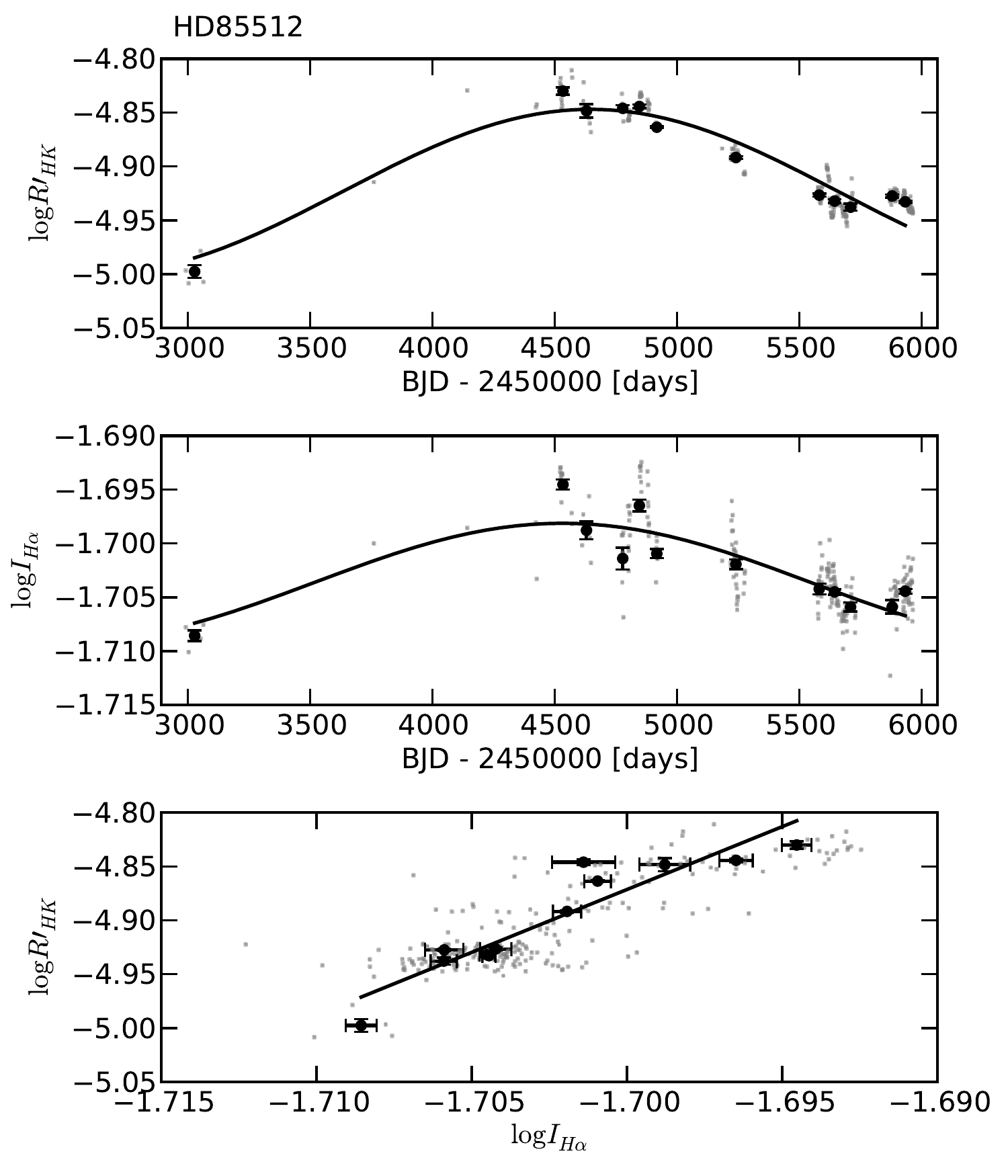}}
\resizebox{\hsize}{!}{\includegraphics{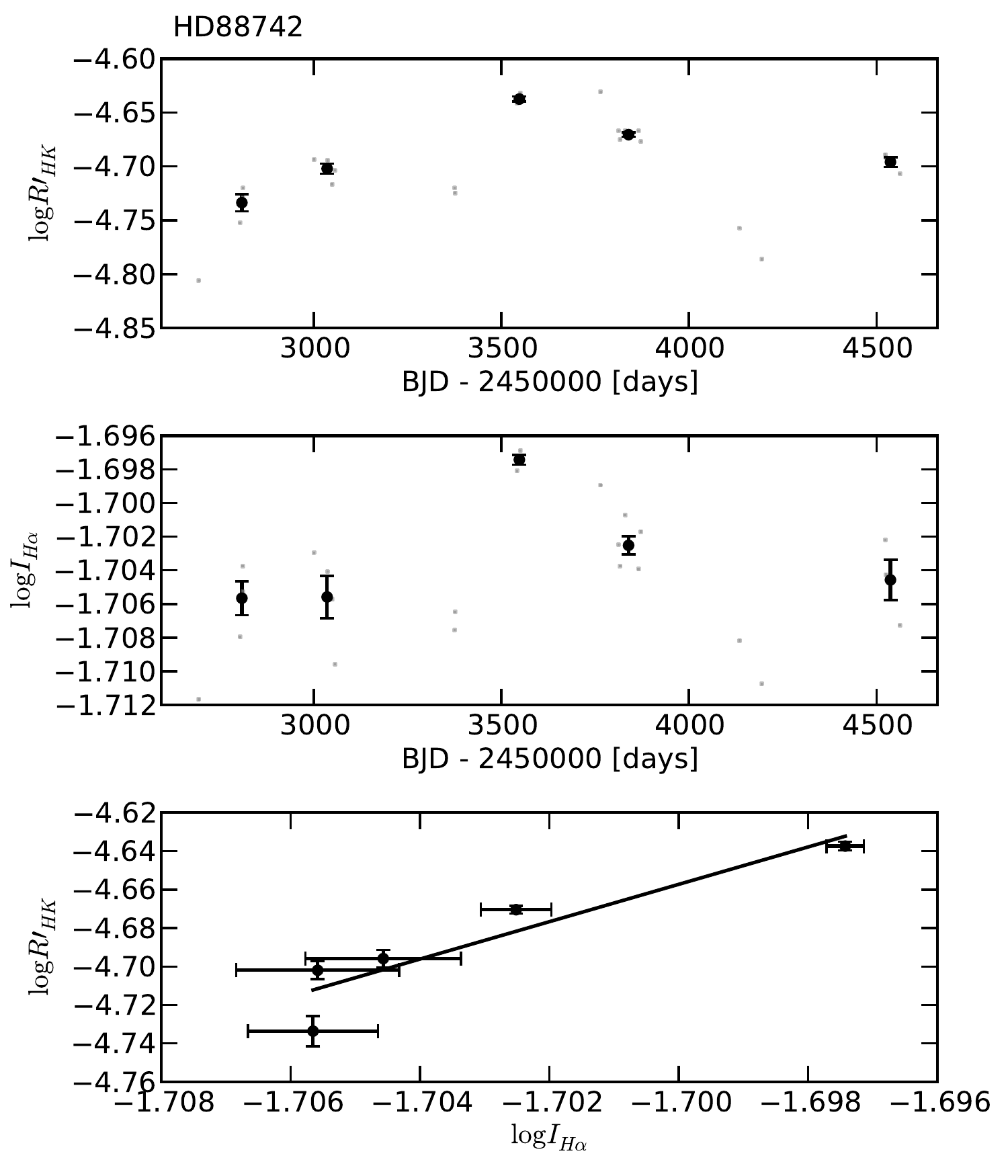}}
\caption{Continued.}
\end{center}
\end{figure}


\begin{table*}[tbp]
\caption{Parameters for the 66 stars with $\lvert \rho \rvert \geq 0.5$ from the nightly averaged 271-star sample..}
\label{table3}
\centering
\begin{tabular}{l c c c c c c c c c c c} \\
\hline
\hline
\multicolumn{1}{l}{Star} &
\multicolumn{1}{c}{$N_{obs}$} &
\multicolumn{1}{c}{$T_{span}$} &
\multicolumn{1}{c}{$\rho$} &
\multicolumn{1}{c}{[Fe/H]} &
\multicolumn{1}{c}{$T_{\textrm{eff}}$} &
\multicolumn{1}{c}{$\langle\log R'_{HK}\rangle$} &
\multicolumn{1}{c}{$\sigma(\log R'_{HK})$} &
\multicolumn{1}{c}{$\langle\log I_{H\alpha}\rangle$} &
\multicolumn{1}{c}{$\sigma(\log I_{H\alpha})$}  \\
\multicolumn{1}{c}{} &
\multicolumn{1}{c}{} &
\multicolumn{1}{c}{[days]} &
\multicolumn{1}{c}{} &
\multicolumn{1}{c}{} &
\multicolumn{1}{c}{[K]} &
\multicolumn{1}{c}{} &
\multicolumn{1}{c}{} &
\multicolumn{1}{c}{} &
\multicolumn{1}{c}{} &
\multicolumn{1}{c}{} \\
\hline
HD105837	&   21	& 2651	&    $ 0.78$	&	$-0.51 \pm 0.01$	&	$5907 \pm  17$	&  $-4.825$	&  0.019	& $-1.7012$	& 0.0024 \\
HD106275	&   17	& 2648	&    $ 0.62$	&	$-0.09 \pm 0.03$	&	$5059 \pm  45$	&  $-4.867$	&  0.065	& $-1.7130$	& 0.0037 \\
HD109200	&  118	& 2866	&    $ 0.60$	&	$-0.31 \pm 0.02$	&	$5134 \pm  38$	&  $-4.938$	&  0.033	& $-1.7079$	& 0.0023 \\
HD110619	&   17	& 2677	&    $ 0.76$	&	$-0.41 \pm 0.01$	&	$5613 \pm  15$	&  $-4.874$	&  0.019	& $-1.7065$	& 0.0019 \\
HD114747	&   32	& 1472	&    $-0.55$	&	$ 0.21 \pm 0.04$	&	$5172 \pm  57$	&  $-4.850$	&  0.073	& $-1.7246$	& 0.0027 \\
HD119638	&   31	& 3276	&    $-0.50$	&	$-0.15 \pm 0.01$	&	$6069 \pm  16$	&  $-4.920$	&  0.015	& $-1.7112$	& 0.0017 \\
HD119782	&   14	& 2194	&    $ 0.69$	&	$-0.07 \pm 0.02$	&	$5160 \pm  34$	&  $-4.691$	&  0.020	& $-1.7067$	& 0.0028 \\
HD124364	&   13	& 2635	&    $ 0.81$	&	$-0.27 \pm 0.01$	&	$5584 \pm  14$	&  $-4.828$	&  0.037	& $-1.7039$	& 0.0025 \\
HD125072	&   24	& 2173	&    $-0.59$	&	$ 0.18 \pm 0.07$	&	$5007 \pm 103$	&  $-4.959$	&  0.052	& $-1.7241$	& 0.0025 \\
HD125455	&   17	& 2164	&    $ 0.63$	&	$-0.18 \pm 0.02$	&	$5162 \pm  41$	&  $-4.897$	&  0.043	& $-1.7156$	& 0.0032 \\
HD13060 	&   18	& 2522	&    $ 0.60$	&	$ 0.02 \pm 0.03$	&	$5255 \pm  45$	&  $-4.825$	&  0.061	& $-1.7073$	& 0.0023 \\
HD130992	&   18	& 1830	&    $ 0.61$	&	$-0.13 \pm 0.06$	&	$4898 \pm  75$	&  $-4.841$	&  0.022	& $-1.7241$	& 0.0037 \\
HD13789 	&   10	&  397	&    $ 0.86$	&	$-0.06 \pm 0.06$	&	$4740 \pm  71$	&  $-4.498$	&  0.026	& $-1.6822$	& 0.0049 \\
HD13808 	&  128	& 2964	&    $ 0.81$	&	$-0.20 \pm 0.03$	&	$5087 \pm  41$	&  $-4.908$	&  0.074	& $-1.7142$	& 0.0033 \\
HD140901	&   24	& 1545	&    $ 0.63$	&	$ 0.09 \pm 0.01$	&	$5610 \pm  21$	&  $-4.711$	&  0.032	& $-1.7103$	& 0.0019 \\
HD14374 	&   17	& 1804	&    $ 0.60$	&	$-0.04 \pm 0.02$	&	$5425 \pm  24$	&  $-4.659$	&  0.031	& $-1.6987$	& 0.0034 \\
HD144585	&   14	& 2538	&    $-0.51$	&	$ 0.33 \pm 0.02$	&	$5914 \pm  22$	&  $-5.073$	&  0.019	& $-1.7188$	& 0.0020 \\
HD145666	&   20	& 1410	&    $ 0.64$	&	$-0.04 \pm 0.01$	&	$5958 \pm  12$	&  $-4.773$	&  0.014	& $-1.7069$	& 0.0015 \\
HD148303	&   25	& 2162	&    $ 0.78$	&	$-0.03 \pm 0.06$	&	$4958 \pm  91$	&  $-4.660$	&  0.041	& $-1.7113$	& 0.0051 \\
HD154577	&  123	& 2606	&    $ 0.81$	&	$-0.70 \pm 0.02$	&	$4900 \pm  37$	&  $-4.888$	&  0.030	& $-1.7029$	& 0.0025 \\
HD157830	&   50	& 2624	&    $ 0.77$	&	$-0.25 \pm 0.01$	&	$5540 \pm  16$	&  $-4.792$	&  0.033	& $-1.7023$	& 0.0031 \\
HD162236	&   14	&  965	&    $ 0.62$	&	$-0.12 \pm 0.02$	&	$5343 \pm  25$	&  $-4.693$	&  0.030	& $-1.6931$	& 0.0032 \\
HD16297 	&   10	& 2052	&    $ 0.66$	&	$-0.01 \pm 0.02$	&	$5422 \pm  22$	&  $-4.706$	&  0.031	& $-1.7088$	& 0.0033 \\
HD172513	&   40	& 1081	&    $ 0.50$	&	$-0.05 \pm 0.01$	&	$5500 \pm  18$	&  $-4.774$	&  0.020	& $-1.7154$	& 0.0019 \\
HD18386 	&   14	&  405	&    $ 0.67$	&	$ 0.14 \pm 0.02$	&	$5457 \pm  29$	&  $-4.616$	&  0.039	& $-1.7152$	& 0.0036 \\
HD18719 	&   12	&  340	&    $ 0.54$	&	$-0.08 \pm 0.02$	&	$5241 \pm  32$	&  $-4.585$	&  0.015	& $-1.7120$	& 0.0044 \\
HD188559	&   25	& 1145	&    $ 0.54$	&	$-0.11 \pm 0.04$	&	$4786 \pm 100$	&  $-4.804$	&  0.055	& $-1.7233$	& 0.0044 \\
HD19034 	&   18	& 2962	&    $ 0.51$	&	$-0.48 \pm 0.01$	&	$5477 \pm  15$	&  $-4.904$	&  0.018	& $-1.7044$	& 0.0018 \\
HD192961	&   11	& 2690	&    $ 0.64$	&	$-0.35 \pm 0.04$	&	$4624 \pm  73$	&  $-4.882$	&  0.033	& $-1.7172$	& 0.0040 \\
HD197210	&   12	& 2218	&    $ 0.75$	&	$-0.03 \pm 0.01$	&	$5577 \pm  20$	&  $-4.890$	&  0.022	& $-1.7133$	& 0.0018 \\
HD197823	&   18	& 1056	&    $ 0.53$	&	$ 0.12 \pm 0.02$	&	$5396 \pm  32$	&  $-4.738$	&  0.051	& $-1.7088$	& 0.0023 \\
HD206172	&   11	& 2550	&    $ 0.53$	&	$-0.24 \pm 0.01$	&	$5608 \pm  14$	&  $-4.860$	&  0.029	& $-1.7074$	& 0.0025 \\
HD20619 	&   26	& 2138	&    $ 0.84$	&	$-0.22 \pm 0.01$	&	$5703 \pm  13$	&  $-4.806$	&  0.032	& $-1.7069$	& 0.0033 \\
HD208272	&   28	&  527	&    $ 0.81$	&	$-0.08 \pm 0.03$	&	$5199 \pm  40$	&  $-4.489$	&  0.020	& $-1.6825$	& 0.0049 \\
HD209100	&   49	& 2949	&    $ 0.58$	&	$-0.20 \pm 0.04$	&	$4754 \pm  89$	&  $-4.782$	&  0.028	& $-1.7155$	& 0.0033 \\
HD209742	&   11	& 2911	&    $ 0.95$	&	$-0.16 \pm 0.03$	&	$5137 \pm  49$	&  $-4.825$	&  0.049	& $-1.7111$	& 0.0031 \\
HD215152	&  194	& 2761	&    $ 0.55$	&	$-0.10 \pm 0.04$	&	$4935 \pm  76$	&  $-4.870$	&  0.033	& $-1.7156$	& 0.0025 \\
HD21749 	&   47	& 2224	&    $ 0.53$	&	$-0.02 \pm 0.08$	&	$4723 \pm 143$	&  $-4.720$	&  0.042	& $-1.7211$	& 0.0046 \\
HD219249	&   26	& 2942	&    $ 0.50$	&	$-0.40 \pm 0.01$	&	$5482 \pm  13$	&  $-4.907$	&  0.017	& $-1.7077$	& 0.0010 \\
HD220339	&   10	& 2122	&    $ 0.82$	&	$-0.35 \pm 0.03$	&	$5029 \pm  52$	&  $-4.798$	&  0.049	& $-1.7011$	& 0.0037 \\
HD222237	&   18	& 1513	&    $ 0.51$	&	$-0.38 \pm 0.04$	&	$4780 \pm  64$	&  $-4.958$	&  0.037	& $-1.7044$	& 0.0027 \\
HD222595	&   26	& 1334	&    $ 0.56$	&	$ 0.01 \pm 0.01$	&	$5648 \pm  16$	&  $-4.813$	&  0.055	& $-1.7135$	& 0.0022 \\
HD224393	&   11	& 2560	&    $ 0.78$	&	$-0.38 \pm 0.01$	&	$5774 \pm  17$	&  $-4.848$	&  0.028	& $-1.7034$	& 0.0021 \\
HD224789	&   33	& 2833	&    $ 0.81$	&	$-0.03 \pm 0.02$	&	$5185 \pm  38$	&  $-4.433$	&  0.020	& $-1.6792$	& 0.0063 \\
HD23356 	&   11	& 2227	&    $ 0.65$	&	$-0.17 \pm 0.03$	&	$5004 \pm  60$	&  $-4.756$	&  0.015	& $-1.7096$	& 0.0038 \\
HD27063 	&   39	& 1575	&    $ 0.75$	&	$ 0.05 \pm 0.01$	&	$5767 \pm  14$	&  $-4.756$	&  0.018	& $-1.7111$	& 0.0014 \\
HD34688 	&   11	& 2713	&    $ 0.78$	&	$-0.20 \pm 0.02$	&	$5169 \pm  39$	&  $-4.895$	&  0.062	& $-1.7109$	& 0.0038 \\
HD40307 	&  193	& 2990	&    $ 0.50$	&	$-0.31 \pm 0.03$	&	$4977 \pm  59$	&  $-4.948$	&  0.056	& $-1.7097$	& 0.0034 \\
HD44573 	&   25	& 2353	&    $ 0.77$	&	$-0.07 \pm 0.03$	&	$5071 \pm  56$	&  $-4.591$	&  0.029	& $-1.7040$	& 0.0038 \\
HD4915  	&   39	& 1822	&    $ 0.86$	&	$-0.21 \pm 0.01$	&	$5658 \pm  13$	&  $-4.796$	&  0.037	& $-1.7036$	& 0.0025 \\
HD63765 	&   46	& 2302	&    $ 0.83$	&	$-0.16 \pm 0.01$	&	$5432 \pm  19$	&  $-4.742$	&  0.039	& $-1.7046$	& 0.0031 \\
HD65277 	&   18	& 2974	&    $ 0.62$	&	$-0.31 \pm 0.04$	&	$4802 \pm  88$	&  $-4.999$	&  0.034	& $-1.7223$	& 0.0022 \\
HD67458 	&   25	& 3158	&    $-0.68$	&	$-0.16 \pm 0.01$	&	$5891 \pm  12$	&  $-4.908$	&  0.018	& $-1.7104$	& 0.0026 \\
HD70889 	&   16	& 1579	&    $ 0.69$	&	$ 0.11 \pm 0.01$	&	$6051 \pm  15$	&  $-4.798$	&  0.033	& $-1.7145$	& 0.0026 \\
HD71835 	&   70	& 2697	&    $ 0.58$	&	$-0.04 \pm 0.02$	&	$5438 \pm  22$	&  $-4.898$	&  0.035	& $-1.7198$	& 0.0018 \\
HD7199  	&   84	& 2872	&    $-0.78$	&	$ 0.28 \pm 0.03$	&	$5386 \pm  45$	&  $-4.988$	&  0.081	& $-1.7257$	& 0.0027 \\
HD72673 	&   66	& 3025	&    $ 0.53$	&	$-0.41 \pm 0.01$	&	$5243 \pm  22$	&  $-4.920$	&  0.027	& $-1.7091$	& 0.0020 \\
HD80883 	&   13	&  526	&    $ 0.78$	&	$-0.25 \pm 0.03$	&	$5233 \pm  35$	&  $-4.670$	&  0.042	& $-1.6913$	& 0.0055 \\
HD8389A 	&   13	& 2998	&    $-0.54$	&	$ 0.34 \pm 0.05$	&	$5283 \pm  64$	&  $-5.040$	&  0.030	& $-1.7242$	& 0.0027 \\
HD85119 	&   19	&  480	&    $ 0.84$	&	$-0.20 \pm 0.02$	&	$5425 \pm  25$	&  $-4.440$	&  0.015	& $-1.6562$	& 0.0043 \\
HD85512 	&  242	& 2973	&    $ 0.82$	&	$-0.32 \pm 0.03$	&	$4715 \pm 102$	&  $-4.905$	&  0.041	& $-1.7026$	& 0.0039 \\
HD8859  	&   16	& 2914	&    $-0.52$	&	$-0.09 \pm 0.01$	&	$5502 \pm  18$	&  $-4.986$	&  0.011	& $-1.7122$	& 0.0021 \\
HD88742 	&   24	& 1868	&    $ 0.91$	&	$-0.02 \pm 0.01$	&	$5981 \pm  13$	&  $-4.699$	&  0.045	& $-1.7042$	& 0.0040 \\
HD90812 	&   17	& 2306	&    $ 0.64$	&	$-0.36 \pm 0.02$	&	$5164 \pm  35$	&  $-4.945$	&  0.039	& $-1.7124$	& 0.0025 \\
HD92719 	&   21	& 2941	&    $ 0.54$	&	$-0.10 \pm 0.01$	&	$5824 \pm  16$	&  $-4.861$	&  0.024	& $-1.7079$	& 0.0014 \\
HD95521 	&   19	& 2968	&    $ 0.79$	&	$-0.15 \pm 0.01$	&	$5773 \pm  18$	&  $-4.875$	&  0.041	& $-1.7103$	& 0.0019 \\
\hline
\end{tabular}
\end{table*}


\begin{longtable}{lccccccccccc}
\caption{\label{table4} Parameters for the 205 stars with $\lvert \rho \rvert \leq 0.5$ from the nightly averaged 271-star sample.} \\
\hline
\hline
\multicolumn{1}{l}{Star} &
\multicolumn{1}{c}{$N_{obs}$} &
\multicolumn{1}{c}{$T_{span}$} &
\multicolumn{1}{c}{$\rho$} &
\multicolumn{1}{c}{[Fe/H]} &
\multicolumn{1}{c}{$T_{\textrm{eff}}$} &
\multicolumn{1}{c}{$\langle\log R'_{HK}\rangle$} &
\multicolumn{1}{c}{$\sigma(\log R'_{HK})$} &
\multicolumn{1}{c}{$\langle\log I_{H\alpha}\rangle$} &
\multicolumn{1}{c}{$\sigma(\log I_{H\alpha})$}  \\
\multicolumn{1}{c}{} &
\multicolumn{1}{c}{} &
\multicolumn{1}{c}{[days]} &
\multicolumn{1}{c}{} &
\multicolumn{1}{c}{} &
\multicolumn{1}{c}{[K]} &
\multicolumn{1}{c}{} &
\multicolumn{1}{c}{} &
\multicolumn{1}{c}{} &
\multicolumn{1}{c}{} &
\multicolumn{1}{c}{} \\
\hline
\endfirsthead
\caption{Continued.}\\
\hline
\hline
\multicolumn{1}{l}{Star} &
\multicolumn{1}{c}{$N_{obs}$} &
\multicolumn{1}{c}{$T_{span}$} &
\multicolumn{1}{c}{$\rho$} &
\multicolumn{1}{c}{[Fe/H]} &
\multicolumn{1}{c}{$T_{\textrm{eff}}$} &
\multicolumn{1}{c}{$\langle\log R'_{HK}\rangle$} &
\multicolumn{1}{c}{$\sigma(\log R'_{HK})$} &
\multicolumn{1}{c}{$\langle\log I_{H\alpha}\rangle$} &
\multicolumn{1}{c}{$\sigma(\log I_{H\alpha})$}  \\
\multicolumn{1}{c}{} &
\multicolumn{1}{c}{} &
\multicolumn{1}{c}{[days]} &
\multicolumn{1}{c}{} &
\multicolumn{1}{c}{} &
\multicolumn{1}{c}{[K]} &
\multicolumn{1}{c}{} &
\multicolumn{1}{c}{} &
\multicolumn{1}{c}{} &
\multicolumn{1}{c}{} &
\multicolumn{1}{c}{} \\
\hline
\endhead
\hline
HD10002 	&   12	&  838	&    $ 0.08$	&	$ 0.17 \pm 0.03$	&	$5313 \pm 44$	& $-5.083$	&  0.013	& $-1.7115$	& 0.0019 \\
HD100508	&   32	& 2283	&    $-0.31$	&	$ 0.39 \pm 0.05$	&	$5449 \pm 61$	& $-5.049$	&  0.030	& $-1.7200$	& 0.0015 \\
HD10180 	&  220	& 2974	&    $-0.08$	&	$ 0.08 \pm 0.01$	&	$5911 \pm 19$	& $-5.006$	&  0.013	& $-1.7173$	& 0.0018 \\
HD102365	&   33	& 2965	&    $-0.03$	&	$-0.29 \pm 0.02$	&	$5629 \pm 29$	& $-4.944$	&  0.010	& $-1.7102$	& 0.0015 \\
HD102438	&   39	& 2243	&    $-0.22$	&	$-0.29 \pm 0.01$	&	$5560 \pm 13$	& $-4.950$	&  0.008	& $-1.7090$	& 0.0018 \\
HD104006	&   24	& 2232	&    $ 0.00$	&	$-0.78 \pm 0.02$	&	$5023 \pm 37$	& $-4.960$	&  0.011	& $-1.6881$	& 0.0016 \\
HD104067	&   86	& 2270	&    $ 0.39$	&	$-0.06 \pm 0.05$	&	$4969 \pm 72$	& $-4.742$	&  0.025	& $-1.7178$	& 0.0030 \\
HD104263	&   27	& 2244	&    $-0.13$	&	$ 0.02 \pm 0.02$	&	$5477 \pm 23$	& $-5.042$	&  0.021	& $-1.7141$	& 0.0023 \\
HD104982	&   40	& 2270	&    $-0.32$	&	$-0.19 \pm 0.01$	&	$5692 \pm 14$	& $-4.954$	&  0.010	& $-1.7101$	& 0.0019 \\
HD106116	&  106	& 2697	&    $-0.31$	&	$ 0.14 \pm 0.01$	&	$5680 \pm 15$	& $-5.023$	&  0.011	& $-1.7159$	& 0.0021 \\
HD10700 	&  230	& 3125	&    $ 0.08$	&	$-0.52 \pm 0.01$	&	$5310 \pm 17$	& $-4.959$	&  0.006	& $-1.7047$	& 0.0012 \\
HD108309	&   21	& 2676	&    $ 0.08$	&	$ 0.12 \pm 0.01$	&	$5775 \pm 14$	& $-5.019$	&  0.017	& $-1.7273$	& 0.0012 \\
HD111031	&   26	& 2244	&    $ 0.08$	&	$ 0.27 \pm 0.02$	&	$5801 \pm 22$	& $-5.067$	&  0.009	& $-1.7166$	& 0.0012 \\
HD11226 	&   28	& 1766	&    $ 0.05$	&	$ 0.04 \pm 0.01$	&	$6098 \pm 14$	& $-5.002$	&  0.005	& $-1.7187$	& 0.0022 \\
HD114853	&   36	& 3031	&    $-0.34$	&	$-0.23 \pm 0.01$	&	$5705 \pm 14$	& $-4.935$	&  0.017	& $-1.7124$	& 0.0032 \\
HD11505 	&   17	& 2944	&    $-0.02$	&	$-0.22 \pm 0.01$	&	$5752 \pm 10$	& $-5.000$	&  0.005	& $-1.7128$	& 0.0012 \\
HD115585	&   17	& 2640	&    $-0.20$	&	$ 0.35 \pm 0.02$	&	$5711 \pm 29$	& $-5.116$	&  0.013	& $-1.7251$	& 0.0014 \\
HD115617	&  142	& 2910	&    $-0.12$	&	$-0.02 \pm 0.01$	&	$5558 \pm 19$	& $-4.990$	&  0.010	& $-1.7113$	& 0.0015 \\
HD115674	&   38	& 2251	&    $ 0.10$	&	$-0.17 \pm 0.01$	&	$5649 \pm 20$	& $-4.900$	&  0.015	& $-1.7101$	& 0.0015 \\
HD117105	&   18	& 2657	&    $ 0.45$	&	$-0.29 \pm 0.01$	&	$5889 \pm 14$	& $-4.947$	&  0.006	& $-1.7113$	& 0.0018 \\
HD117207	&   16	& 2680	&    $ 0.22$	&	$ 0.22 \pm 0.02$	&	$5667 \pm 21$	& $-5.060$	&  0.004	& $-1.7157$	& 0.0013 \\
HD122862	&   17	& 2294	&    $ 0.09$	&	$-0.12 \pm 0.01$	&	$5982 \pm 13$	& $-5.015$	&  0.007	& $-1.7198$	& 0.0009 \\
HD123265	&   16	& 2641	&    $-0.02$	&	$ 0.19 \pm 0.03$	&	$5338 \pm 44$	& $-5.097$	&  0.011	& $-1.7143$	& 0.0022 \\
HD12345 	&   15	& 2472	&    $ 0.21$	&	$-0.21 \pm 0.02$	&	$5395 \pm 29$	& $-4.992$	&  0.010	& $-1.7139$	& 0.0011 \\
HD12387 	&   15	& 2838	&    $ 0.06$	&	$-0.24 \pm 0.01$	&	$5700 \pm 18$	& $-4.980$	&  0.009	& $-1.7114$	& 0.0013 \\
HD124292	&   28	& 3100	&    $-0.12$	&	$-0.13 \pm 0.02$	&	$5443 \pm 22$	& $-4.995$	&  0.011	& $-1.7120$	& 0.0029 \\
HD125881	&   24	& 3002	&    $ 0.02$	&	$ 0.06 \pm 0.01$	&	$6036 \pm 17$	& $-4.873$	&  0.024	& $-1.7137$	& 0.0015 \\
HD126525	&   48	& 2805	&    $-0.24$	&	$-0.10 \pm 0.01$	&	$5638 \pm 13$	& $-4.981$	&  0.007	& $-1.7120$	& 0.0018 \\
HD128674	&   19	& 2262	&    $ 0.41$	&	$-0.38 \pm 0.01$	&	$5551 \pm 15$	& $-4.916$	&  0.007	& $-1.7082$	& 0.0014 \\
HD129642	&   45	& 1085	&    $ 0.07$	&	$-0.06 \pm 0.04$	&	$5026 \pm 76$	& $-4.962$	&  0.019	& $-1.7126$	& 0.0018 \\
HD130930	&   14	& 1781	&    $ 0.30$	&	$ 0.01 \pm 0.03$	&	$5027 \pm 61$	& $-5.017$	&  0.015	& $-1.7077$	& 0.0010 \\
HD1320  	&   13	& 1957	&    $-0.09$	&	$-0.27 \pm 0.01$	&	$5679 \pm 14$	& $-4.874$	&  0.016	& $-1.7100$	& 0.0009 \\
HD132648	&   27	& 2623	&    $ 0.47$	&	$-0.37 \pm 0.01$	&	$5418 \pm 16$	& $-4.841$	&  0.033	& $-1.7064$	& 0.0035 \\
HD134060	&  105	& 2897	&    $-0.13$	&	$ 0.14 \pm 0.01$	&	$5966 \pm 14$	& $-5.000$	&  0.009	& $-1.7150$	& 0.0018 \\
HD134606	&  121	& 2448	&    $-0.07$	&	$ 0.27 \pm 0.02$	&	$5633 \pm 28$	& $-5.082$	&  0.008	& $-1.7150$	& 0.0016 \\
HD134664	&   27	& 1066	&    $ 0.41$	&	$ 0.10 \pm 0.01$	&	$5865 \pm 19$	& $-4.881$	&  0.027	& $-1.7178$	& 0.0016 \\
HD136352	&  148	& 2809	&    $-0.23$	&	$-0.34 \pm 0.01$	&	$5664 \pm 14$	& $-4.949$	&  0.005	& $-1.7080$	& 0.0017 \\
HD136713	&   41	& 2202	&    $-0.33$	&	$ 0.07 \pm 0.05$	&	$4994 \pm 74$	& $-4.795$	&  0.038	& $-1.7220$	& 0.0022 \\
HD136894	&   37	& 2044	&    $-0.28$	&	$-0.10 \pm 0.02$	&	$5412 \pm 22$	& $-4.995$	&  0.005	& $-1.7051$	& 0.0016 \\
HD13724 	&   26	& 2134	&    $ 0.01$	&	$ 0.23 \pm 0.02$	&	$5868 \pm 27$	& $-4.760$	&  0.026	& $-1.7172$	& 0.0025 \\
HD137388	&   30	& 2148	&    $ 0.36$	&	$ 0.18 \pm 0.03$	&	$5240 \pm 53$	& $-4.894$	&  0.049	& $-1.7257$	& 0.0024 \\
HD138549	&   22	& 2621	&    $ 0.18$	&	$ 0.00 \pm 0.01$	&	$5582 \pm 19$	& $-4.828$	&  0.044	& $-1.7152$	& 0.0021 \\
HD1388  	&   64	& 3025	&    $ 0.09$	&	$-0.01 \pm 0.01$	&	$5954 \pm 10$	& $-4.979$	&  0.007	& $-1.7140$	& 0.0016 \\
HD142709	&   13	& 2523	&    $ 0.18$	&	$-0.35 \pm 0.03$	&	$4728 \pm 65$	& $-4.999$	&  0.051	& $-1.7155$	& 0.0022 \\
HD143114	&   19	& 1789	&    $-0.19$	&	$-0.41 \pm 0.01$	&	$5775 \pm 18$	& $-4.946$	&  0.005	& $-1.7088$	& 0.0012 \\
HD144628	&   51	& 2105	&    $ 0.28$	&	$-0.41 \pm 0.02$	&	$5085 \pm 34$	& $-4.952$	&  0.022	& $-1.7114$	& 0.0020 \\
HD145598	&   32	& 2107	&    $ 0.30$	&	$-0.78 \pm 0.02$	&	$5417 \pm 21$	& $-4.916$	&  0.011	& $-1.7010$	& 0.0018 \\
HD1461  	&  193	& 3027	&    $-0.06$	&	$ 0.19 \pm 0.01$	&	$5765 \pm 18$	& $-5.020$	&  0.013	& $-1.7128$	& 0.0014 \\
HD146233	&   51	& 2602	&    $-0.20$	&	$ 0.04 \pm 0.01$	&	$5818 \pm 13$	& $-4.928$	&  0.025	& $-1.7142$	& 0.0014 \\
HD14747 	&   14	& 2860	&    $-0.06$	&	$-0.39 \pm 0.01$	&	$5516 \pm 16$	& $-4.945$	&  0.018	& $-1.7070$	& 0.0020 \\
HD147512	&   29	& 1734	&    $ 0.08$	&	$-0.08 \pm 0.01$	&	$5530 \pm 15$	& $-4.990$	&  0.006	& $-1.7121$	& 0.0014 \\
HD150433	&   58	& 2223	&    $-0.05$	&	$-0.36 \pm 0.01$	&	$5665 \pm 12$	& $-4.961$	&  0.005	& $-1.7041$	& 0.0016 \\
HD151504	&   14	& 1897	&    $ 0.01$	&	$ 0.06 \pm 0.02$	&	$5457 \pm 31$	& $-5.038$	&  0.004	& $-1.7139$	& 0.0019 \\
HD15337 	&   29	& 1975	&    $-0.35$	&	$ 0.06 \pm 0.03$	&	$5179 \pm 44$	& $-4.916$	&  0.038	& $-1.7214$	& 0.0017 \\
HD154088	&  124	& 2014	&    $ 0.04$	&	$ 0.28 \pm 0.03$	&	$5374 \pm 43$	& $-5.064$	&  0.015	& $-1.7129$	& 0.0020 \\
HD154363	&   19	& 2529	&    $-0.24$	&	$-0.62 \pm 0.04$	&	$4723 \pm 89$	& $-4.820$	&  0.050	& $-1.6810$	& 0.0043 \\
HD157172	&   82	& 2266	&    $ 0.06$	&	$ 0.11 \pm 0.02$	&	$5451 \pm 27$	& $-4.996$	&  0.036	& $-1.7182$	& 0.0020 \\
HD157338	&   24	& 1460	&    $ 0.04$	&	$-0.08 \pm 0.01$	&	$6027 \pm 13$	& $-4.969$	&  0.010	& $-1.7154$	& 0.0017 \\
HD157347	&   20	& 2892	&    $ 0.26$	&	$ 0.02 \pm 0.01$	&	$5676 \pm 16$	& $-5.014$	&  0.006	& $-1.7132$	& 0.0017 \\
HD1581  	&  130	& 2624	&    $ 0.12$	&	$-0.18 \pm 0.01$	&	$5977 \pm 12$	& $-4.936$	&  0.007	& $-1.7095$	& 0.0009 \\
HD161098	&   75	& 2015	&    $ 0.42$	&	$-0.27 \pm 0.01$	&	$5560 \pm 15$	& $-4.911$	&  0.021	& $-1.7099$	& 0.0017 \\
HD161612	&   31	& 2154	&    $ 0.09$	&	$ 0.16 \pm 0.02$	&	$5616 \pm 22$	& $-5.032$	&  0.006	& $-1.7182$	& 0.0015 \\
HD162396	&   39	& 1884	&    $-0.17$	&	$-0.35 \pm 0.01$	&	$6090 \pm 19$	& $-4.973$	&  0.010	& $-1.7114$	& 0.0014 \\
HD165920	&   18	& 2326	&    $ 0.04$	&	$ 0.29 \pm 0.04$	&	$5339 \pm 55$	& $-5.085$	&  0.014	& $-1.7191$	& 0.0021 \\
HD166724	&   19	& 2567	&    $ 0.46$	&	$-0.09 \pm 0.03$	&	$5127 \pm 52$	& $-4.734$	&  0.026	& $-1.7077$	& 0.0035 \\
HD16714 	&   24	& 2089	&    $ 0.22$	&	$-0.20 \pm 0.01$	&	$5518 \pm 18$	& $-4.965$	&  0.008	& $-1.7133$	& 0.0018 \\
HD168871	&   25	& 2951	&    $-0.01$	&	$-0.09 \pm 0.01$	&	$5983 \pm 13$	& $-4.980$	&  0.009	& $-1.7149$	& 0.0018 \\
HD170493	&   12	& 2101	&    $-0.17$	&	$ 0.14 \pm 0.11$	&	$4751 \pm 08$	& $-4.814$	&  0.061	& $-1.7216$	& 0.0029 \\
HD171665	&   12	& 1730	&    $ 0.34$	&	$-0.05 \pm 0.01$	&	$5655 \pm 12$	& $-4.906$	&  0.017	& $-1.7146$	& 0.0018 \\
HD174545	&   13	& 2574	&    $-0.23$	&	$ 0.22 \pm 0.04$	&	$5216 \pm 57$	& $-4.929$	&  0.041	& $-1.7198$	& 0.0017 \\
HD176986	&   79	& 2625	&    $ 0.06$	&	$ 0.00 \pm 0.03$	&	$5018 \pm 59$	& $-4.835$	&  0.024	& $-1.7201$	& 0.0026 \\
HD177409	&   16	& 1979	&    $ 0.37$	&	$-0.04 \pm 0.01$	&	$5898 \pm 10$	& $-4.863$	&  0.028	& $-1.7083$	& 0.0012 \\
HD177565	&   26	& 1684	&    $ 0.33$	&	$ 0.08 \pm 0.01$	&	$5627 \pm 19$	& $-4.939$	&  0.041	& $-1.7159$	& 0.0026 \\
HD177758	&   10	& 2834	&    $ 0.39$	&	$-0.58 \pm 0.02$	&	$5862 \pm 23$	& $-4.929$	&  0.003	& $-1.7049$	& 0.0015 \\
HD17970 	&   19	& 2962	&    $ 0.07$	&	$-0.45 \pm 0.04$	&	$5040 \pm 48$	& $-5.008$	&  0.012	& $-1.7015$	& 0.0023 \\
HD180409	&   11	& 2834	&    $-0.12$	&	$-0.17 \pm 0.01$	&	$6013 \pm 18$	& $-4.925$	&  0.004	& $-1.7103$	& 0.0015 \\
HD181433	&  123	& 3011	&    $ 0.07$	&	$ 0.33 \pm 0.13$	&	$4962 \pm 34$	& $-5.144$	&  0.014	& $-1.7130$	& 0.0023 \\
HD183658	&   16	& 2171	&    $ 0.14$	&	$ 0.03 \pm 0.01$	&	$5803 \pm 17$	& $-4.987$	&  0.008	& $-1.7128$	& 0.0022 \\
HD183783	&   11	& 2676	&    $-0.09$	&	$-0.20 \pm 0.07$	&	$4595 \pm 73$	& $-4.907$	&  0.049	& $-1.7163$	& 0.0020 \\
HD185615	&   19	& 2297	&    $-0.10$	&	$ 0.08 \pm 0.02$	&	$5570 \pm 20$	& $-5.043$	&  0.014	& $-1.7156$	& 0.0016 \\
HD188748	&   20	& 2133	&    $-0.26$	&	$-0.12 \pm 0.01$	&	$5623 \pm 17$	& $-4.967$	&  0.013	& $-1.7088$	& 0.0018 \\
HD189567	&  174	& 2941	&    $ 0.41$	&	$-0.24 \pm 0.01$	&	$5726 \pm 15$	& $-4.916$	&  0.016	& $-1.7142$	& 0.0014 \\
HD189625	&   16	& 1743	&    $ 0.42$	&	$ 0.18 \pm 0.02$	&	$5846 \pm 22$	& $-4.810$	&  0.041	& $-1.7133$	& 0.0025 \\
HD190248	&  136	& 2942	&    $ 0.02$	&	$ 0.33 \pm 0.03$	&	$5604 \pm 38$	& $-5.095$	&  0.010	& $-1.7168$	& 0.0010 \\
HD190954	&   11	& 2204	&    $-0.29$	&	$-0.41 \pm 0.02$	&	$5430 \pm 24$	& $-4.969$	&  0.012	& $-1.7070$	& 0.0022 \\
HD192031	&   11	&  519	&    $ 0.45$	&	$-0.84 \pm 0.01$	&	$5215 \pm 16$	& $-4.954$	&  0.006	& $-1.7030$	& 0.0019 \\
HD192310	&  206	& 3082	&    $ 0.14$	&	$-0.04 \pm 0.03$	&	$5166 \pm 49$	& $-4.991$	&  0.040	& $-1.7136$	& 0.0017 \\
HD193193	&   30	& 1134	&    $-0.09$	&	$-0.05 \pm 0.01$	&	$5979 \pm 13$	& $-4.933$	&  0.016	& $-1.7158$	& 0.0018 \\
HD19467 	&   21	& 2941	&    $ 0.15$	&	$-0.14 \pm 0.01$	&	$5720 \pm 10$	& $-5.002$	&  0.015	& $-1.7131$	& 0.0019 \\
HD196761	&   10	& 2822	&    $ 0.17$	&	$-0.31 \pm 0.01$	&	$5415 \pm 16$	& $-4.918$	&  0.025	& $-1.7102$	& 0.0026 \\
HD199190	&   23	&  762	&    $ 0.23$	&	$ 0.15 \pm 0.01$	&	$5926 \pm 17$	& $-5.052$	&  0.014	& $-1.7198$	& 0.0014 \\
HD199288	&   16	& 2602	&    $ 0.02$	&	$-0.63 \pm 0.01$	&	$5765 \pm 19$	& $-4.895$	&  0.005	& $-1.7042$	& 0.0013 \\
HD199960	&   28	&  854	&    $-0.37$	&	$ 0.28 \pm 0.02$	&	$5973 \pm 26$	& $-5.012$	&  0.019	& $-1.7217$	& 0.0013 \\
HD20003 	&  104	& 2280	&    $-0.25$	&	$ 0.04 \pm 0.02$	&	$5494 \pm 27$	& $-4.988$	&  0.040	& $-1.7203$	& 0.0014 \\
HD203432	&   33	& 1405	&    $ 0.19$	&	$ 0.29 \pm 0.02$	&	$5645 \pm 25$	& $-4.858$	&  0.057	& $-1.7201$	& 0.0022 \\
HD20407 	&   20	& 3062	&    $ 0.04$	&	$-0.44 \pm 0.01$	&	$5866 \pm 14$	& $-4.899$	&  0.007	& $-1.7078$	& 0.0010 \\
HD204313	&   70	& 2026	&    $ 0.03$	&	$ 0.18 \pm 0.02$	&	$5776 \pm 22$	& $-5.019$	&  0.017	& $-1.7202$	& 0.0015 \\
HD204385	&   13	& 2889	&    $-0.13$	&	$ 0.07 \pm 0.01$	&	$6033 \pm 16$	& $-4.976$	&  0.009	& $-1.7164$	& 0.0014 \\
HD204941	&   38	& 2546	&    $ 0.31$	&	$-0.19 \pm 0.03$	&	$5056 \pm 52$	& $-4.952$	&  0.030	& $-1.7146$	& 0.0020 \\
HD205536	&   22	& 2218	&    $ 0.33$	&	$-0.05 \pm 0.02$	&	$5442 \pm 23$	& $-5.016$	&  0.006	& $-1.7152$	& 0.0020 \\
HD207129	&   79	& 1875	&    $ 0.32$	&	$ 0.00 \pm 0.01$	&	$5937 \pm 13$	& $-4.903$	&  0.030	& $-1.7112$	& 0.0016 \\
HD207700	&   13	& 2111	&    $ 0.30$	&	$ 0.04 \pm 0.01$	&	$5666 \pm 18$	& $-5.006$	&  0.010	& $-1.7207$	& 0.0015 \\
HD20781 	&  124	& 2966	&    $-0.14$	&	$-0.11 \pm 0.02$	&	$5256 \pm 29$	& $-5.035$	&  0.011	& $-1.7171$	& 0.0022 \\
HD20782 	&   55	& 2983	&    $ 0.30$	&	$-0.06 \pm 0.01$	&	$5774 \pm 14$	& $-4.919$	&  0.015	& $-1.7149$	& 0.0019 \\
HD20794 	&  279	& 3033	&    $-0.12$	&	$-0.40 \pm 0.01$	&	$5401 \pm 17$	& $-4.981$	&  0.006	& $-1.7034$	& 0.0018 \\
HD207970	&   12	& 2955	&    $-0.44$	&	$ 0.07 \pm 0.02$	&	$5556 \pm 25$	& $-5.028$	&  0.010	& $-1.7148$	& 0.0016 \\
HD20807 	&   39	& 2308	&    $ 0.12$	&	$-0.23 \pm 0.01$	&	$5866 \pm 11$	& $-4.881$	&  0.013	& $-1.7082$	& 0.0017 \\
HD208704	&   12	& 2638	&    $ 0.02$	&	$-0.09 \pm 0.01$	&	$5826 \pm 11$	& $-4.957$	&  0.011	& $-1.7194$	& 0.0019 \\
HD210752	&   14	& 2534	&    $-0.03$	&	$-0.57 \pm 0.01$	&	$5923 \pm 23$	& $-4.874$	&  0.004	& $-1.7045$	& 0.0018 \\
HD210918	&   31	& 1857	&    $-0.25$	&	$-0.09 \pm 0.01$	&	$5755 \pm 12$	& $-5.002$	&  0.015	& $-1.7157$	& 0.0013 \\
HD211415	&   13	& 2997	&    $ 0.24$	&	$-0.21 \pm 0.01$	&	$5850 \pm 14$	& $-4.919$	&  0.023	& $-1.7117$	& 0.0012 \\
HD21209A	&   12	& 3022	&    $-0.05$	&	$-0.41 \pm 0.04$	&	$4671 \pm 65$	& $-4.840$	&  0.022	& $-1.7082$	& 0.0024 \\
HD212708	&   30	& 1058	&    $-0.34$	&	$ 0.27 \pm 0.02$	&	$5681 \pm 27$	& $-5.076$	&  0.014	& $-1.7187$	& 0.0020 \\
HD213628	&   12	& 2608	&    $ 0.26$	&	$ 0.01 \pm 0.01$	&	$5555 \pm 20$	& $-4.957$	&  0.010	& $-1.7145$	& 0.0025 \\
HD213941	&   26	& 2217	&    $ 0.35$	&	$-0.46 \pm 0.01$	&	$5532 \pm 18$	& $-4.909$	&  0.013	& $-1.7095$	& 0.0023 \\
HD214385	&   11	& 2911	&    $-0.04$	&	$-0.34 \pm 0.01$	&	$5654 \pm 15$	& $-4.924$	&  0.016	& $-1.7094$	& 0.0016 \\
HD21693 	&  141	& 2951	&    $ 0.02$	&	$ 0.00 \pm 0.02$	&	$5430 \pm 26$	& $-4.909$	&  0.055	& $-1.7161$	& 0.0021 \\
HD21938 	&   18	& 3019	&    $ 0.10$	&	$-0.47 \pm 0.01$	&	$5778 \pm 18$	& $-4.939$	&  0.007	& $-1.7076$	& 0.0021 \\
HD220256	&   22	& 1217	&    $-0.08$	&	$-0.10 \pm 0.03$	&	$5144 \pm 48$	& $-5.022$	&  0.016	& $-1.7088$	& 0.0013 \\
HD220507	&   48	& 2220	&    $ 0.02$	&	$ 0.01 \pm 0.01$	&	$5698 \pm 17$	& $-5.052$	&  0.010	& $-1.7186$	& 0.0019 \\
HD221356	&   23	& 2941	&    $-0.29$	&	$-0.20 \pm 0.03$	&	$6112 \pm 37$	& $-4.919$	&  0.004	& $-1.7062$	& 0.0010 \\
HD222669	&   46	& 1403	&    $ 0.13$	&	$ 0.05 \pm 0.01$	&	$5894 \pm 17$	& $-4.863$	&  0.022	& $-1.7118$	& 0.0022 \\
HD224619	&   15	& 2992	&    $-0.10$	&	$-0.20 \pm 0.01$	&	$5436 \pm 16$	& $-4.975$	&  0.013	& $-1.7148$	& 0.0012 \\
HD22879 	&   50	& 2686	&    $ 0.05$	&	$-0.83 \pm 0.02$	&	$5857 \pm 27$	& $-4.908$	&  0.007	& $-1.6987$	& 0.0015 \\
HD23456 	&   20	& 2200	&    $ 0.10$	&	$-0.32 \pm 0.01$	&	$6178 \pm 18$	& $-4.909$	&  0.010	& $-1.7079$	& 0.0019 \\
HD26965A	&   24	& 2998	&    $ 0.35$	&	$-0.31 \pm 0.03$	&	$5153 \pm 38$	& $-4.944$	&  0.034	& $-1.7040$	& 0.0027 \\
HD283   	&   11	& 2592	&    $-0.46$	&	$-0.54 \pm 0.02$	&	$5157 \pm 28$	& $-4.949$	&  0.011	& $-1.7085$	& 0.0018 \\
HD28471 	&   17	& 2683	&    $ 0.04$	&	$-0.05 \pm 0.01$	&	$5745 \pm 14$	& $-4.991$	&  0.021	& $-1.7140$	& 0.0018 \\
HD28701 	&   17	& 1263	&    $-0.08$	&	$-0.32 \pm 0.01$	&	$5710 \pm 12$	& $-4.986$	&  0.012	& $-1.7114$	& 0.0011 \\
HD28821 	&   18	& 2904	&    $ 0.08$	&	$-0.12 \pm 0.01$	&	$5660 \pm 13$	& $-4.975$	&  0.013	& $-1.7151$	& 0.0025 \\
HD30278 	&   21	& 1513	&    $-0.05$	&	$-0.17 \pm 0.02$	&	$5394 \pm 29$	& $-5.006$	&  0.013	& $-1.7124$	& 0.0019 \\
HD30306 	&   21	& 2904	&    $ 0.43$	&	$ 0.17 \pm 0.02$	&	$5529 \pm 26$	& $-5.074$	&  0.013	& $-1.7156$	& 0.0020 \\
HD31527 	&  182	& 3011	&    $ 0.01$	&	$-0.17 \pm 0.01$	&	$5898 \pm 13$	& $-4.955$	&  0.006	& $-1.7131$	& 0.0018 \\
HD31822 	&   43	& 2352	&    $ 0.03$	&	$-0.19 \pm 0.01$	&	$6042 \pm 16$	& $-4.865$	&  0.007	& $-1.7113$	& 0.0018 \\
HD32724 	&   21	& 2954	&    $ 0.02$	&	$-0.17 \pm 0.01$	&	$5818 \pm 13$	& $-5.032$	&  0.013	& $-1.7150$	& 0.0011 \\
HD33725 	&   17	& 3011	&    $-0.10$	&	$-0.17 \pm 0.02$	&	$5274 \pm 30$	& $-4.972$	&  0.035	& $-1.7154$	& 0.0015 \\
HD34449 	&   13	& 2626	&    $ 0.12$	&	$-0.09 \pm 0.01$	&	$5848 \pm 17$	& $-4.883$	&  0.013	& $-1.7102$	& 0.0017 \\
HD35854 	&   17	& 2958	&    $ 0.47$	&	$-0.13 \pm 0.03$	&	$4928 \pm 56$	& $-4.799$	&  0.037	& $-1.7110$	& 0.0038 \\
HD36003 	&   57	& 1494	&    $-0.08$	&	$-0.20 \pm 0.06$	&	$4647 \pm 88$	& $-4.872$	&  0.040	& $-1.7080$	& 0.0032 \\
HD36108 	&   23	& 3275	&    $ 0.25$	&	$-0.21 \pm 0.01$	&	$5916 \pm 12$	& $-4.992$	&  0.006	& $-1.7156$	& 0.0030 \\
HD36379 	&   45	& 2341	&    $-0.22$	&	$-0.17 \pm 0.01$	&	$6030 \pm 14$	& $-4.976$	&  0.006	& $-1.7174$	& 0.0018 \\
HD37986 	&   19	& 2975	&    $-0.24$	&	$ 0.26 \pm 0.03$	&	$5507 \pm 38$	& $-5.082$	&  0.013	& $-1.7245$	& 0.0021 \\
HD3823  	&   33	& 2265	&    $-0.00$	&	$-0.28 \pm 0.01$	&	$6022 \pm 14$	& $-4.988$	&  0.007	& $-1.7137$	& 0.0010 \\
HD38277 	&   10	& 3019	&    $-0.12$	&	$-0.07 \pm 0.01$	&	$5871 \pm 10$	& $-5.019$	&  0.007	& $-1.7228$	& 0.0013 \\
HD38858 	&   66	& 3009	&    $-0.01$	&	$-0.22 \pm 0.01$	&	$5733 \pm 12$	& $-4.918$	&  0.013	& $-1.7102$	& 0.0015 \\
HD38973 	&   22	& 2353	&    $-0.07$	&	$ 0.05 \pm 0.01$	&	$6016 \pm 17$	& $-4.972$	&  0.013	& $-1.7138$	& 0.0010 \\
HD39194 	&  156	& 3008	&    $ 0.18$	&	$-0.61 \pm 0.02$	&	$5205 \pm 23$	& $-4.951$	&  0.014	& $-1.7003$	& 0.0015 \\
HD40397 	&   21	& 3060	&    $ 0.42$	&	$-0.13 \pm 0.01$	&	$5527 \pm 20$	& $-5.013$	&  0.009	& $-1.7104$	& 0.0014 \\
HD44120 	&   18	& 3019	&    $ 0.33$	&	$ 0.12 \pm 0.01$	&	$6052 \pm 15$	& $-5.070$	&  0.017	& $-1.7206$	& 0.0010 \\
HD44420 	&   17	& 2898	&    $ 0.14$	&	$ 0.29 \pm 0.02$	&	$5818 \pm 22$	& $-5.036$	&  0.024	& $-1.7188$	& 0.0012 \\
HD44447 	&   23	& 2989	&    $ 0.23$	&	$-0.22 \pm 0.01$	&	$5999 \pm 14$	& $-4.977$	&  0.015	& $-1.7117$	& 0.0015 \\
HD44594 	&   21	& 2903	&    $-0.01$	&	$ 0.15 \pm 0.01$	&	$5840 \pm 14$	& $-5.004$	&  0.020	& $-1.7175$	& 0.0024 \\
HD45184 	&  102	& 3013	&    $ 0.24$	&	$ 0.04 \pm 0.01$	&	$5869 \pm 14$	& $-4.905$	&  0.026	& $-1.7126$	& 0.0013 \\
HD45289 	&   16	& 3023	&    $ 0.44$	&	$-0.02 \pm 0.01$	&	$5717 \pm 18$	& $-5.033$	&  0.008	& $-1.7169$	& 0.0024 \\
HD45364 	&   62	& 2917	&    $-0.16$	&	$-0.17 \pm 0.01$	&	$5434 \pm 20$	& $-4.959$	&  0.022	& $-1.7092$	& 0.0021 \\
HD47186 	&  104	& 2879	&    $-0.28$	&	$ 0.23 \pm 0.02$	&	$5675 \pm 21$	& $-5.051$	&  0.009	& $-1.7131$	& 0.0014 \\
HD50590 	&   12	& 2193	&    $-0.05$	&	$-0.22 \pm 0.04$	&	$4870 \pm 67$	& $-4.974$	&  0.032	& $-1.7150$	& 0.0021 \\
HD51608 	&  126	& 2966	&    $ 0.12$	&	$-0.07 \pm 0.01$	&	$5358 \pm 22$	& $-4.982$	&  0.020	& $-1.7141$	& 0.0023 \\
HD55693 	&   27	& 3278	&    $-0.43$	&	$ 0.29 \pm 0.02$	&	$5914 \pm 26$	& $-4.999$	&  0.020	& $-1.7221$	& 0.0028 \\
HD59468 	&  141	& 2754	&    $-0.17$	&	$ 0.03 \pm 0.01$	&	$5618 \pm 20$	& $-4.996$	&  0.012	& $-1.7101$	& 0.0010 \\
HD59711A	&   16	& 2911	&    $ 0.22$	&	$-0.12 \pm 0.01$	&	$5722 \pm 13$	& $-4.946$	&  0.010	& $-1.7107$	& 0.0013 \\
HD65562 	&   15	& 2945	&    $ 0.43$	&	$-0.10 \pm 0.03$	&	$5076 \pm 47$	& $-4.954$	&  0.035	& $-1.7069$	& 0.0018 \\
HD65907A	&   61	& 2608	&    $-0.30$	&	$-0.31 \pm 0.01$	&	$5945 \pm 16$	& $-4.914$	&  0.010	& $-1.7059$	& 0.0013 \\
HD66221 	&   17	& 2350	&    $-0.04$	&	$ 0.17 \pm 0.02$	&	$5635 \pm 25$	& $-5.058$	&  0.020	& $-1.7158$	& 0.0028 \\
HD6735  	&   17	& 2960	&    $ 0.26$	&	$-0.06 \pm 0.01$	&	$6082 \pm 15$	& $-4.877$	&  0.014	& $-1.7141$	& 0.0012 \\
HD68607 	&   29	& 1209	&    $ 0.26$	&	$ 0.07 \pm 0.03$	&	$5215 \pm 45$	& $-4.728$	&  0.036	& $-1.7166$	& 0.0024 \\
HD68978A	&   60	& 2339	&    $-0.03$	&	$ 0.04 \pm 0.02$	&	$5965 \pm 22$	& $-4.879$	&  0.015	& $-1.7169$	& 0.0018 \\
HD69655 	&   21	& 2905	&    $ 0.06$	&	$-0.18 \pm 0.01$	&	$5961 \pm 12$	& $-4.943$	&  0.009	& $-1.7112$	& 0.0014 \\
HD71334 	&   21	& 2973	&    $-0.29$	&	$-0.09 \pm 0.01$	&	$5694 \pm 13$	& $-4.987$	&  0.010	& $-1.7076$	& 0.0016 \\
HD7134  	&   16	& 1767	&    $ 0.09$	&	$-0.29 \pm 0.01$	&	$5940 \pm 14$	& $-4.949$	&  0.005	& $-1.7142$	& 0.0015 \\
HD71479 	&   21	& 2974	&    $-0.14$	&	$ 0.24 \pm 0.01$	&	$6026 \pm 18$	& $-5.040$	&  0.015	& $-1.7264$	& 0.0020 \\
HD72579 	&   23	& 2974	&    $-0.35$	&	$ 0.20 \pm 0.02$	&	$5449 \pm 30$	& $-5.087$	&  0.009	& $-1.7187$	& 0.0018 \\
HD72769 	&   21	& 2703	&    $-0.07$	&	$ 0.30 \pm 0.02$	&	$5640 \pm 27$	& $-5.090$	&  0.015	& $-1.7183$	& 0.0022 \\
HD73121 	&   19	& 2708	&    $ 0.07$	&	$ 0.09 \pm 0.01$	&	$6091 \pm 16$	& $-5.061$	&  0.013	& $-1.7211$	& 0.0021 \\
HD73524 	&   64	& 2975	&    $-0.12$	&	$ 0.16 \pm 0.01$	&	$6017 \pm 13$	& $-5.002$	&  0.017	& $-1.7137$	& 0.0014 \\
HD74014 	&   17	& 2963	&    $-0.29$	&	$ 0.22 \pm 0.02$	&	$5561 \pm 27$	& $-5.072$	&  0.010	& $-1.7189$	& 0.0008 \\
HD7449  	&   84	& 2926	&    $ 0.38$	&	$-0.11 \pm 0.01$	&	$6024 \pm 13$	& $-4.850$	&  0.015	& $-1.7089$	& 0.0015 \\
HD78429 	&   57	& 1960	&    $-0.24$	&	$ 0.09 \pm 0.01$	&	$5760 \pm 19$	& $-4.927$	&  0.029	& $-1.7194$	& 0.0023 \\
HD78558 	&   20	& 2969	&    $ 0.04$	&	$-0.44 \pm 0.01$	&	$5711 \pm 18$	& $-4.974$	&  0.009	& $-1.7108$	& 0.0023 \\
HD78612 	&   22	& 2968	&    $ 0.37$	&	$-0.24 \pm 0.01$	&	$5834 \pm 14$	& $-5.005$	&  0.017	& $-1.7155$	& 0.0025 \\
HD78747 	&   42	& 2553	&    $-0.07$	&	$-0.67 \pm 0.01$	&	$5778 \pm 18$	& $-4.921$	&  0.008	& $-1.7019$	& 0.0008 \\
HD81639 	&   18	& 2706	&    $ 0.03$	&	$-0.17 \pm 0.02$	&	$5522 \pm 20$	& $-4.990$	&  0.013	& $-1.7143$	& 0.0019 \\
HD82342 	&   27	& 3008	&    $ 0.01$	&	$-0.54 \pm 0.03$	&	$4820 \pm 61$	& $-4.943$	&  0.032	& $-1.6983$	& 0.0030 \\
HD82516 	&   53	& 2319	&    $-0.04$	&	$ 0.01 \pm 0.04$	&	$5104 \pm 60$	& $-4.955$	&  0.044	& $-1.7299$	& 0.0026 \\
HD83529 	&   22	& 2706	&    $ 0.40$	&	$-0.22 \pm 0.01$	&	$5902 \pm 12$	& $-4.970$	&  0.009	& $-1.7133$	& 0.0020 \\
HD8406  	&   14	& 3002	&    $ 0.12$	&	$-0.10 \pm 0.01$	&	$5726 \pm 12$	& $-4.856$	&  0.009	& $-1.7113$	& 0.0010 \\
HD85390 	&   63	& 2965	&    $ 0.06$	&	$-0.07 \pm 0.03$	&	$5186 \pm 54$	& $-4.959$	&  0.026	& $-1.7145$	& 0.0022 \\
HD86140 	&   10	& 2699	&    $ 0.40$	&	$-0.25 \pm 0.04$	&	$4903 \pm 59$	& $-4.806$	&  0.020	& $-1.7042$	& 0.0017 \\
HD8638  	&   33	& 1826	&    $-0.15$	&	$-0.38 \pm 0.02$	&	$5507 \pm 26$	& $-4.953$	&  0.006	& $-1.7034$	& 0.0016 \\
HD88084 	&   18	& 2692	&    $ 0.20$	&	$-0.10 \pm 0.01$	&	$5766 \pm 11$	& $-4.973$	&  0.009	& $-1.7124$	& 0.0016 \\
HD8828  	&   45	& 2223	&    $ 0.38$	&	$-0.16 \pm 0.02$	&	$5403 \pm 25$	& $-4.996$	&  0.010	& $-1.7080$	& 0.0020 \\
HD89454 	&   48	& 1467	&    $ 0.46$	&	$ 0.12 \pm 0.01$	&	$5728 \pm 17$	& $-4.701$	&  0.029	& $-1.7182$	& 0.0024 \\
HD90156 	&   83	& 2937	&    $ 0.01$	&	$-0.24 \pm 0.01$	&	$5599 \pm 12$	& $-4.947$	&  0.006	& $-1.7066$	& 0.0016 \\
HD90711 	&   17	& 2305	&    $ 0.38$	&	$ 0.24 \pm 0.03$	&	$5444 \pm 39$	& $-5.004$	&  0.034	& $-1.7215$	& 0.0021 \\
HD93385 	&  136	& 2908	&    $-0.07$	&	$ 0.02 \pm 0.01$	&	$5977 \pm 18$	& $-4.988$	&  0.007	& $-1.7133$	& 0.0019 \\
HD94151 	&   21	& 2724	&    $ 0.02$	&	$ 0.04 \pm 0.01$	&	$5583 \pm 19$	& $-4.974$	&  0.031	& $-1.7168$	& 0.0015 \\
HD95456 	&   77	& 2338	&    $-0.08$	&	$ 0.16 \pm 0.02$	&	$6276 \pm 22$	& $-4.982$	&  0.019	& $-1.7208$	& 0.0014 \\
HD96423 	&   22	& 2722	&    $ 0.05$	&	$ 0.10 \pm 0.01$	&	$5711 \pm 18$	& $-5.035$	&  0.013	& $-1.7142$	& 0.0021 \\
HD96700 	&  168	& 3270	&    $-0.22$	&	$-0.18 \pm 0.01$	&	$5845 \pm 13$	& $-4.948$	&  0.011	& $-1.7148$	& 0.0017 \\
HD97037 	&   18	& 2668	&    $ 0.35$	&	$-0.07 \pm 0.01$	&	$5883 \pm 14$	& $-4.998$	&  0.009	& $-1.7155$	& 0.0013 \\
HD97343 	&   21	& 2695	&    $ 0.03$	&	$-0.06 \pm 0.01$	&	$5410 \pm 20$	& $-5.015$	&  0.009	& $-1.7132$	& 0.0019 \\
HD9782  	&   27	& 2216	&    $-0.35$	&	$ 0.09 \pm 0.01$	&	$6023 \pm 19$	& $-4.974$	&  0.007	& $-1.7146$	& 0.0011 \\
HD9796  	&   15	& 2573	&    $ 0.42$	&	$-0.25 \pm 0.02$	&	$5179 \pm 28$	& $-4.874$	&  0.025	& $-1.6902$	& 0.0017 \\
HD97998 	&   17	& 2351	&    $ 0.28$	&	$-0.42 \pm 0.01$	&	$5716 \pm 21$	& $-4.902$	&  0.007	& $-1.7085$	& 0.0016 \\
HD98281 	&   54	& 2226	&    $ 0.47$	&	$-0.26 \pm 0.02$	&	$5381 \pm 23$	& $-4.887$	&  0.027	& $-1.7071$	& 0.0023 \\
\hline
\end{longtable}

\listofobjects

\end{document}